\documentclass[usenatbib]{mnras}
\usepackage{multirow}
\usepackage{rotating}
\usepackage{colortbl}
\usepackage{xcolor}
\usepackage[fleqn]{amsmath}
\usepackage{amssymb}
\usepackage{amsfonts}
\usepackage{verbatim}
\usepackage{scalefnt}
\usepackage[percent]{overpic}

\usepackage{etoolbox}
\makeatletter
\patchcmd\@combinedblfloats{\box\@outputbox}{\unvbox\@outputbox}{}{%
  \errmessage{\noexpand\@combinedblfloats could not be patched}%
}%
\makeatother

\newcommand{\comments}[1]{} 
\newcommand\T{\rule{0pt}{2.6ex}}       
\newcommand\B{\rule[-1.2ex]{0pt}{0pt}} 

\newcommand{\krome}{\textsc{krome} }
\newcommand{\kromes}{\textsc{krome}}
\newcommand{\gasoline}{\textsc{gasoline{2}} }
\newcommand{\gasolines}{\textsc{gasoline{2}}}

\newcommand{\mC}{\mathrm C}

\newcommand{\mH}{\mathrm H}

\newcommand{\me}{\mathrm e}
\newcommand{\mSi}{\mathrm Si}
\newcommand{\mO}{\mathrm O}

\newcommand{\arl}[1]{\url{#1}}

\newcommand{\Hp}{\rm{H}^{+}}

\newcommand{\Hep}{\rm{He}^{+}}
\newcommand{\He}{\rm{He}}

\newcommand{\cp}{\rm{C^{+}}}
\newcommand{\op}{\rm{O}^{+}}

\newcommand{\sip}{\rm{Si^{+}}}
\newcommand{\sipp}{\rm{Si^{++}}}
\newcommand{\expf}[3]{\exp \left(#1\frac{#2}{#3}\right)}
\def\simless{\mathbin{\lower 3pt\hbox
   {$\rlap{\raise 5pt\hbox{$\char'074$}}\mathchar"7218$}}}   
\def\simgreat{\mathbin{\lower 3pt\hbox  
   {$\rlap{\raise 5pt\hbox{$\char'076$}}\mathchar"7218$}}}

\title[Non-equilibrium metal cooling in the ISM]{The effect of non-equilibrium metal cooling on the interstellar medium}

\author[P.~R. Capelo et al.]{Pedro R. Capelo,$^{1}$\thanks{E-mail: pcapelo@physik.uzh.ch} Stefano Bovino,$^{2,3}$ Alessandro Lupi,$^{4}$ \newauthor Dominik~R.~G.~Schleicher$^{2}$ and Tommaso Grassi$^{5,6,7}$\\
$^1$Center for Theoretical Astrophysics and Cosmology, Institute for Computational Science, University of Zurich,\\
Winterthurerstrasse 190, CH-8057 Z{\"u}rich, Switzerland\\
$^2$Departamento de Astronom{\'i}a, Faculdad Ciencias F{\'i}sicas y Matem{\'a}ticas, Universidad de Concepci{\'o}n, Av. Esteban Iturra\\
s/n Barrio Universitario, Casilla 160, Concepci{\'o}n, Chile\\
$^3$Hamburger Sternwarte, Universit{\"a}t Hamburg, Gojenbergsweg 112, DE-21029 Hamburg, Germany\\
$^4$Institut d'Astrophysique de Paris, Sorbonne Universit\'{e}s, UPMC Univ. Paris 06 et CNRS, UMR 7095,\\ 98bis boulevard Arago, FR-75014 Paris, France\\
$^5$University Observatory Munich, Scheinerstr. 1, DE-81679 Munich, Germany\\
$^6$Excellence Cluster Origin and Structure of the Universe, Boltzmannstr. 2, DE-85748 Garching, Germany\\
$^7$Centre for Star and Planet Formation, Niels Bohr Institute \& Natural History Museum of Denmark, University of Copenhagen,\\
{\O}ster Voldgade 5--7, DK-1350 Copenhagen K.
}

\begin{document}

\date{Accepted 2017 December 23. Received 2017 December 07; in original form 2017 October 02}

\label{firstpage}


\maketitle

\begin{abstract}
By using a novel interface between the modern smoothed particle hydrodynamics code \gasoline and the chemistry package \kromes, we follow the hydrodynamical and chemical evolution of an isolated galaxy. In order to assess the relevance of different physical parameters and prescriptions, we constructed a suite of ten simulations, in which we vary the chemical network (primordial and metal species), how metal cooling is modelled (non-equilibrium versus equilibrium; optically thin versus thick approximation), the initial gas metallicity (from ten to hundred per cent solar), and how molecular hydrogen forms on dust. This is the first work in which metal injection from supernovae, turbulent metal diffusion, and a metal network with non-equilibrium metal cooling are self-consistently included in a galaxy simulation. We find that properly modelling the chemical evolution of several metal species and the corresponding non-equilibrium metal cooling has important effects on the thermodynamics of the gas, the chemical abundances, and the appearance of the galaxy: the gas is typically warmer, has a larger molecular gas mass fraction, and has a smoother disc. We also conclude that, at relatively high metallicity, the choice of molecular-hydrogen formation rates on dust is not crucial. Moreover, we confirm that a higher initial metallicity produces a colder gas and a larger fraction of molecular gas, with the low-metallicity simulation best matching the observed molecular Kennicutt--Schmidt relation. Finally, our simulations agree quite well with observations which link star formation rate to metal emission lines.
\end{abstract}

\begin{keywords}
astrochemistry -- molecular processes -- galaxies: evolution -- galaxies: ISM -- methods: numerical -- ISM: molecules.
\end{keywords}


\section{Introduction}\label{gaso2krome_isogal:sec:Introduction}

Numerical simulations of galaxy formation and evolution, both cosmological and idealised, have been extremely important in shaping our understanding of structure formation, for both the dark matter (DM) component and the baryonic collapsing objects within. In particular, DM-only cosmological simulations have been very successful at reproducing observations of the large-scale structure of the Universe, helping the cold DM model with a cosmological constant to become the most widely accepted paradigm for structure formation \citep[e.g.][]{Springel_et_al_2005a}.

Simulating the baryonic content of the Universe, however, has proven a lot more difficult, as hydrodynamical simulations have, for a long time, failed to produce consistent results and to reproduce galaxies that match observations \citep[see, e.g.][]{Schaye_et_al_2010,Scannapieco_et_al_2012,Schaye_et_al_2015}.

This was due to a variety of factors, linked in part to the use of different hydrodynamics algorithms (e.g. smoothed particle hydrodynamics -- SPH -- and adaptive mesh refinement -- AMR -- methods; see, e.g. \citealt{Agertz_et_al_2007}) and, most importantly, to how small-scale physical processes were modelled. Indeed, due to the huge dynamic range involved in the simulations and to limitations in computer power, many processes have to be modelled with so-called `subgrid' recipes, which heavily depend on resolution. Large uncertainties come from, e.g. modelling stellar evolution and the initial mass function (IMF) in stellar population synthesis models \citep[e.g.][]{Conroy_et_al_2009,Gonzales-Perez_et_al_2014}, supernova (SN) and supermassive black hole (BH) feedback \citep[e.g.][]{Bellovary_et_al_2010,DallaVecchia_Schaye_2008,DallaVecchia_Schaye_2012,Durier_DallaVecchia_2012}, star formation \citep[SF; e.g.][]{Hopkins_et_al_2013,Braun_Schmidt_2015}, and magnetic fields \citep[e.g.][]{Pakmor_Springel_2013,Rodenbeck_Schleicher_2016}.

In recent years, there has been much improvement, due in part to the introduction of modern-SPH codes, which has alleviated several of the issues raised in \citet{Agertz_et_al_2007}, and to a strong effort in developing subgrid recipes \citep[e.g.][]{Keller_et_al_2014,Kimm_et_al_2015,Hopkins_et_al_2017}.

A crucial aspect of the modelling of galaxies is the proper modelling of the gas chemistry and cooling which, so far, has received relatively limited attention. The cooling rate of gas has important dynamical effects, as it controls, e.g. the accretion mode of gas in DM haloes, the amount of SF in galaxies, and accretion on to BHs.

Several galaxy simulations assume that gas is in (collisional and/or photo-) ionisation equilibrium and employ cooling tables or functions (often constructed with {\textsc{cloudy}}; \citealt{Ferland_et_al_2013}) to compute gas cooling rates, based on gas density, temperature, and redshift. A few studies follow a few atomic primordial species to compute non-equilibrium cooling, but still employ the use of tables for molecular hydrogen and metals \citep[e.g.][]{Shen_et_al_2010}.

Only during the past decade, non-equilibrium chemical networks with molecular hydrogen have been incorporated in galaxy simulations. \citeauthor{Gnedin_et_al_2009} (\citeyear{Gnedin_et_al_2009}; see also \citealt{Gnedin_Kravtsov_2011} for a revised version) were the first to self-consistently implement an eight-species primordial chemical model (with HI, HII, H2I, H2II, H$^-$, HeI, HeII, and HeIII)\footnote{In this work, we adopt the astronomy notation in which, e.g. H, HI, and H2I indicate total hydrogen, atomic neutral hydrogen, and molecular neutral hydrogen, respectively.} to follow the rate of formation and destruction of molecular hydrogen, applying it to cosmological AMR simulations of galaxies down to redshift $z = 4$. This model, in which SF is directly linked to the abundance of molecular hydrogen, was successfully used to study in detail the Kennicutt--Schmidt (KS; \citealt{Schmidt_1959,Schmidt_1963,Kennicutt_1989,Kennicutt_1998}) relation \citep[][]{Gnedin_Kravtsov_2010,Feldmann_et_al_2011,Gnedin_Kravtsov_2011}.

\citet{Christensen_et_al_2012} later implemented a model related to that of \citet{Gnedin_et_al_2009} to run isolated SPH simulations of a Milky Way-sized galaxy and cosmological SPH simulations of a dwarf galaxy to $z = 0$, finding a colder, denser, and clumpier interstellar medium (ISM) with respect to `standard' simulations. \citet{Tomassetti_et_al_2015} also implemented a similar model and ran cosmological AMR simulations of a Milky Way-sized galaxy to $z = 2$, to assess the effect of using different SF recipes.

In all the above studies, only primordial species are evolved. Metal cooling is therefore still modelled assuming equilibrium. \citet{Richings_Schaye_2016} perform SPH simulations of isolated galaxies with a static DM potential, using the {\textsc{chimes}} chemistry solver module described in \citet{Richings_et_al_2014a,Richings_et_al_2014b}. They follow the evolution of 157 chemical species and the non-equilibrium cooling from the ionization states of 11 elements, to assess the effects of metallicity and ultraviolet (UV) radiation in galaxies. In their implementation, however, they do not model metal injection from SNae nor turbulent metal diffusion, therefore keeping the gas metallicity fixed throughout the entire simulation. This assumption, though, is likely to fail in star-forming galaxies, wherein the gas metallicity can increase even by one order of magnitude (see Section~\ref{gaso2krome_isogal:sec:dependence_on_metallicity}).

\citeauthor{Hu_et_al_2016} (\citeyear{Hu_et_al_2016}; see also \citealt{Hu_et_al_2017}) go a step further in that direction, by modelling metal injection from SNae in their isolated SPH simulations of dwarf galaxies. However, they track only six species (H2I, HII, CO, HI, CII, and OI), of which only the first three are explicitly followed.

Further investigations are clearly needed to explore the treatment of metal line cooling (following the evolution of more metal species), the comparison of equilibrium versus non-equilibrium cooling, the effect of UV radiation models, and dust physics. \citet{Lupi_et_al_2017}, using initial conditions and settings similar to ours, study in detail the effect of having a local UV field from young stars and several ways to model it \citep[see also][]{Nickerson_et_al_2017}. For this reason, in this work, we focus more on the role of metals as coolants and on how molecular hydrogen forms on dust.

By using a novel interface between a modern-SPH code (\gasolines; \citealt{Wadsley_et_al_2017}) and a chemistry package (\kromes; \citealt{Grassi_et_al_2014}), we explicitly follow nine primordial and seven metal species, model the corresponding non-equilibrium cooling, compute both metal injection from SNae and turbulent metal diffusion, and explore models of H2I formation on dust.

Properly modelling the gas chemistry is not only important to accurately compute gas cooling, but is also relevant in the context of observational efforts: computing the abundance of molecular hydrogen is important to compare the simulation results to observed relations between SF rate (SFR) and H2I surface densities (the so-called molecular KS relation; \citealt{Bigiel_et_al_2008,Bigiel_et_al_2010}). Moreover, evolving a chemical network which includes metals allows us to use CI, CII, and OI as tracers of the ISM \citep[e.g.][]{Malhotra_et_al_2001,Cormier_et_al_2015,Michalowski_et_al_2016,Pallottini_et_al_2017a,Pallottini_et_al_2017b} and of SFR \citep[e.g.][]{DeLooze_et_al_2014}, giving us a better theoretical understanding on how all these species trace different components of the ISM and of star-forming regions. This is particularly timely, given the wealth of far-infrared data coming from recent (e.g. the Herschel Space Observatory; \citealt{Pilbratt_et_al_2010}; low redshift) and current (e.g. the Atacama Large Millimeter/submillimeter Array -- ALMA; \citealt{ALMA_2015}; high redshift) facilities.

In Section~\ref{gaso2krome_isogal:sec:Numerical_setup}, we describe in detail the new code interface, the initial conditions, and the suite of simulations. In Section~\ref{gaso2krome_isogal:sec:Results}, we present the results from our suite, focussing on the role of non-equilibrium metal cooling, the dependence on metallicity and on the H2I formation model, and on the role of metal species as tracers of the ISM and SFR. We summarise and conclude in Section~\ref{gaso2krome_isogal:sec:Conclusions}.


\section{Numerical Setup}\label{gaso2krome_isogal:sec:Numerical_setup}

For all the simulations presented in this work, we interfaced two publicly available codes\footnote{Unless otherwise stated, when we simply write `the code', we refer to the ensemble of the two codes: \gasoline (\url{http://gasoline-code.com/}) and \krome (\url{http://kromepackage.org/}).} -- the $N$-body SPH code \gasoline (\citealt{Wadsley_et_al_2017}; an extension of the $N$-body tree code {\textsc{pkdgrav}}; \citealt{Stadel_2001}) and the chemistry package \krome \citep[][]{Grassi_et_al_2014} -- to accurately follow the hydrodynamical and chemical evolution of an isolated galaxy.

The modern-SPH \gasoline code differs from the original {\textsc{gasoline}} code \citep[][]{Wadsley_et_al_2004} in many respects (see \citealt{Wadsley_et_al_2017} for a thorough description). The main additions employed in the simulations of this work, compared to, e.g. those performed in \citet{Capelo_et_al_2015} with a previous version of the code for galaxy simulations, are the use of an improved SPH smoothing kernel (the Wendland C$^2$ kernel; \citealt{Wendland_1995,Dehnen_Aly_2012,Keller_et_al_2014}) that does not suffer from the pairing/clumping instability \citep[][]{Schuessler_Schmitt_1981}, an increased number of neighbours (64), the geometric-density-average force expression (\citealt{Keller_et_al_2014}; see also \citealt{Monaghan_1992,Ritchie_Thomas_2001}), found to minimise numerical surface tension, and the use of a pressure floor \citep[][]{Robertson_Kravtsov_2008,Brook_et_al_2012,Roskar_et_al_2015}, to ensure the correct fragmentation behaviour of the gas.

The chemistry package \krome \citep[][]{Grassi_et_al_2014} is a flexible code which can be embedded as a library in any hydrodynamics code and that, given any chemical network, generates an optimised {\textsc{fortran}} code to solve the corresponding system of coupled ordinary differential equations (ODEs) and follow the time-dependent evolution of the chemical species as well as the gas temperature. Additionally, the package provides a variety of modules to model several physical processes, including radiative and thermochemical cooling/heating, photochemistry, and dust physics (e.g. formation of molecular hydrogen by catalysis on dust, gas cooling by dust, and dust evaporation; see also \citealt{Grassi_et_al_2017}). \krome has already been successfully interfaced with a variety of hydrodynamics codes, such as the AMR codes {\textsc{enzo}} (\citealt{Bryan_et_al_2014}; see, e.g. \citealt{Bovino_et_al_2014}), {\textsc{ramses}} (\citealt{Teyssier_2002}; see, e.g. \citealt{Katz_et_al_2015}), and {\textsc{flash}} (\citealt{Fryxell_2000}; see, e.g. \citealt{Seifried_Walch_2016}; \citealt{Kortgen_et_al_2017}), and, very recently, the mesh-less finite-mass code {\textsc{gizmo}} (\citealt{Hopkins_2015}; see \citealt{Lupi_et_al_2017}). An updated list of applications and interfaced codes can be found at the \krome website.

In this work, the \textit{first} application of a working interface between \krome and an SPH implementation of the equations of hydrodynamics is presented.


\subsection{Initial Conditions}\label{gaso2krome_isogal:sec:Initial_conditions}

The initial conditions were built using the {\textsc{makedisk}} code \citep[][]{Springel_White_1999,Springel_et_al_2005b} and are identical for all the runs, except for the initial metallicity of the gas which, depending on the simulation, can be $0.1\,Z_{\odot}$, $0.5\,Z_{\odot}$, or $Z_{\odot}$, where $Z_{\odot} = 0.0134$ \citep[][]{Asplund_et_al_2009}.

The galaxy is composed of a DM halo and of a baryonic disc and bulge. The DM halo is described by a spherical Navarro--Frenk--White \citep[NFW;][]{Navarro_et_al_1996} profile up to the virial radius, $r_{\rm vir}$, and by an exponentially decaying NFW profile outside $r_{\rm vir}$ \citep[][]{Springel_White_1999}. The DM spin and concentration parameters are 0.04 \citep[][]{Vitvitska_et_al_2002} and 4 \citep[][]{Dutton_Maccio_2014,Diemer_Kravtsov_2015}, respectively, with the concentration parameter setting both the DM scale radius and the exponential decay pace. The baryonic disc is composed of a stellar and a gaseous exponential disc with an isothermal sheet \citep[][]{Spitzer_1942,Camm_1950}, which share the same initial disc scale radius, $r_{\rm d}$, obtained from imposing conservation of specific angular momentum of the material that forms the disc \citep[][]{Mo_et_al_1998}, and disc scale height $z_{\rm d} = 0.1r_{\rm d}$. The virial mass fraction of the baryonic disc is 0.02, of which 60 per cent is gaseous and 40 per cent is stellar, typical of high-redshift galaxies \citep[][]{Tacconi_et_al_2010}. The vertical structure of the gas is governed by hydrostatic equilibrium, which sets the initial gas temperature to be between $\sim$$10^2$ and $\sim$$10^6$~K, depending on the gas density, and we additionally impose a global gas temperature floor of 10~K. The stellar bulge is described by a spherical \citet{Hernquist_1990} profile, with a virial mass fraction of 0.004 and a scale radius $r_{\rm b} = 0.2r_{\rm d}$. The virial mass and radius of the galaxy are $2 \times 10^{11}$~M$_{\odot}$ \citep[][]{Adelberger_et_al_2005} and 45~kpc, respectively, and $r_{\rm d} = 1.28$~kpc.

The (constant) gravitational softenings of the stellar, gas, and DM particles are $\epsilon_{\rm star} = 10$, $\epsilon_{\rm gas} = 20$, and $\epsilon_{\rm DM} = 30$~pc, respectively. The smoothing length $h$, defined as half the distance to the furthest of the 64 neighbours, cannot be smaller than $0.1 \epsilon_{\rm gas}$. The initial particle mass for stars, gas, and DM is $3.3 \times 10^3$, $4.6 \times 10^3$, and $1.1 \times 10^5$~M$_{\odot}$, respectively, making the total initial number of particles $\sim$$4 \times 10^6$.

The galaxy's structure is similar to that of the main galaxy in the gas-rich galaxy merger of \citet{Capelo_et_al_2015}, where only equilibrium chemistry was followed, and of the galaxy in \citet[][]{Lupi_et_al_2017}, where the non-equilibrium chemistry of a primordial network and several radiative-transfer methods were studied. The structural parameters are typical of high redshift ($z = 3$) galaxies, but the same system can be interpreted as a representative of relatively small, gas-rich low-redshift galaxies \citep[see also][]{Capelo_et_al_2017}.


\subsection{Star formation, feedback, and metal injection}\label{gaso2krome_isogal:sec:SF_and_SN}

SF and stellar feedback (energy, mass, and metal injection) are modelled within \gasoline and follow the recipes described in \citeauthor{Stinson_et_al_2006} (\citeyear{Stinson_et_al_2006}; see also \citealt{Katz_1992,Christensen_et_al_2010,Shen_et_al_2010}), which we summarise here.

At the beginning of the simulation, the existing stars are all set to be 2~Gyr old. New stars are allowed to form from gas with temperature $T_{\rm gas} < T_{\rm SF} = 10^3$~K and density $\rho_{\rm gas} > \rho_{\rm SF} = 100 \,m_{\rm H}$~cm$^{-3}$ \citep[][]{Capelo_et_al_2015}, where $m_{\rm H}$ is the hydrogen mass. When the above conditions are met, gas particles are stochastically selected to form stars according to the probability $p = (m_{\rm gas}/m_{\rm star})[1-\exp(-\epsilon_* \Delta t/t_{\rm dyn})]$, where $m_{\rm gas}$ is the mass of the (potentially) star-forming gas particle, $m_{\rm star}$ is the mass of the (potentially) formed star (set to half the initial gas particle mass), $\epsilon_*$ is the SF efficiency parameter, $\Delta t = 10^6$~yr is the SF time-scale (i.e. how often SF is computed), $t_{\rm dyn} = (4 \pi G \rho_{\rm gas})^{-1/2}$ is the local dynamical time, and $G$ is the gravitational constant.

The above probability function implies that, on average, $dM_*/dt = \epsilon_* m_{\rm gas}/t_{\rm dyn}$, effectively ensuring that SF approximately follows the slope of the KS relation between SFR and gas surface densities, whereas the normalisation can be matched by tuning the SF efficiency parameter. In our simulations, we first `relax' the galaxy, i.e. we run the simulation for 0.05~Gyr using $\epsilon_* = 0.005$, followed by another 0.05~Gyr using $\epsilon_* = 0.01$, in order to counteract the fact that, at the beginning of the simulation, there is no heating from SNae which could prevent excessive gas cooling and consequent (unphysically high) bursts of SF. After the `relaxation' period of 0.1~Gyr, we run the simulation with $\epsilon_* = 0.015$ \citep[][]{Krumholz_et_al_2012}, found to match the normalisation of the KS law fairly well and to be consistent with the average SF efficiency computed as a function of $\rho_{\rm SF}$ in \citet{Lupi_et_al_2017}.

Every stellar particle, being much more massive than real individual stars, represents a stellar population, assumed to cover the entire IMF described in \citet{Kroupa_2001}. Stellar masses are then converted into stellar lifetimes according to \citet{Raiteri_et_al_1996}, in order to compute when and if a star explodes as a SN.

Stars with masses between 8 and 40~M$_{\odot}$ can explode as a Type II SN (SNII), each injecting $E_{\rm SN} = 10^{51}$~erg into the surrounding gas particles (using the SPH smoothing kernel) as thermal energy, according to the `blastwave model' of \citet{Stinson_et_al_2006}, in which, in order to avoid the gas from radiating away the SN energy due to limited resolution, radiative cooling is disabled during the survival time of the hot low-density shell of the SN \citep[][]{McKee_Ostriker_1977}. Together with energy, for each SN event, a given amount of total, iron, and oxygen mass, dependent on the progenitor mass $m_{\rm star}$,

\begin{equation}\label{gaso2krome_isogal:eq:SNII_injection}
\begin{aligned}
&M_{\rm ejected,SNII} = 0.7682 \,(m_{\rm star}/{\rm M}_{\odot})^{1.056} {\rm M}_{\odot}, \\
&M_{\rm Fe,SNII} = 2.802 \times 10^{-4} (m_{\rm star}/{\rm M}_{\odot})^{1.864} {\rm M}_{\odot}, \\
&M_{\rm O,SNII} = 4.586 \times 10^{-4} (m_{\rm star}/{\rm M}_{\odot})^{2.721} {\rm M}_{\odot},
\end{aligned}
\end{equation}
\\
\noindent is injected into the surrounding gas \citep[][]{Woosley_Weaver_1995,Raiteri_et_al_1996}.

Type Ia SNae (SNIa), occurring in evolved binary systems, are also modelled in the code, injecting into the surrounding gas $E_{\rm SN} = 10^{51}$~erg and a fixed amount of mass and metals, independent of the progenitor mass \citep[][]{Thielemann_et_al_1986,Raiteri_et_al_1996}:\\

\begin{equation}\label{gaso2krome_isogal:eq:SNIa_injection}
\begin{aligned}
&M_{\rm ejected,SNIa} = 1.4 \,{\rm M}_{\odot}, \\
&M_{\rm Fe,SNIa} = 0.63 \,{\rm M}_{\odot}, \\
&M_{\rm O,SNIa} = 0.13 \,{\rm M}_{\odot}.
\end{aligned}
\end{equation}
\indent Since SNIa stars should not be clustered and should not produce large blastwaves, shutting the cooling in this case would produce too large an effect.

Stars with masses $1 < m_{\rm star} < 8$~M$_{\odot}$ do not explode as SNae but release part of their mass as stellar winds, according to \citet{Kennicutt_et_al_1994} and \citet{Weidemann_1987}, with the returned gas having the same metallicity of the low-mass stars.

The injection of metals from SNae changes the metallicity of the gas, which is computed as

\begin{equation}\label{gaso2krome_isogal:eq:metallicity}
\begin{aligned}
&M_{\rm metals} =  1.06 \,M_{\rm Fe} + 2.09 \,M_{\rm O}.
\end{aligned}
\end{equation}
\\
\indent Another phenomenon that can change the metallicity of the gas is turbulent diffusion, modelled in the code by estimating the particle diffusion coefficients as $d = C_{\rm turb}|\textbf{S}|h^2$, where $\textbf{S}$ is the trace-free local shear tensor and $C_{\rm turb} = 0.05$ \citep[][]{Wadsley_et_al_2008,Shen_et_al_2010}. The same model and coefficient are applied to thermal diffusion.

The change in metallicity, $f_{\rm metals}$, naturally implies a change in the mass fraction of primordial elements ($f_{\rm H}$ and $f_{\rm He}$ for hydrogen and helium,\footnote{In this work, we use $f_{\rm species}$ and $n_{\rm species}$ to signify the mass fraction and number density of such species, respectively. We translate from the astronomy notation, $f_{\rm H} = X$ and $f_{\rm He} = Y$, but use interchangeably $Z$ and $f_{\rm metals}$, where `metals' includes all species that are not H and He.} respectively), and we follow the implementation of \citeauthor{Shen_et_al_2010} (\citeyear{Shen_et_al_2010}; see also \citealt{Jimenez_et_al_2003}), who adopt

\begin{equation}\label{gaso2krome_isogal:eq:Jimenez}
\begin{aligned}
&f_{\rm He} = 0.236+2.1f_{\rm metals} \;\;({\rm for} \;f_{\rm metals} \le 0.1), \\
&f_{\rm He} = 0.446[1-10(f_{\rm metals}-0.1)/9] \;\;({\rm for} \;f_{\rm metals} > 0.1), \\
&f_{\rm H} = 1-f_{\rm He}-f_{\rm metals}.
\end{aligned}
\end{equation}
\\
We note that, in our simulations, we never have $f_{\rm metals} > 0.1$.


\subsection{Thermodynamics, radiation, and chemistry}\label{gaso2krome_isogal:sec:chemistry}

Chemistry and radiative cooling/heating are computed within \krome and follow the models described in \citet{Bovino_et_al_2016}. Depending on the simulation, we either follow a system of nine primordial species (`eq-metals runs': primordial species, i.e. HI, HII, H2I, H2II, H$^-$, HeI, HeII, HeIII, and e$^-$) and dust (to treat the formation of H2I on dust grains), as well as metal line cooling via an effective cooling function (see below), or a system with the same nine primordial species and dust plus seven metal species (`non-eq-metals runs': primordial species, dust, CI, CII, OI, OII, SiI, SiII, and SiIII) and non-equilibrium metal cooling at low temperature. In the latter case, a linear system for the individual metal excitation levels is solved on-the-fly for the most important coolants in the ISM \citep[C, O, and Si; see also][]{Maio_et_al_2007,Glover_Jappsen_2007}. \krome employs the implicit high-order ODE solver {\textsc{dlsodes}} \citep[][]{Hindmarsh_1983} to integrate the system of differential equations corresponding to nine species and 40 reactions (eq-metals runs) or 16 species and 65 reactions (non-eq-metals runs). No three-body reactions were included, because of the relatively low gas density achieved in the simulations. The reactions are listed in Appendix~\ref{gaso2krome_isogal:sec:Reactions}.

We assume a uniform extragalactic UV background \citep{Haardt_Madau_2012} at $z = 3$, using $5 \times 10^3$ energy bins ranging from $1.237 \times 10^{-1}$ to $4.997 \times 10^7$~eV. We note that we did not include any radiative feedback from the local stellar population. However, we refer the reader to \citet{Lupi_et_al_2017}, in which the effects of the local radiation, together with different methods to account for radiative feedback, were studied, using a galaxy model almost identical to that of this work. In that case, the number of energy bins had to be reduced to 10, as the coupling to an on-the-fly radiative-transfer module is computationally very expensive.

To account for the attenuation of the radiation in high-density regions, where the radiation background is known to be shielded, we compute the optical depth $\tau = \Sigma_i \sigma_i N_i$, as the sum over every species $i$ of the product of the photo cross-section $\sigma$ and the column density $N$. We approximate the column density as $N_i \sim n_i L_{\rm Jeans}$, where $n_i$ is the species number density and $L_{\rm Jeans}$ is the Jeans length \citep[][]{Jeans_1902},\footnote{In the case of molecular-hydrogen self-shielding, we followed the model of \citet{Wolcott-Green_et_al_2011} but used as the characteristic length-scale a fixed length of 20~pc (i.e. the gravitational softening of the gas) instead of the Jeans length. We re-ran 0.2~Gyr of Run~01 using the model of \citet{Richings_et_al_2014b} and the Jeans length and obtained indistinguishable results.} which was shown in \citet{Safranek-Shrader_et_al_2017} to be a reasonable approximation for the shielding length.

In our simulations we follow the formation of H2I both in the gas phase and through catalysis on dust. In particular, H2I is formed through gas-phase channels (e.g. reactions~7--8 and 9--10 in Appendix~\ref{gaso2krome_isogal:sec:Reactions}) but, given the relatively high values of metallicity and dust-to-gas ratio considered in this work, the predominant formation channel \citep{Hollenbach_McKee_1979} is by catalysis on dust grains (reaction~31). We assume that the dust-to-gas ratio, $D \equiv \rho_{\rm dust}/\rho_{\rm gas}$, scales linearly with metallicity as $D = D_{\odot} Z/Z_{\odot}$, where $D_{\odot} = 0.00934$ \citep[][]{Yamasawa_et_al_2011}.

We employ two different models of H2I formation on dust. The first model (`J75'; from \citealt{Jura_1975}) assumes fixed dust properties and no dependence on gas temperature, yielding the rate

\begin{equation}\label{gaso2krome_isogal:eq:J75}
\begin{aligned}
&\frac{{\rm d}n_{\rm H2I}}{{\rm d}t} = 3 \times 10^{-17} n_{\rm HI}n_{\rm H}\frac{Z}{Z_{\odot}} C_{\rho} \;\;{\rm cm}^{-3}~{\rm s}^{-1},
\end{aligned}
\end{equation}
\\
\noindent where $n_{\rm H} = n_{\rm HI} + n_{\rm HII} + n_{\rm H^-} + 2n_{\rm H2I} + 2n_{\rm H2II}$ and $C_{\rho}$ is the clumping factor, which accounts for the missed H2I formation in the high-density regions due to limited resolution. In this work, we use a constant clumping factor which, depending on the simulation, can be either $C_{\rho} = 1$ (i.e. no clumping factor) or 10 (see also \citealt{Lupi_et_al_2017}, where a variable clumping factor model is implemented).

The second model (`CS09'; from \citealt{Cazaux_Spaans_2009}) computes the formation rate as a function of dust type and size, dust and gas number density and temperature, and gas thermal velocity $v_{\rm gas}$ (\citealt{Hollenbach_McKee_1979}; CS09). If one assumes only one type of grain for simplicity, the dust density can be written as

\begin{equation}\label{gaso2krome_isogal:eq:CS09dust}
\begin{aligned}
&\rho_{\rm dust} = C_{\rm dust}\int_{a_{\rm min}}^{a_{\rm max}} m(a)\varphi(a)da,
\end{aligned}
\end{equation}
\\
where $m(a) = 4 \pi \rho_0 a^3/3$ is the mass of a grain particle of size $a$, with $a_{\rm min} \le a \le a_{\rm max}$, $\varphi(a) = {\rm d}n(a)/{\rm d}a$, where $n(a)$ is the number density of grains of size $a$, and $\rho_0$ is the bulk density of the grains (2.25 and 3.13~g~cm$^{-3}$ for carbonaceous and silicates, respectively; \citealt{Zhukovska_et_al_2008}). The constant $C_{\rm dust}$ is a normalisation factor constrained by $\rho_{\rm dust} = D \rho_{\rm gas}$. The equation can be extended to a mix of grain types \citep[see][]{Grassi_et_al_2017}. For this work, we assumed two grain types (carbonaceous and silicates), $\varphi(a) = a^{-3.5}$ \citep[][]{Mathis_et_al_1977,Draine_Lee_1984}, $a_{\rm min} = 5 \times 10^{-7}$~cm, and $a_{\rm max} = 2.5 \times 10^{-5}$~cm; divided the size range into $N_{\rm dust} = 20$ logarithmically spaced bins; and computed the rate of H2I formation on dust as

\begin{equation}\label{gaso2krome_isogal:eq:CS09}
\begin{aligned}
&\frac{{\rm d}n_{\rm H2I}}{{\rm d}t} = \frac{\pi}{2} n_{\rm HI} v_{\rm gas} C_{\rm dust} \sum_{j} \sum_{i} n_{{\rm dust,}ij} a_{ij}^2 \epsilon_{ij}S_i,
\end{aligned}
\end{equation}
\\
where the summations are over the grain types $j = ({\rm C}, {\rm Si})$ and size bins $i = (1$--$N_{\rm dust}$), $n_{{\rm dust},ij}$ is the number density of dust type $j$ in bin $i$, and the efficiency factors $\epsilon_{ij}$ and sticking coefficients $S_i$ both depend on $T_{\rm gas}$ and on the size-dependent dust temperature $T_{\rm dust}$ (\citealt{Hollenbach_McKee_1979}; CS09; \citealt{Grassi_et_al_2017}).

Following the methodology proposed by \citet{Grassi_et_al_2017}, we tabulated the reaction rates of H2I formation on dust as a function of gas number density\footnote{We compute the gas number density within \gasoline as $n_{\rm gas} = \rho_{\rm gas}/(\mu m_{\rm H})$, with $\mu$, the mean molecular weight of the gas, calculated as $\mu^{-1} = (f_{\rm e}m_{\rm H}/m_{\rm e}) + f_{\rm HI} + f_{\rm H^-} + f_{\rm HII} + 0.5(f_{\rm H2I} + f_{\rm H2II}) + 0.25(f_{\rm HeI} + f_{\rm HeII} + f_{\rm HeIII}) + Z/\langle A \rangle_{\rm metals}$, where $m_{\rm e}$ is the electron mass and $\langle A \rangle_{\rm metals} = 17.6003$ (see also \citealt{Bovino_et_al_2016}; there is a typo in their Eqs 28 and 30).} ($10^{-6} \le n_{\rm gas} \le 10^6$~cm$^{-3}$) and temperature ($3 \le T_{\rm gas} \le 10^9$~K), adopting a constant dust-evaporation temperature of $1.8 \times 10^3$~K, computing the dust opacity as in \citet{Omukai_et_al_2005}, and assuming zero gas opacity, since grains are the dominant contribution. Our results are rather insensitive to the specific radiation background employed for calculating the dust temperatures \citep[][]{Bovino_et_al_2016}, therefore here we only considered the cosmic microwave background (CMB) radiation from $z = 3$ for definiteness. Using the same methodology, we also created similar tables for the averaged dust temperature and for the gas cooling by dust.\footnote{Since usually $T_{\rm dust} < T_{\rm gas}$, this is typically called cooling. However, one can have gas heating by dust when $T_{\rm dust} > T_{\rm gas}$.} All tables were created assuming solar metallicity, therefore, cooling by dust and the rate of H2I formation on dust are then (linearly) rescaled with $Z/Z_{\odot}$ (and, in the case of the formation rate, multiplied by $C_{\rho}$) within \kromes. We note that, when we employ the first method of H2I formation on dust (i.e. J75), we do not compute the dust temperature nor the gas cooling by dust. However, it was shown in \citet[][]{Grassi_et_al_2017} that the effect of gas cooling by dust at low gas density ($\lesssim 10^7$~cm$^{-3}$) is not relevant \citep[see also][]{Bovino_et_al_2016}.

Gas cooling is an extremely important physical process that has profound effects on the evolution of a galaxy, since it has direct consequences on SF (and on other phenomena, like e.g. accretion on to BHs, not modelled here; e.g. \citealt{Bellovary_et_al_2013}). In this work, we include several cooling processes (see \citealt{Grassi_et_al_2014,Bovino_et_al_2016} for a thorough explanation): HI, HeI, and HeII collisional excitation and ionisation; HII, HeII, and HeIII recombination; HeI dielectronic recombination; Compton cooling from the CMB; Bremsstrahlung \citep[][]{Cen_1992}; H2I roto-vibrational cooling \citep[][]{Bovino_et_al_2016,Glover_Abel_2008,Glover_2015} and collisional dissociation \citep[][]{Omukai_2000}; grain-surface recombination \citep[][]{Bakes_Tielens_1994}; gas cooling by dust (\citealt{Hollenbach_McKee_1979}; except in the J75 runs); and metal line cooling, as explained below.\footnote{We also include the following heating processes: photoheating of atoms and molecules due to ionisations and photodissociations \citep[][]{Grassi_et_al_2014}, including H2I UV pumping and H2I direct photodissociation heating \citep[][]{Burton_et_al_1990}; heating due to the formation of H2I on dust and in the gas phase \citep[][]{Hollenbach_McKee_1979,Omukai_2000}; and photoelectric heating by dust grains \citep[][]{Bakes_Tielens_1994}. In this work, we do not include any ionisation and heating from cosmic rays, as they become important only at very high densities \citep[e.g.][]{Papadopoulos_Thi_2013}.}

Metal line cooling is dominant for gas of relatively high metallicity and low density, such as that simulated in this work. Ideally, it should be computed in non equilibrium, as it was shown that gas cooling (isochorically or isobarically) from $10^{7-8}$~K departs from equilibrium already at $\sim$$10^6$~K, either in the absence \citep[][]{Gnat_Sternberg_2007} or presence \citep[][]{Oppenheimer_Schaye_2013a} of a photo-ionizing background. This was confirmed also when the background is fluctuating, both in idealised \citep[][]{Oppenheimer_Schaye_2013b} and cosmological simulations \citep[][]{Segers_et_al_2017}. However, at high temperatures, given the large number of metal species, their ionisation states, transitions, and collisional processes, it becomes expensive to properly follow a metal network (even just considering the most important metal coolants like, e.g. carbon, oxygen, and silicon) and its related non-equilibrium cooling. This is why most authors employ cooling metal tables, such as collisional ionisation equilibrium tables \citep[CIE; e.g.][]{Sutherland_Dopita_1993} or, when assuming a given photoionising radiation background, photoionization equilibrium (PIE) tables (see \citealt{Bovino_et_al_2016} for a summary of the types of tables used in the literature). The usage, however, is common also at low temperatures (i.e. $T_{\rm gas} < 10^4$~K), where the departure from equilibrium is larger.

In this work, for $T_{\rm gas} \ge 10^4$~K, we utilise the (appropriately translated to be used within \kromes) PIE table computed by \citeauthor{Shen_et_al_2013} (\citeyear{Shen_et_al_2013}; see also \citealt{Shen_et_al_2010}) using {\textsc{cloudy}} \citep[][]{Ferland_et_al_1998}, under the assumption of an extragalactic radiation background by \citet{Haardt_Madau_2012}. The tables are provided as a function of gas temperature ($10 \le T_{\rm gas} \le 10^9$~K), density ($10^{-9} \le n_{\rm H} \le 10^4$~cm$^{-3}$), and redshift ($0 \le z \le 15.1$), although in this work we consider a fixed redshift ($z = 3$), and assume the default {\textsc{cloudy}} solar composition -- containing the first 30 elements in the periodic table -- except for H and He. We then linearly rescale the cooling rates with metallicity, as done for the cooling by dust. We caution that the PIE tables were constructed under the optically thin approximation (i.e. with no shielding), therefore, when using them, we over-estimate metal cooling at $T_{\rm gas} < 10^4$~K.

For $T_{\rm gas} < 10^4$~K, we either use these same tables (eq-metals runs) or non-equilibrium cooling by CI, CII, OI, OII, SiI, and SiII (non-eq-metals runs). We chose these elements because they are the most important metal coolants in the ISM \citep[e.g.][]{Wolfire_et_al_2003}. In this case, we solved a linear system for the individual metal excitation levels of the most important atoms and ions time-dependently. In all cases, we impose a temperature floor of 10~K.


\subsection{Evolving the galaxy}\label{gaso2krome_isogal:sec:Evolving_the_galaxy}

\begin{table} \centering
\vspace{-3.5pt}
\caption[Simulations specs]{Main simulations parameters. (1) Run name. (2) Metallicity of all the gas at the beginning of the relaxation. (3) Model of H2I formation on dust: CS09 or J75. (4) Clumping factor. (5) A chemical network of nine primordial and seven metal species is (is not) solved, together with non-equilibrium metal cooling, for the gas at low temperatures ($T < 10^4$~K) for the non-eq-metals (eq-metals) runs.
\label{gaso2krome_isogal:tab:simulations_specs}}
\vspace{10pt}
\begin{tabular*}{0.43\textwidth}{m{25pt}m{25pt}m{25pt}m{25pt}m{60pt}}
\hline
Run	& $Z/Z_{\odot}$ & H2I & $C_{\rho}$ & Non-eq. Metals \T \B \\
\hline
01		& 0.5		& CS09	& 1	& No		\T \B \\
01m		& 0.5		& CS09	& 1	& Yes	\T \B \\
02		& 0.5		& CS09	& 10	& No		\T \B \\
02m		& 0.5		& CS09	& 10	& Yes	\T \B \\
03		& 0.5		& J75	& 1	& No		\T \B \\
03m		& 0.5		& J75	& 1	& Yes	\T \B \\
04		& 1.0		& CS09	& 1	& No		\T \B \\
04m		& 1.0		& CS09	& 1	& Yes	\T \B \\
05		& 0.1		& CS09	& 1	& No		\T \B \\
05m		& 0.1		& CS09	& 1	& Yes	\T \B \\
\hline
\end{tabular*}
\vspace{0.0pt}
\end{table}

From the initial metallicity of the gas ($0.1 Z_{\odot}$, $0.5 Z_{\odot}$, or $Z_{\odot}$), we first compute $f_{\rm H}$ and $f_{\rm He}$ based on Eqs~\eqref{gaso2krome_isogal:eq:Jimenez}, to be consistent with what done by \gasoline during the subsequent evolution. In the case of the non-eq-metals runs, we also impose the total abundances $f_{\rm C} = 0.18 f_{\rm metals}$, $f_{\rm O} = 0.43 f_{\rm metals}$, and $f_{\rm Si} = 0.05 f_{\rm metals}$ \citep[][]{Asplund_et_al_2009}. We make an initial guess of the abundances of the nine or 16 species, assuming a mostly neutral atomic gas. We then run the \krome code on all the gas particles in the initial conditions, for a time that depends on the density of each gas particle (shorter for denser gas), to evolve the system towards chemical equilibrium, keeping $T_{\rm gas}$ constant.

With these new initial conditions, we run the code for 0.1~Gyr (roughly three times the local dynamical time of the galaxy, when using a typical $\rho_{\rm gas} = 10^{-24}$~g~cm$^{-3}$) using low values of the SF efficiency parameter (as explained in Section~\ref{gaso2krome_isogal:sec:SF_and_SN}), during which the galaxy is relaxed. The `real' runs start at the end of this `chemical equilibrium and relaxation' process, when we reset the time to $t = 0$, and last for 0.4~Gyr. We chose this final time as a compromise between the wish to have a truly in-equilibrium system (free of any numerical artefacts stemming from the initial conditions) and the notion that a system can be realistically considered in isolation for no more than a few $10^8$~yr, before cosmological gas inflows \citep[][]{Dekel_et_al_2009} and/or mergers \citep[][]{Genel_et_al_2009} become important. We also note that, even if there is no gas replenishment from cosmological gas inflows, this problem is made less severe by the stellar winds from low-mass stars \citep[][]{Stinson_et_al_2006}.

\begin{figure}
\centering
\vspace{-6.0pt}
\includegraphics[width=1.12\columnwidth,angle=0]{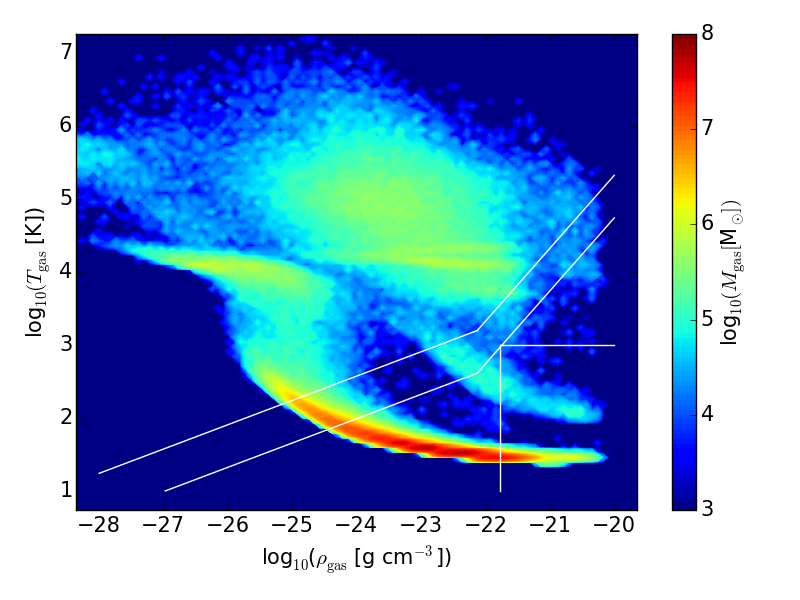}
\includegraphics[width=1.12\columnwidth,angle=0]{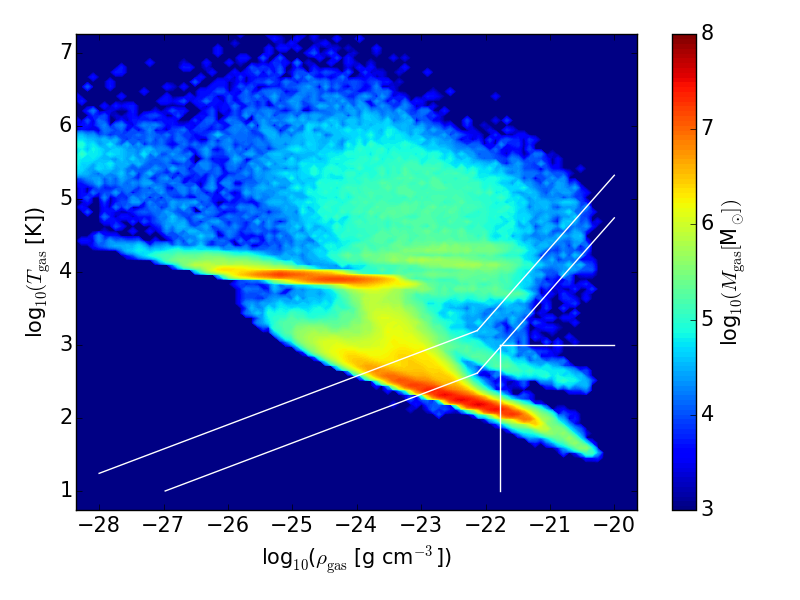}
\vspace{-0.0pt}
\caption[]{Dependence on the modelling of metals. Gas $\rho$--$T$ diagrams at 0.4~Gyr, for the eq-metals Run~01 (upper panel) and the non-eq-metals Run~01m (lower panel). The horizontal and vertical lines show the SF temperature and density thresholds, respectively. The diagonal lines show the temperature corresponding to the pressure floor, assuming an ideal gas and $\mu = 0.60$ (lower line) or 2.30 (upper line). For all the gas phase diagrams in this work, we chose the limits of the colour bars such that a dark blue pixel means there is no gas in that phase-space region, since the (initial) mass of a gas particle is $4.6 \times 10^3$~M$_{\odot}$. The number of logarithmic bins in the $x$- and $y$-axis is always $10^2$. The eq-metals gas is on average colder than the eq-metals gas.
}
\label{gaso2krome_isogal:fig:rhoT_run01vs01m}
\end{figure}

In order to ensure the correct fragmentation behaviour of the gas, it is important that the Jeans mass ($M_{\rm Jeans}$) and length of the gas are always resolved. In \citet{Capelo_et_al_2015}, where the mass resolution and SF density threshold were identical to those of this work, a temperature floor of 500~K was imposed, so that the Jeans mass of gas with $\rho_{\rm gas} \lesssim \rho_{\rm SF}$ would be resolved by at least 64 particles (see \citealt{Capelo_et_al_2017} for a more accurate explanation). In this work, we adopt instead a pressure floor, as described in \citeauthor{Roskar_et_al_2015} (\citeyear{Roskar_et_al_2015}; see also \citealt{Agertz_et_al_2009}). In \gasolines, the minimum gas pressure is set to $P_{\rm gas,min} = \alpha \,G \,[{\rm max}(\epsilon_{\rm gas},h)]^2 \rho_{\rm gas}^2$, where $\alpha$ is a parameter that sets the minimum number $N_{\rm Jeans}$ of resolution elements, ${\rm max}(\epsilon_{\rm gas},h)$, that must resolve the Jeans length. If we define $L_{\rm Jeans} = [\pi \gamma P_{\rm gas}/(G \rho_{\rm gas}^2)]^{1/2}$, where $\gamma = 5/3$ is the gas adiabatic index, imposing $L_{\rm Jeans} > N_{\rm Jeans} \,{\rm max}(\epsilon_{\rm gas},h)$ is equivalent to imposing $P_{\rm gas} > [\rho_{\rm gas} N_{\rm Jeans} \,{\rm max}(\epsilon_{\rm gas},h)]^2 G/(\pi \gamma)$. Therefore, the number of resolution elements that resolve $L_{\rm Jeans}$ is related to $\alpha$ by $N_{\rm Jeans} = (\alpha \pi \gamma)^{1/2}$. We impose $\alpha = 3$ and obtain $N_{\rm Jeans} = 4$ \citep[see][]{Truelove_et_al_1997}. A similar floor, within {\textsc{ramses}}, has already been tested in \citet{Gabor_et_al_2016} when modelling a galaxy with similar resolution and SF density threshold as in this work. As a sanity check, we also impose $L_{\rm Jeans} \ge \epsilon_{\rm gas}$. We note that the presence of a pressure floor effectively makes the gas non ideal when $P_{\rm gas,min}$ is reached, since the gas temperature is allowed to cool below the ideal-gas equivalent $T_{\rm gas,min}$ (see Fig.~\ref{gaso2krome_isogal:fig:rhoT_run01vs01m}), as done in, e.g. \citet{Richings_Schaye_2016}.

Metal injection from stellar evolution and turbulent metal diffusion imply that $f_{\rm H}$, $f_{\rm He}$, and $f_{\rm metals}$ all evolve with time [see Eqs~\eqref{gaso2krome_isogal:eq:Jimenez}]. At each time-step, before computing the chemical evolution, we rescale all species abundances according to

\begin{equation}\label{gaso2krome_isogal:eq:rescaling}
\begin{aligned}
&f_{\rm H \;species, \,new} = f_{\rm H \;species, \,old} (f_{\rm H,new}/f_{\rm H,old}),
\end{aligned}
\end{equation}
\\
\noindent where $f_{\rm H \;species, \,new/old}$ and $f_{\rm H,new/old}$ are the new/old mass fraction of a hydrogen species (e.g. HI, or HII, etc.) and of hydrogen, respectively. The same is done for He and, in the non-eq-metals runs, for metals. This way, the abundance ratios are conserved. Moreover, the mass fractions of C and Si are always approximately consistent with the metal ejection from SNae (we do not follow the evolution of Fe).

We emphasise here that this is the \textit{first} work in which a metal network, its related non-equilibrium metal cooling, metal injection from SNae, and turbulent metal diffusion are all self-consistently modelled in a galaxy simulation. We remind the reader that SNae can change the metallicity of the gas by more than an order of magnitude in less than half Gyr (see Section~\ref{gaso2krome_isogal:sec:dependence_on_metallicity}).


\subsubsection{The suite of simulations}\label{gaso2krome_isogal:sec:The_suite_of_simulations}

\begin{figure}
\centering
\vspace{-6.0pt}
\includegraphics[width=1.12\columnwidth,angle=0]{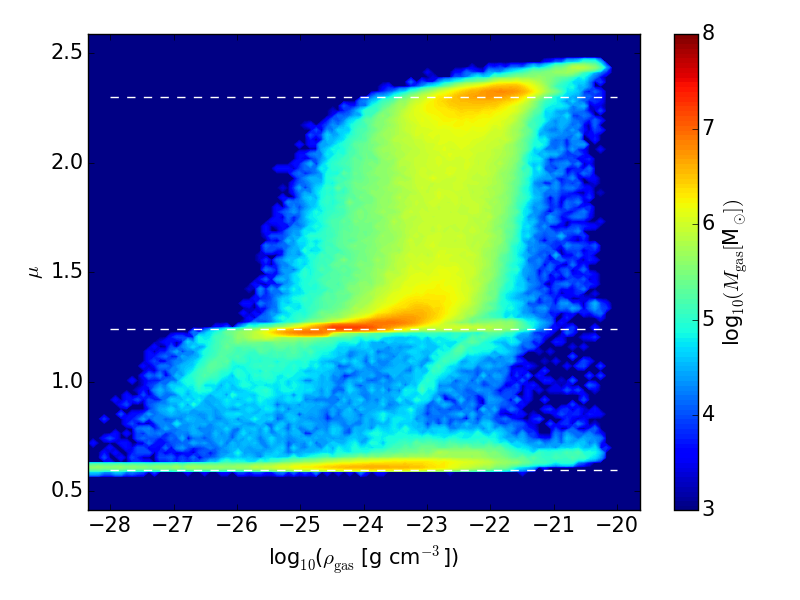}
\includegraphics[width=1.12\columnwidth,angle=0]{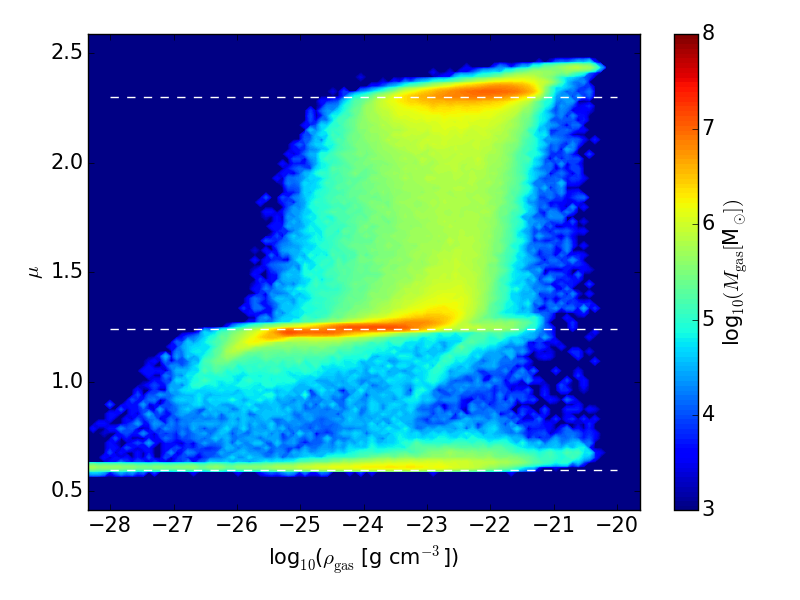}
\vspace{-0.0pt}
\caption[]{Dependence on the modelling of metals. Gas $\rho$--$\mu$ diagrams at 0.4~Gyr, for the eq-metals Run~01 (upper panel) and the non-eq-metals Run~01m (lower panel). The horizontal lines show the values of $\mu$ for (from top to bottom) a fully neutral molecular gas ($f_{\rm H} = f_{\rm H2I}$, $f_{\rm He} = f_{\rm HeI}$), a fully neutral atomic gas ($f_{\rm H} = f_{\rm HI}$, $f_{\rm He} = f_{\rm HeI}$), and a fully ionized atomic gas ($f_{\rm H} = f_{\rm HII}$, $f_{\rm He} = f_{\rm HeIII}$), computed for a gas with $Z = 0.5Z_{\odot}$. The mean molecular weight can get higher than the upper horizontal line because of the range in metallicity (see the upper middle panel of Fig.~\ref{gaso2krome_isogal:fig:phasediagrams_run05vs01vs04} for Run~01; the equivalent plot for Run~01m is very similar), since $\mu$ increases with $Z$. The non-eq-metals gas is more molecular than the eq-metals gas.
}
\label{gaso2krome_isogal:fig:rhomu_run01vs01m}
\end{figure}

The suite of simulations is composed of two distinct sub-sets, the eq-metals and the non-eq-metals set, which differ only in how metals are modelled (see Section~\ref{gaso2krome_isogal:sec:chemistry}). Within each sub-set, we vary (a) the method of H2I formation on dust (and include/exclude gas cooling by dust), (b) the clumping factor of such formation mechanism, and (c) the initial metallicity. This brings the total number of simulations to ten.

In Run~01, a chemical network of nine primordial species (HI, HII, H2I, H2II, H$^-$, HeI, HeII, HeIII, and e$^-$) is followed and several photo- and thermal processes are modelled during the evolution of an isolated gas-rich galaxy in which the initial metallicity of the gas is 50 per cent solar; H2I is formed on dust following the model of CS09, with no clumping factor (i.e. $C_{\rho} = 1$); and a shielded \citet{Haardt_Madau_2012} radiation background at $z = 3$ is assumed.

Run~01m is the same as Run~01, except for additional seven metal species (CI, CII, OI, OII, SiI, SiII, and SiIII) in the chemical network; and CI, CII, OI, OII, SiI, and SiII non-equilibrium cooling at low temperatures (i.e. $T_{\rm gas} < 10^4$~K).

The other eight runs are variations of the first two runs, in which we vary one parameter at a time. The main parameters of the simulations are listed in Table~\ref{gaso2krome_isogal:tab:simulations_specs}.


\section{Results}\label{gaso2krome_isogal:sec:Results}

\begin{figure}
\centering
\vspace{-14.5pt}
\includegraphics[width=1.12\columnwidth,angle=0]{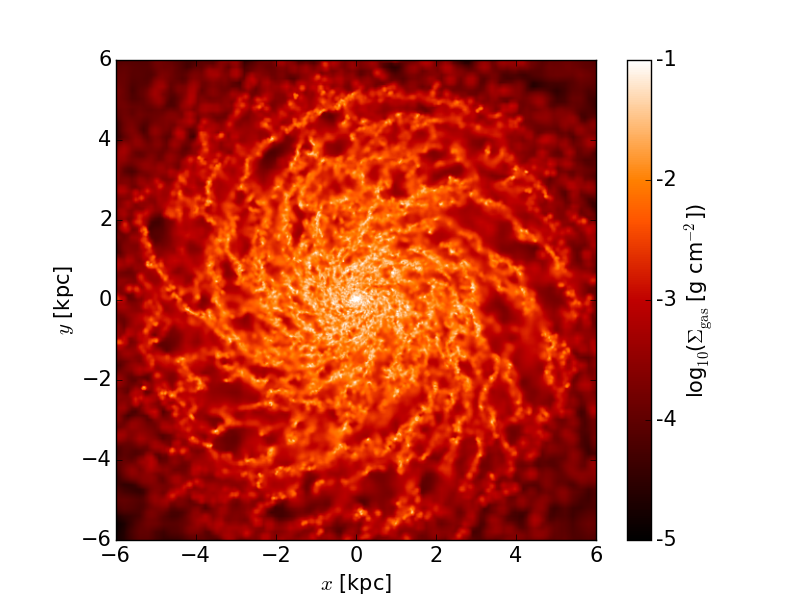}
\includegraphics[width=1.12\columnwidth,angle=0]{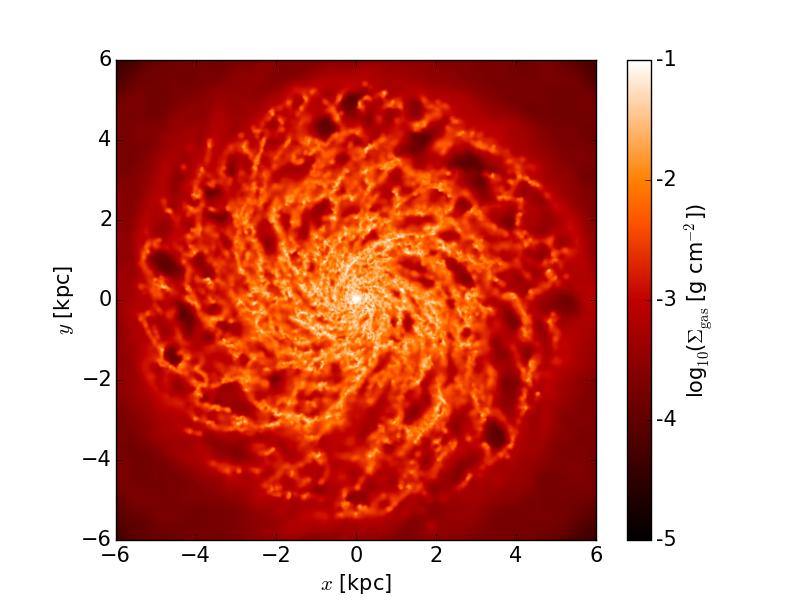}
\vspace{8.0pt}
\caption[]{Dependence on the modelling of metals. Gas surface density maps at 0.4~Gyr for the eq-metals Run~01 (upper panel) and the non-eq-metals Run~01m (lower panel). The eq-metals gas is more fragmented than the non-eq-metals gas.
}
\label{gaso2krome_isogal:fig:gassurfdenmaps_run01vs01m}
\end{figure}

\begin{figure}
\centering
\vspace{-14.5pt}
\includegraphics[width=1.12\columnwidth,angle=0]{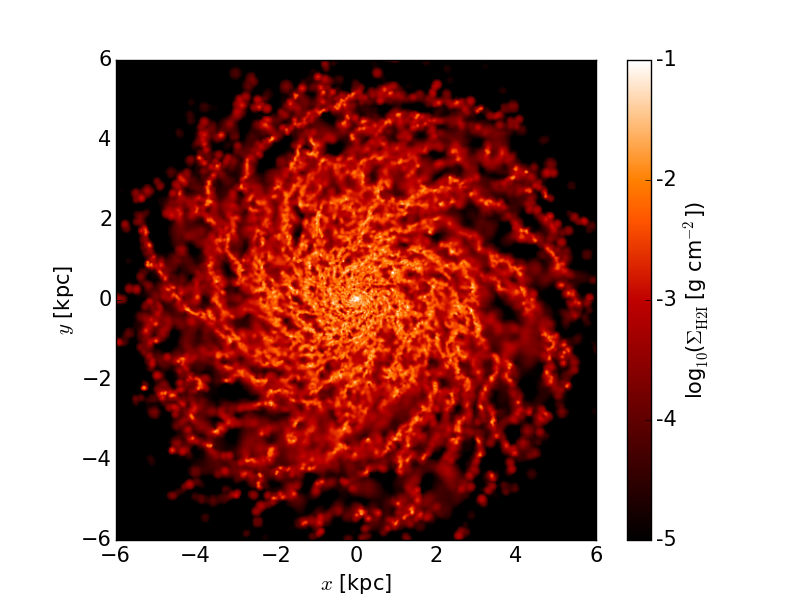}
\includegraphics[width=1.12\columnwidth,angle=0]{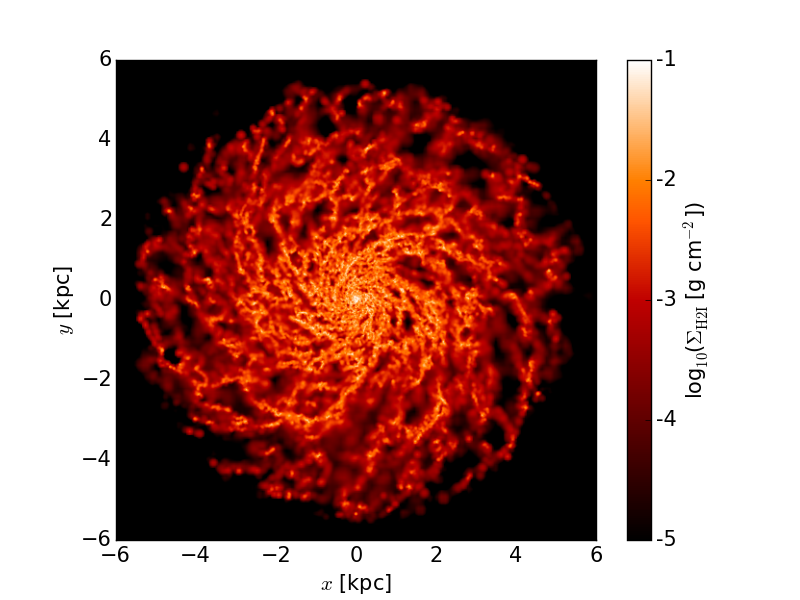}
\vspace{8.0pt}
\caption[]{Dependence on the modelling of metals. H2I surface density maps at 0.4~Gyr for the eq-metals Run~01 (upper panel) and the non-eq-metals Run~01m (lower panel). The eq-metals molecular disc is slightly more extended than the non-eq-metals counterpart.
}
\label{gaso2krome_isogal:fig:H2Isurfdenmaps_run01vs01m}
\end{figure}

\begin{figure}
\centering
\vspace{-5.0pt}
\includegraphics[width=1.02\columnwidth,angle=0]{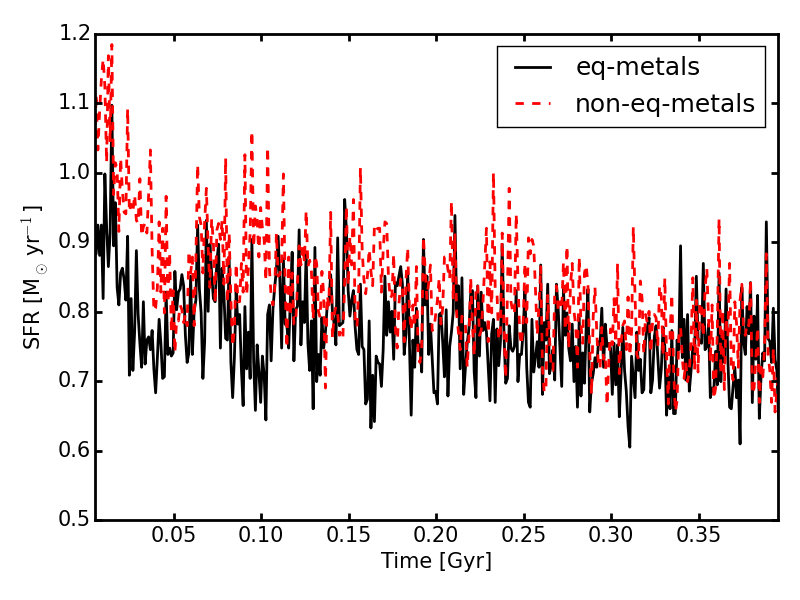}
\vspace{-0.0pt}
\caption[]{Dependence on the modelling of metals. SFR versus time for the eq-metals Run~01 (black solid line) and the non-eq-metals Run~01m (red dashed line). The two runs exhibit very similar SF histories.
}
\label{gaso2krome_isogal:fig:sfr}
\end{figure}

\begin{figure}
\centering
\vspace{-6.0pt}
\includegraphics[width=1.12\columnwidth,angle=0]{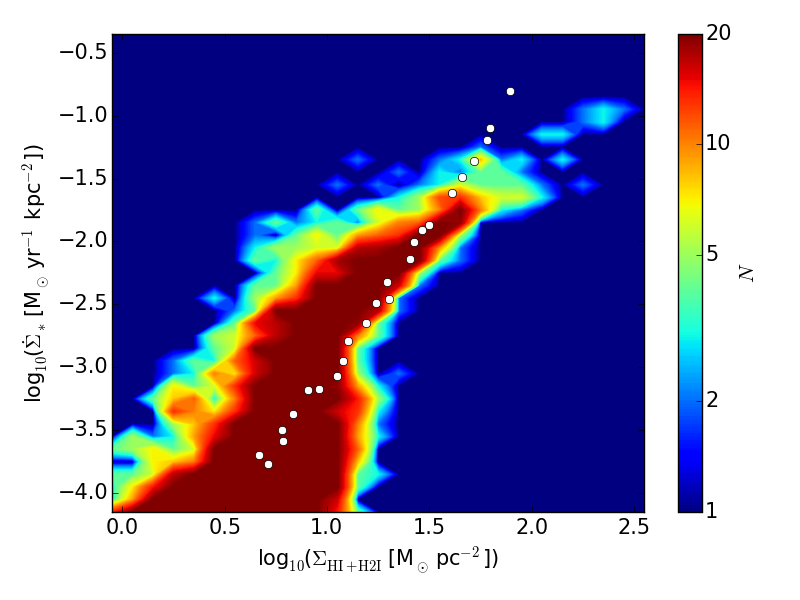}
\includegraphics[width=1.12\columnwidth,angle=0]{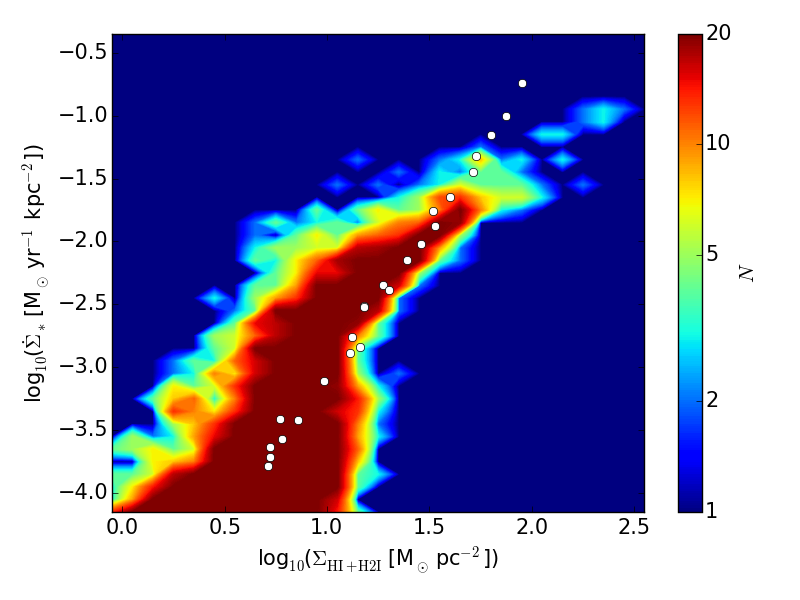}
\vspace{0.0pt}
\caption[]{Dependence on the modelling of metals. KS diagrams at 0.4~Gyr between HI\texttt{+}H2I gas and SFR (in the last 100~Myr) surface density, for the 50 innermost 200-pc-thick concentric cylindrical bins, for the eq-metals Run~01 (upper panel) and the non-eq-metals Run~01m (lower panel). The data from the simulation (white points) are compared to the contours by \citet{Bigiel_et_al_2010}. The two runs display very similar classic KS diagrams and both agree very well with the observations.
}
\label{gaso2krome_isogal:fig:ks_HIH2I_bigiel_run01vs01m}
\end{figure}

\begin{figure}
\centering
\vspace{-6.0pt}
\includegraphics[width=1.12\columnwidth,angle=0]{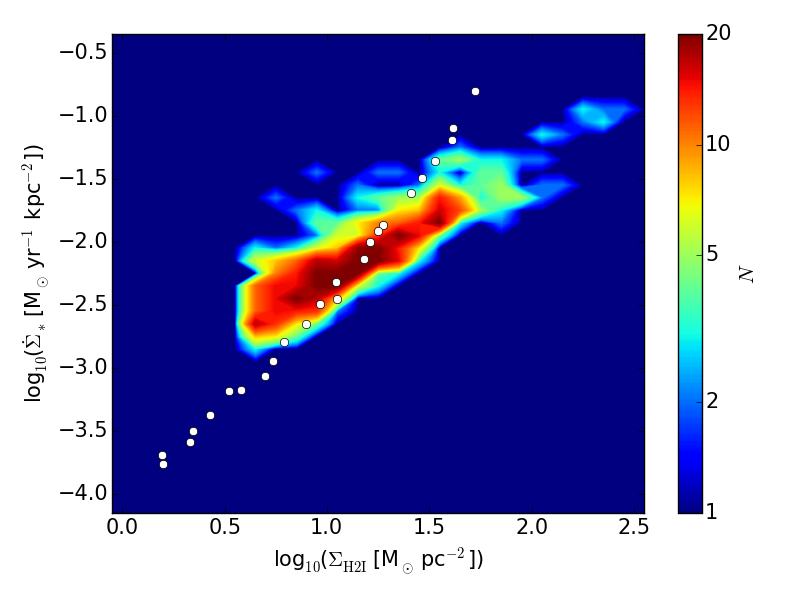}
\includegraphics[width=1.12\columnwidth,angle=0]{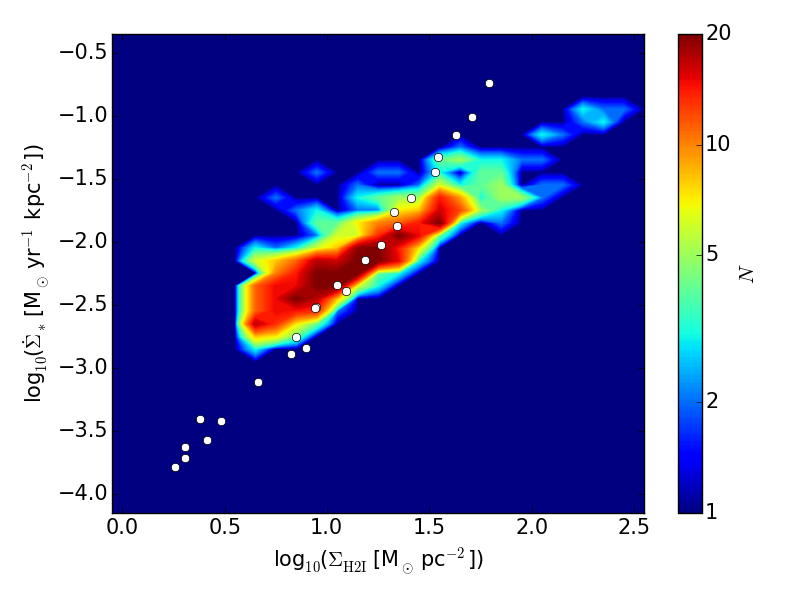}
\vspace{0.0pt}
\caption[]{Dependence on the modelling of metals. Molecular KS diagrams at 0.4~Gyr between H2I gas and SFR (in the last 100~Myr) surface density, for the 50 innermost 200-pc-thick concentric cylindrical bins, for the eq-metals Run~01 (upper panel) and the non-eq-metals Run~01m (lower panel). The data from the simulation (white points) are compared to the contours by \citet{Bigiel_et_al_2010}. The two runs feature very similar molecular KS diagrams.
}
\label{gaso2krome_isogal:fig:ks_H2I_bigiel_run01vs01m}
\end{figure}

\begin{figure*}
\centering
\vspace{-3.0pt}
\includegraphics[width=0.69\columnwidth,angle=0]{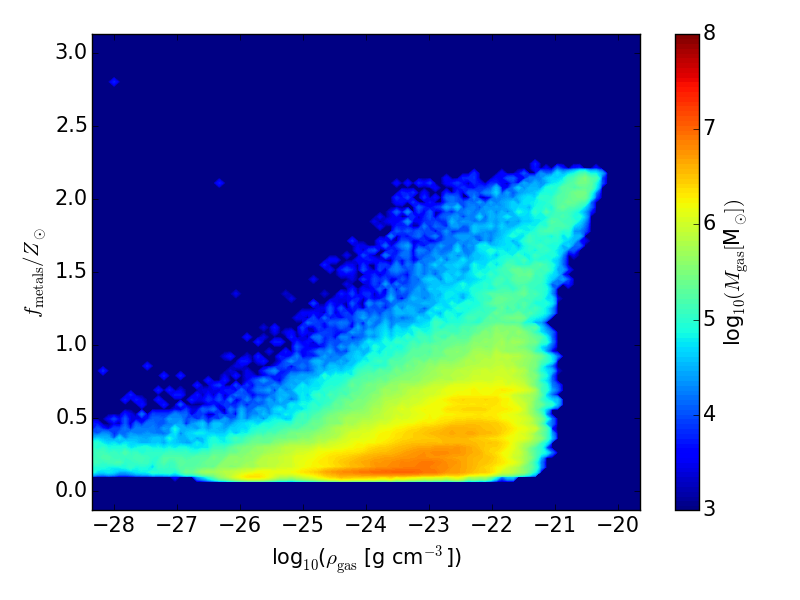}
\includegraphics[width=0.69\columnwidth,angle=0]{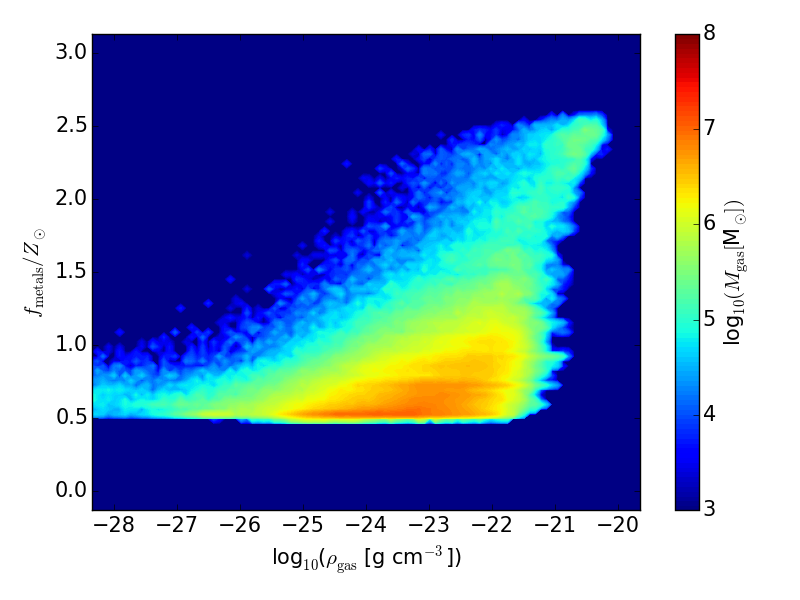}
\includegraphics[width=0.69\columnwidth,angle=0]{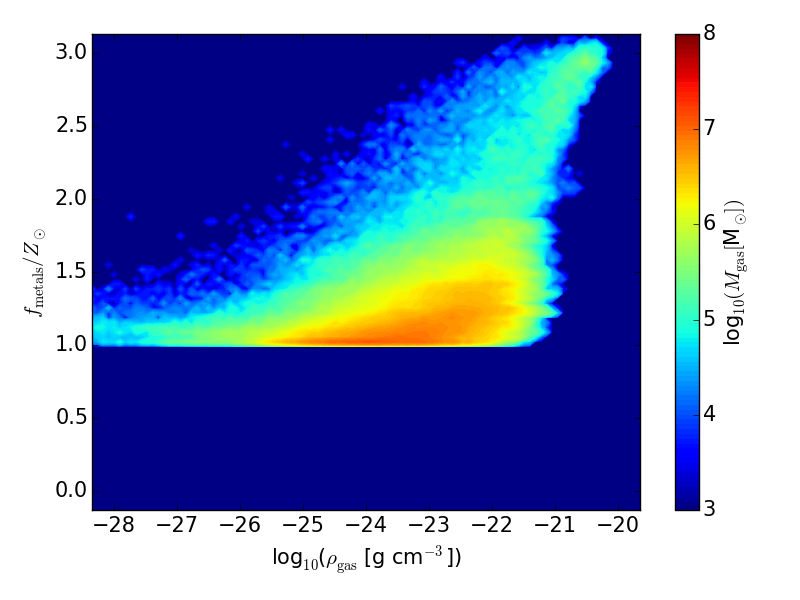}
\includegraphics[width=0.69\columnwidth,angle=0]{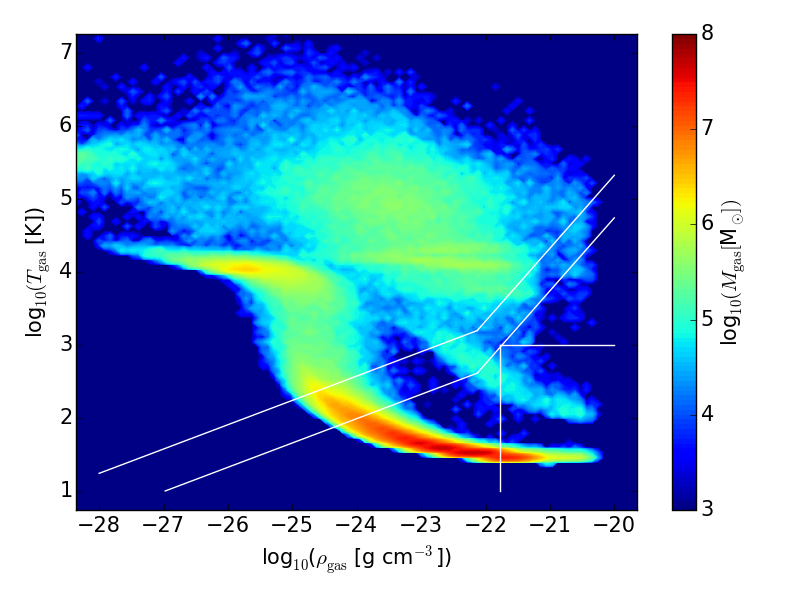}
\includegraphics[width=0.69\columnwidth,angle=0]{inpaper/isogal_hr_z3_gf0_6_0_5ZSol_phot_nomet_dustcazcool_C1_UVz3_shield_05100_rhoT.png}
\includegraphics[width=0.69\columnwidth,angle=0]{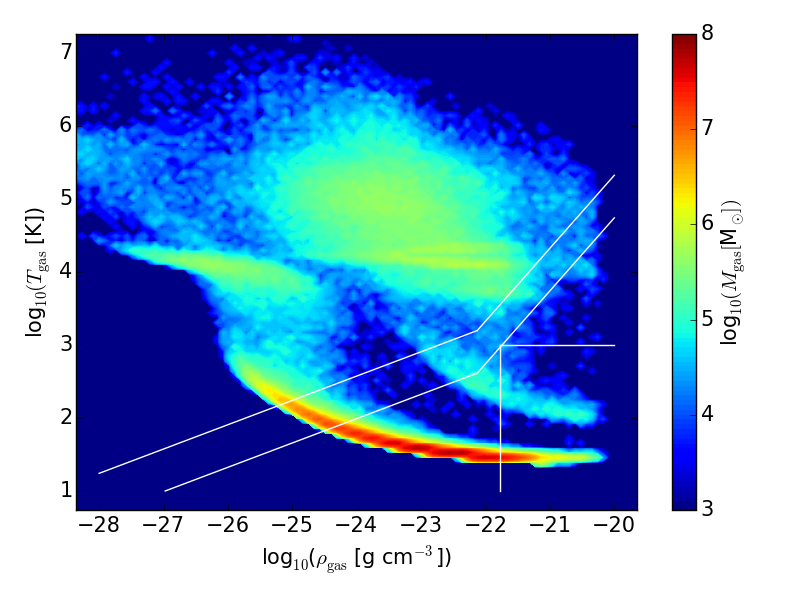}
\includegraphics[width=0.69\columnwidth,angle=0]{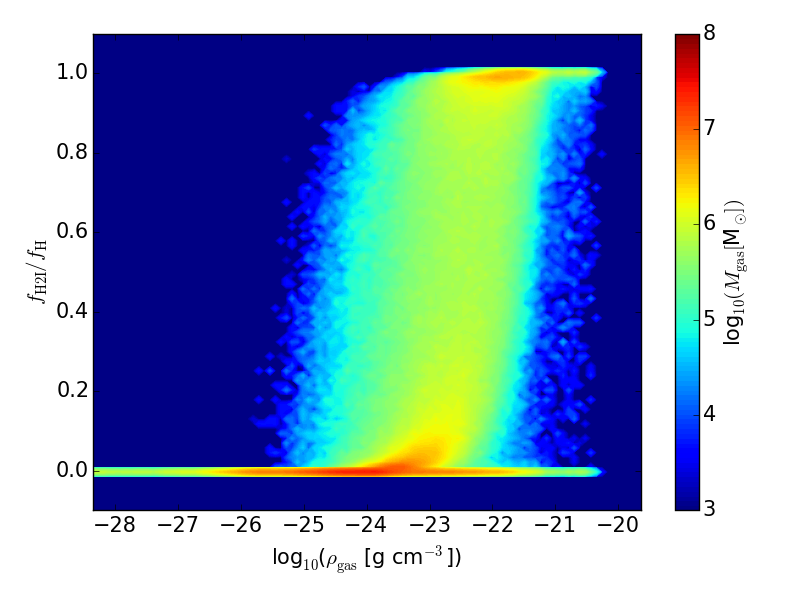}
\includegraphics[width=0.69\columnwidth,angle=0]{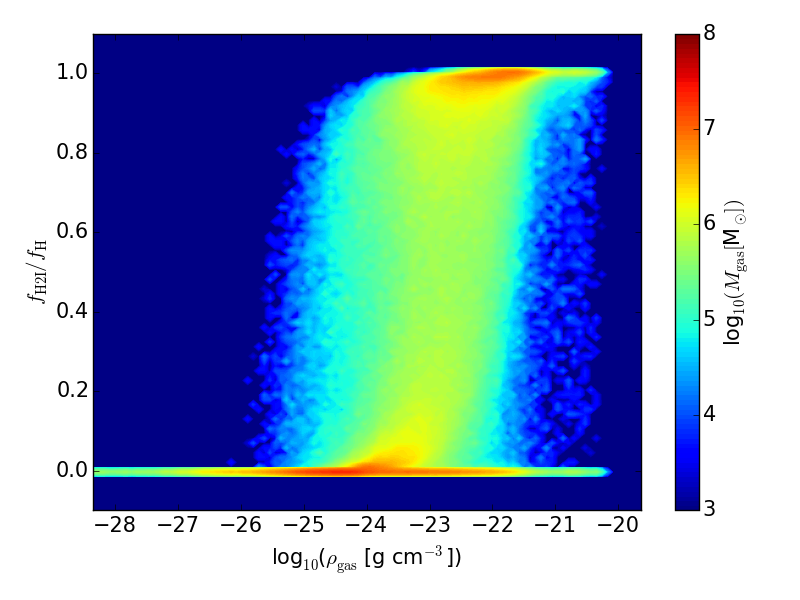}
\includegraphics[width=0.69\columnwidth,angle=0]{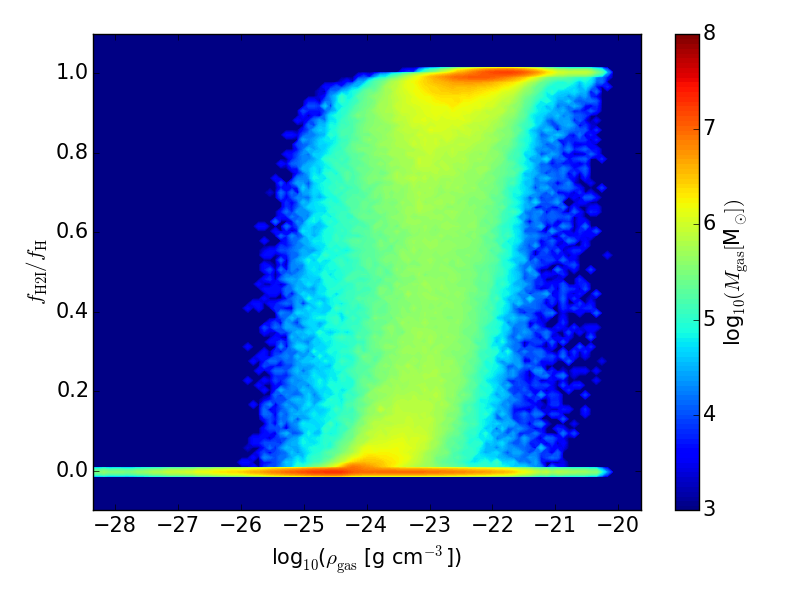}
\vspace{-0.0pt}
\caption[]{Dependence on the initial gas metallicity. Gas phase diagrams at 0.4~Gyr for eq-metals runs of different initial metallicity [Run~05 ($0.1Z_{\odot}$; left-hand panels), Run~01 ($0.5Z_{\odot}$; central panels), and Run~04 ($Z_{\odot}$; right-hand panels)]: metallicity (upper panels), temperature (middle panels), and H2I hydrogen mass fraction (lower panels), all as a function of gas density. The spread of metallicity and the gas $\rho$--$T$ structure are similar amongst the three runs, whereas the amount of H2I increases with $f_{\rm metals}$.
}
\label{gaso2krome_isogal:fig:phasediagrams_run05vs01vs04}
\end{figure*}

\begin{figure*}
\centering
\vspace{-8.0pt}
\includegraphics[width=0.69\columnwidth,angle=0]{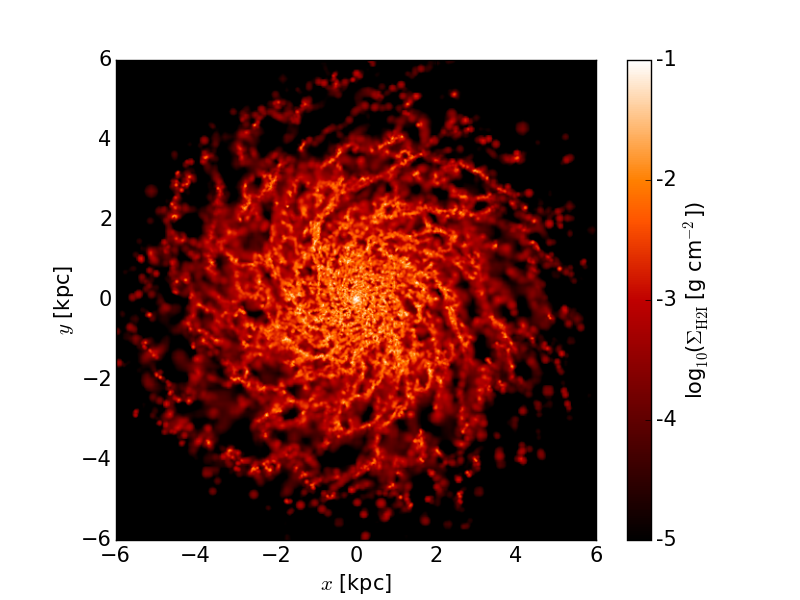}
\includegraphics[width=0.69\columnwidth,angle=0]{inpaper/isogal_hr_z3_gf0_6_0_5ZSol_phot_nomet_dustcazcool_C1_UVz3_shield_05100_H2Imap_12kpc_faceon.png}
\includegraphics[width=0.69\columnwidth,angle=0]{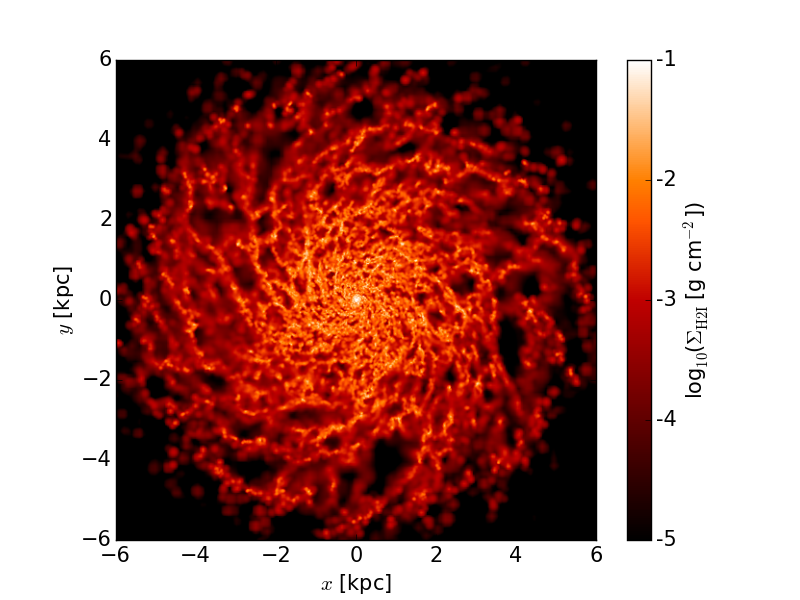}
\includegraphics[width=0.68\columnwidth,angle=0]{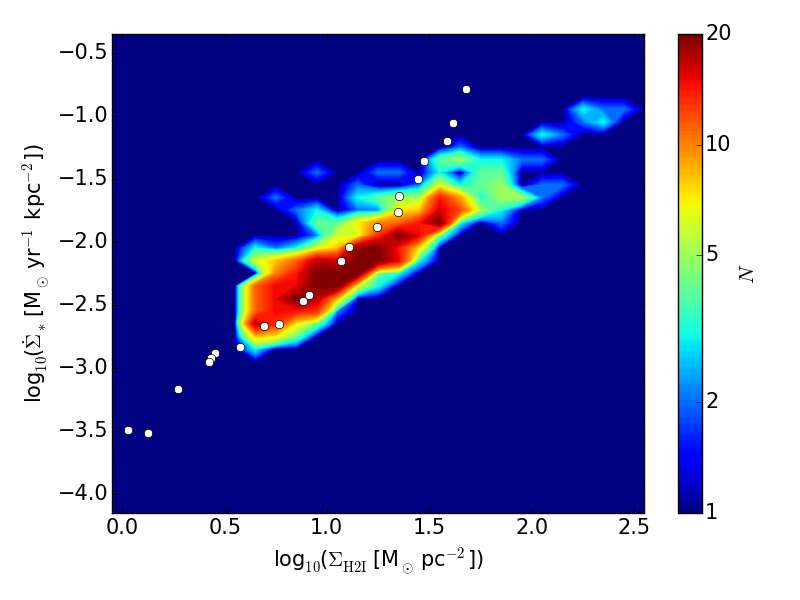}$\;\;$
\includegraphics[width=0.68\columnwidth,angle=0]{inpaper/isogal_hr_z3_gf0_6_0_5ZSol_phot_nomet_dustcazcool_C1_UVz3_shield_05100_ks_H2I_Bigiel.png}$\;$
\includegraphics[width=0.68\columnwidth,angle=0]{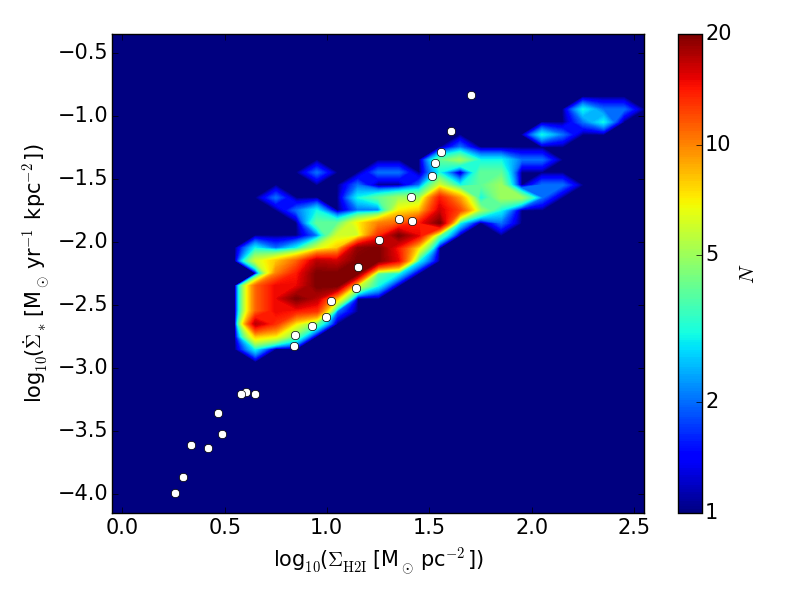}
\includegraphics[width=0.68\columnwidth,angle=0]{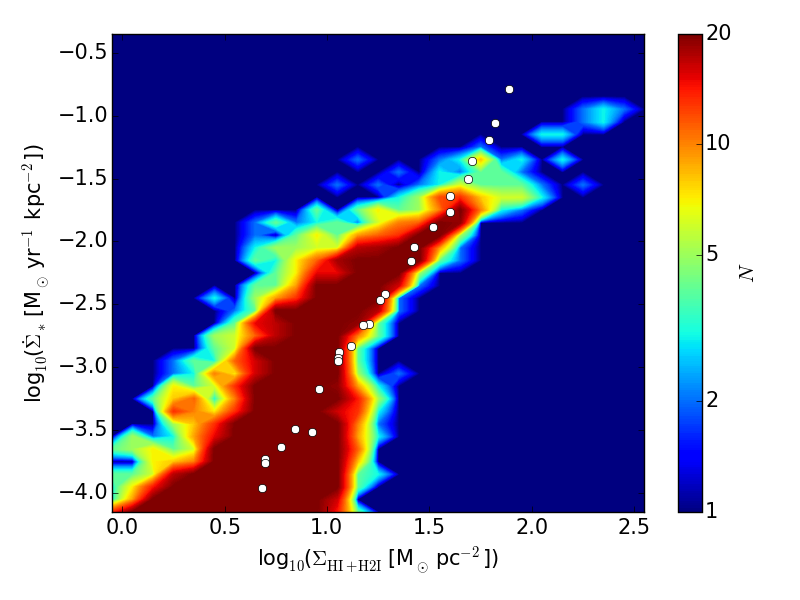}$\;\;$
\includegraphics[width=0.68\columnwidth,angle=0]{inpaper/isogal_hr_z3_gf0_6_0_5ZSol_phot_nomet_dustcazcool_C1_UVz3_shield_05100_ks_HI_and_H2I_Bigiel.png}$\;$
\includegraphics[width=0.68\columnwidth,angle=0]{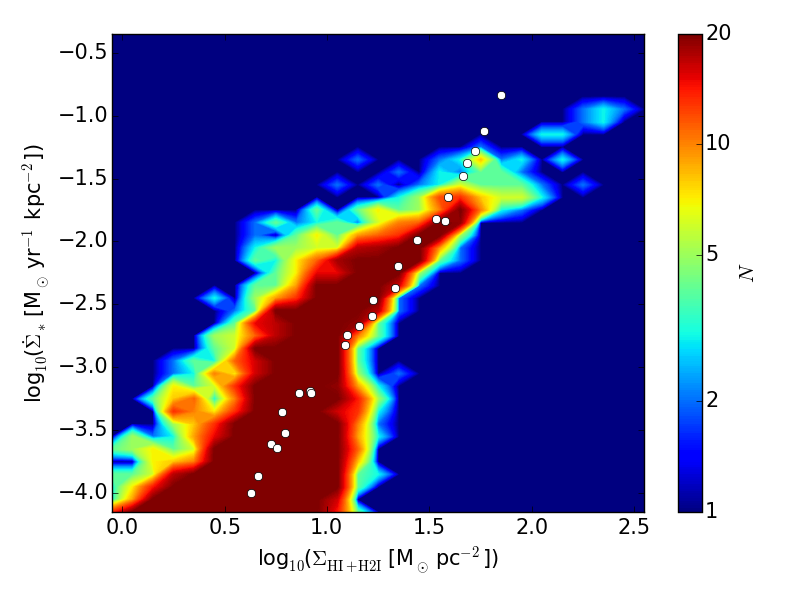}
\vspace{-0.0pt}
\caption[]{Dependence on the initial gas metallicity. Face-on surface density maps of the H2I gas (upper panels) and KS diagrams for the H2I gas (middle panels) and HI\texttt{+}H2I gas (lower panels) at 0.4~Gyr, for eq-metals runs of different initial metallicity: Run~05 ($0.1Z_{\odot}$; left-hand panels), Run~01 ($0.5Z_{\odot}$; central panels), and Run~04 ($Z_{\odot}$; right-hand panels). The decrease of H2I with decreasing metallicity implies a less steep slope in the molecular KS diagram, which better matches observations.
}
\label{gaso2krome_isogal:fig:gasandH2IsurfdenmapsandKS_run05vs01vs04}
\end{figure*}

In this section, we describe the results of our ten simulations, mostly by means of comparing different sub-sets of runs. In Section~\ref{gaso2krome_isogal:sec:eqmetvsnoneqmet}, we assess the effects of including the chemical evolution and non-equilibrium cooling (for $T_{\rm gas} < 10^4$~K) of seven metal species. In Section~\ref{gaso2krome_isogal:sec:dependence_on_metallicity}, we study the dependence on the (initial) gas metallicity. In Section~\ref{gaso2krome_isogal:sec:H2Iformation}, we evaluate the dependence of the results on how the formation of H2I on dust is modelled. In Section~\ref{gaso2krome_isogal:sec:metals_as_tracers}, we focus on the role of metals as tracers of the ISM and SFR. Unless otherwise stated, all results are shown at 0.4~Gyr, but we note that, being the system an isolated galaxy, all results hold also at earlier times ($t \gtrsim 0.15$~Gyr).


\subsection{Equilibrium versus non-equilibrium metal cooling}\label{gaso2krome_isogal:sec:eqmetvsnoneqmet}

In this section, we compare the results of two simulations, which are identical in everything  except in the treatment of metal chemistry and cooling. We focus here on the comparison between Runs~01 and 01m, but we note that similar comparisons (i.e. Runs~02 and 02m, Runs~03 and 03m, Runs~04 and 04m, and Runs~05 and 05m) yield similar results, unless otherwise stated. We remind the reader that we use eq-metal -- PIE -- cooling in both runs for $T_{\rm gas}\, \ge \,10^4$~K, since we assume that gas at those temperatures is in equilibrium. Therefore, we expect to see differences only for $T_{\rm gas} \lesssim 10^4$~K.

We summarise the main differences in the modelling of metals between Runs~01 and 01m. In addition to the obvious fact that (i) metal cooling is computed assuming equilibrium in Run~01 and lack thereof in Run~01m, we note that (ii) the network utilised for the PIE tables contains the first 30 elements in the periodic table, whereas the non-eq-metal computations include seven metal species. Additionally, we stress that (iii) the PIE tables were constructed assuming an optically thin gas, whereas the non-equilibrium metal cooling explicitly assumes a shielded UV background.

Assuming an optically thin or thick gas has potentially considerable implications. As a reference, at $T_{\rm gas} < 10^4$~K, the PIE \citep[UV at $z = 3$;][]{Haardt_Madau_2012} cooling is two orders of magnitude stronger than the CIE (no UV) cooling (\citealt{Bovino_et_al_2016}; see also \citealt{Shen_et_al_2010}). Therefore, the PIE tables likely over-estimate the metal cooling at low temperatures.

The different number of species can also in principle produce different values of metal cooling. For example, \citeauthor{Richings_et_al_2014a} (\citeyear{Richings_et_al_2014a}; see also \citealt{Richings_et_al_2014b}) claim that Fe, which we have not included in our network, could be potentially relevant. On the other hand, Fe is believed to be strongly depleted into solid form \citep[][]{Jenkins_2009}, and there is general agreement that C, O, and Si (the metal elements we do follow in our simulations) are by far the dominant metal coolants in the ISM \citep[e.g.][]{Tielens_Hollenbach_1985,Wolfire_et_al_1995,Wolfire_et_al_2003,Shen_et_al_2010}.

Using the same networks employed in this work, \citet{Bovino_et_al_2016} performed isochoric one-zone tests with constant gas temperature to compute the ratio of equilibrium to non-equilibrium metal cooling (in both cases in the optically thin approximation; see their Fig.~16), showing that gas at both low (0.1~cm$^{-3}$) and high ($10^2$~cm$^{-3}$) densities experiences two evolution regimes: during the first part of the evolution, the non-equilibrium metal cooling is stronger than the PIE cooling (even by orders of magnitude), because of the larger number of electrons (due to incomplete recombinations) and consequent enhanced collisions; during the second part, the system evolves towards an equilibrium state and the differences between equilibrium and non-equilibrium metal cooling tend to diminish, with the ratio between the eq-metal to non-eq-metal cooling rates (both at equilibrium) varying between $\sim$0.6 (for low $T_{\rm gas}$) and $\sim$3 (for $T_{\rm gas} = 9 \times 10^3$~K).

In Fig.~\ref{gaso2krome_isogal:fig:rhoT_run01vs01m}, we show the gas density--temperature diagrams for the two runs. In both diagrams, one can observe three distinct regions in the phase space: (1) the gas with $T_{\rm gas} \gtrsim 10^4$~K and $10^{-25} \lesssim \rho_{\rm gas} \lesssim 10^{-22}$~g~cm$^{-3}$ is the ISM heated by SNae;\footnote{We note that this relatively dense, hot gas is kept artificially hot for some time because of the cooling shut-off SN model (see Section~\ref{gaso2krome_isogal:sec:SF_and_SN} and \citealt{Stinson_et_al_2006} for more details), hence it only shares the gas temperature with the observed hot ionized medium \citep[][]{Ferriere_2001}.} (2) this same gas eventually cools down and fills a region of the phase diagram at $\rho_{\rm gas} \lesssim 10^{-25}$~g~cm$^{-3}$ and $T_{\rm gas} \sim 10^4$~K (the so-called warm neutral medium, WNM; e.g. \citealt{Ferriere_2001}), under which temperature the magnitude of the cooling function drops dramatically; (3) part of this gas, together with the gas that was never heated by SNae, manages to cool down to $T_{\rm gas} < 10^4$~K, with the shape of this third region heavily depending on the low-temperature cooling function.

The phase diagrams exhibit differences according to the above expectations. The eq-metals gas is on average colder than the non-eq-metals gas at all densities, with the difference in temperature increasing as the density decreases, ranging from a factor of a few at high densities to $\sim$one order of magnitude at low densities. In particular, the gas mass fraction of WNM (defined here as the gas with $3.8 < \log_{10}{(T_{\rm gas} [K]) < 4.2}$ and $\rho_{\rm gas} < 10^{-24}$~g~cm$^{-3}$) is much larger in the non-eq-metals case than in the eq-metals case (16 versus 2 per cent). Most importantly, the gas mass fraction of gas with $T_{\rm gas} < 10^2$~K (the upper temperature limit for the cold neutral medium; e.g. \citealt{Ferriere_2001}) is 1.5 and 76.8 per cent for the non-eq-metals and eq-metals gas, respectively.

Because of the substantial differences between CIE and PIE cooling for $T_{\rm gas} < 10^4$~K observed in, e.g. \citet{Shen_et_al_2010} and \citet{Bovino_et_al_2016}, we attribute the differences between Runs~01 and 01m mostly to the optically thin/thick approximation in the metal cooling. To assert this, we ran a version of Runs~01 and 01m in which we assumed the UV background to not be shielded and found that the gas thermodynamics of the two no-shielding runs is very similar (see Appendix~\ref{gaso2krome_isogal:sec:The_importance_of_shielding}), with a couple of minor differences: (a) in the density range $\rho_{\rm gas} \lesssim 10^{-23}$~g~cm$^{-3}$, the slope is slightly different (constant temperature versus marginally decreasing temperature); (b) in the density range $10^{-23} \lesssim \rho_{\rm gas} \lesssim 10^{-22}$~g~cm$^{-3}$, the eq-metals gas is slightly colder than the non-eq-metals gas. The differences in the two gas $\rho$--$T$ diagrams of the original Runs~01 and 01m are therefore likely due to the implementation of shielding (or lack thereof), rather than to the differences in the modelling of metals.

In order to understand the causes of the remaining, minor differences and disentangle non-equilibrium effects from those due to different metal networks, we once again consider the isochoric tests at constant temperature performed by \citet{Bovino_et_al_2016}. We note that the non-eq-metals models reach equilibrium after $\sim$$10^3$--$10^5$~yr, depending on the density. These time-scales are much shorter than the dynamical time-scales at those densities, implying that, in our simulations of an isolated galaxy, the gas can be considered somewhat in equilibrium. We caution, however, that the comparison in \citet{Bovino_et_al_2016} is on metal cooling rather than on net cooling. Therefore, we conservatively state that the minor differences that are not attributable to shielding are due to a combination of different metal networks and non-equilibrium effects.

The different $T_{\rm gas}$ has direct consequences on the chemical composition of the gas. In Fig.~\ref{gaso2krome_isogal:fig:rhomu_run01vs01m}, we show the dependence of the mean molecular weight on gas density. In both runs, the vast majority of the gas lies around (and in between) the two phase-space regions of neutral molecular ($\mu \sim 2.30$) and neutral atomic ($\mu \sim 1.24$) gas, with only a small fraction as ionized atomic ($\mu \sim 0.60$) gas. The difference between the two runs lies predominantly in the neutral molecular region: there is more H2I in the non-eq-metals gas (23 per cent of the gas mass has $\mu > 2.26$) than in the eq-metals gas (13 per cent). This is because formation of H2I on dust, according to the CS09 model, peaks at $\log_{10}(T_{\rm gas}) \sim 2.2$ (owing to the opposing dependences on $T_{\rm gas}$ of dust sticking and gas thermal velocity), and there is much more non-eq-metals gas at or around the peak temperature [36 per cent of the gas mass has $2 < \log_{10}(T_{\rm gas} [K]) < 2.4$] than eq-metals gas (6 per cent). This interpretation is corroborated by the fact that the gas mass fractions of gas with $\mu > 2.26$ for the two runs using the J75 model of H2I formation on dust (in which there is no dependence on gas temperature) are closer to each other: 8 and 13 per cent for Runs~03 and 03m, respectively. Moreover, even though they are less important than catalysis on dust, we note that the gas-phase reactions of H2I formation peak at $10^{4}$--$10^{5}$~K, where the non-eq-metals run has much more gas than in the eq-metals run (see also the difference in WNM gas mass stated above).

A colder gas is also prone to more fragmentation \citep[][]{Toomre_1964}. In Fig.~\ref{gaso2krome_isogal:fig:gassurfdenmaps_run01vs01m}, we show the surface density maps, viewed face-on, for the total gas. The central ($\lesssim 5$~kpc), denser regions of the galaxy are very similar in the two runs. On the other hand, the outer regions -- where the gas density is lower and therefore the temperature difference between the two runs is larger (see Fig.~\ref{gaso2krome_isogal:fig:rhoT_run01vs01m}) -- are quite different. The eq-metals gas map shows a clumpy outer region, whereas the outer gas in the non-eq-metals run is very smooth. The H2I maps in Fig.~\ref{gaso2krome_isogal:fig:H2Isurfdenmaps_run01vs01m} show a slightly more extended disc in the eq-metals case. This is linked to the enhanced fragmentation in Run~01, where the local H2I fraction is increased within clumps. We note, however, that, despite this, the total amount of H2I is larger in the non-eq-metals run.

\begin{figure*}
\centering
\vspace{-3.0pt}
\includegraphics[width=0.69\columnwidth,angle=0]{inpaper/isogal_hr_z3_gf0_6_0_5ZSol_phot_nomet_dustcazcool_C1_UVz3_shield_05100_rhoXH2I.png}
\includegraphics[width=0.69\columnwidth,angle=0]{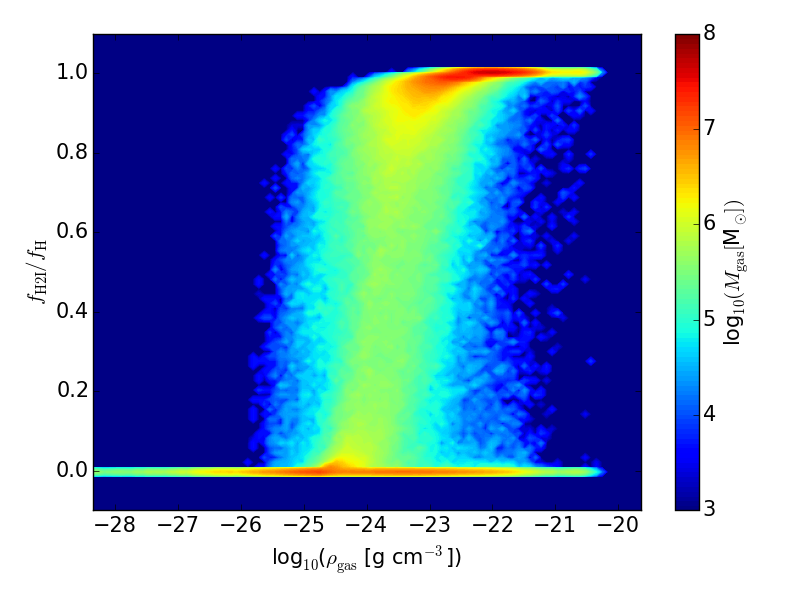}
\includegraphics[width=0.69\columnwidth,angle=0]{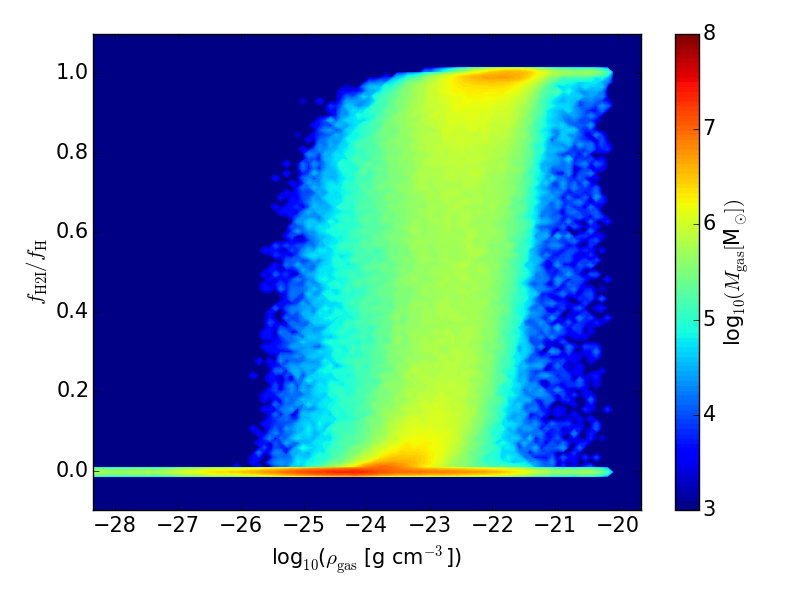}
\vspace{-0.0pt}
\caption[]{Dependence on the model of H2I formation on dust. We compare the H2I hydrogen mass fraction for Run~01 (CS09 and $C_{\rho} = 1$; left-hand panel), Run~02 (CS09 and $C_{\rho} = 10$; central panel), and Run~03 (J75 and $C_{\rho} = 1$; right-hand panel) at 0.4~Gyr.
}
\label{gaso2krome_isogal:fig:run01vs02vs03}
\end{figure*}

The large difference in the density--temperature diagrams does not translate into large differences in SF, as it would be naively expected (since SF occurs in cold dense gas). In Fig.~\ref{gaso2krome_isogal:fig:sfr}, we show the SFR as a function of time for Runs~01 and 01m. In both simulations, the SFR varies around 0.7--0.9~M$_{\odot}$~yr$^{-1}$, with the non-eq-metals run forming slightly more stars than the eq-metals run. The difference, however, is minimal and approaches zero in the last $\sim$100~Myr of the simulation, with the final stellar mass equal to $2.7 \times 10^9$~M$_{\odot}$ in both runs.

In Figs~\ref{gaso2krome_isogal:fig:ks_HIH2I_bigiel_run01vs01m} and \ref{gaso2krome_isogal:fig:ks_H2I_bigiel_run01vs01m}, we show the classic and molecular KS diagrams, respectively. To construct these diagrams, we positioned the galaxy face-on, divided it into 50 concentric cylindrical bins of thickness 200~pc between 0 and 10~kpc, and computed the surface density of the HI\texttt{+}H2I gas (for the classic KS diagram) or H2I gas (for the molecular KS diagram), and of the SFR, defined here as the mass of new stars formed in the past 100~Myr, divided by 100~Myr. In all the diagrams, we over-plotted the data by \citet{Bigiel_et_al_2010}.\footnote{From their online data, we combined the Spiral2 sample of Table~2 (applying the filter $\log_{10}(\dot{\Sigma}_{*} [{\rm M}_{\odot}~{\rm yr}^{-1}~{\rm kpc}^{-2}]) > -3.7$) and the Spirals sample of Table~3.} We note that our simulations agree very well with the observed classic KS relation, both in slope and in normalisation (the only discrepancy occurring at high gas surface densities, due to small number statistics in the observed data). This was expected, since the SF recipe and SF efficiency parameter were constructed and tuned to match observations.

Our data also agree reasonably well with the normalisation of the observed molecular KS relation -- in the sense that the simulated and observed curves intersect in the middle of the gas surface density range ($10 \lesssim \Sigma_{\rm H2I} \lesssim 30$~M$_{\odot}$~pc$^{-2}$) observed by \citet{Bigiel_et_al_2010} -- although they have a steeper slope than the observed one. As noted in the next sections, however, the slope discrepancy becomes less severe when we decrease the gas metallicity (see Section~\ref{gaso2krome_isogal:sec:dependence_on_metallicity}) or change the method of H2I formation on dust (see Section~\ref{gaso2krome_isogal:sec:H2Iformation}). We stress here that, in our SF recipe, we did not directly link the abundance of molecular hydrogen to SF, as done in other works \citep[e.g.][]{Gnedin_et_al_2009,Christensen_et_al_2012,Tomassetti_et_al_2015,Pallottini_et_al_2017b}. The correlation between H2I and SFR surface densities emerges naturally from the simulations, hinting to the idea that SF occurs in cold, dense gas, where molecular hydrogen is generally more abundant \citep[see also discussion in][]{Lupi_et_al_2017}.

\begin{figure*}
\centering
\vspace{-6.0pt}
\includegraphics[width=1.04\columnwidth,angle=0]{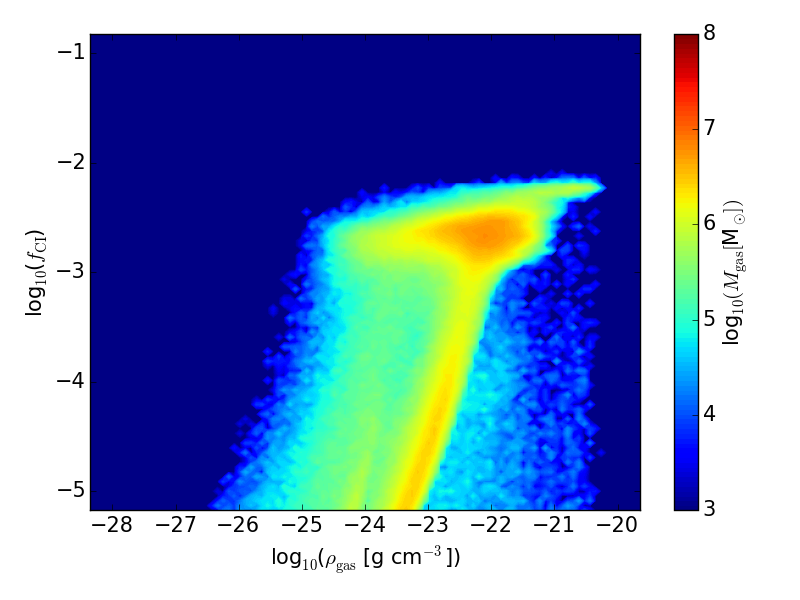}
\includegraphics[width=1.04\columnwidth,angle=0]{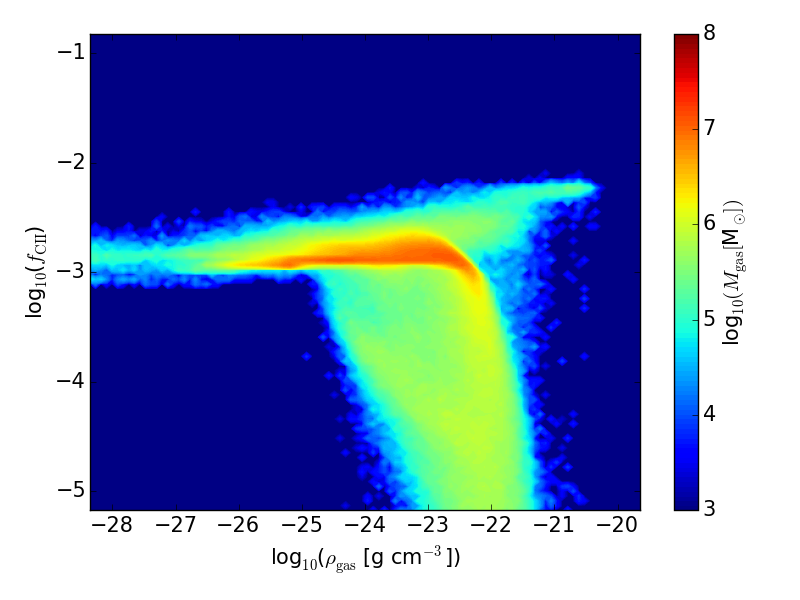}
\includegraphics[width=1.04\columnwidth,angle=0]{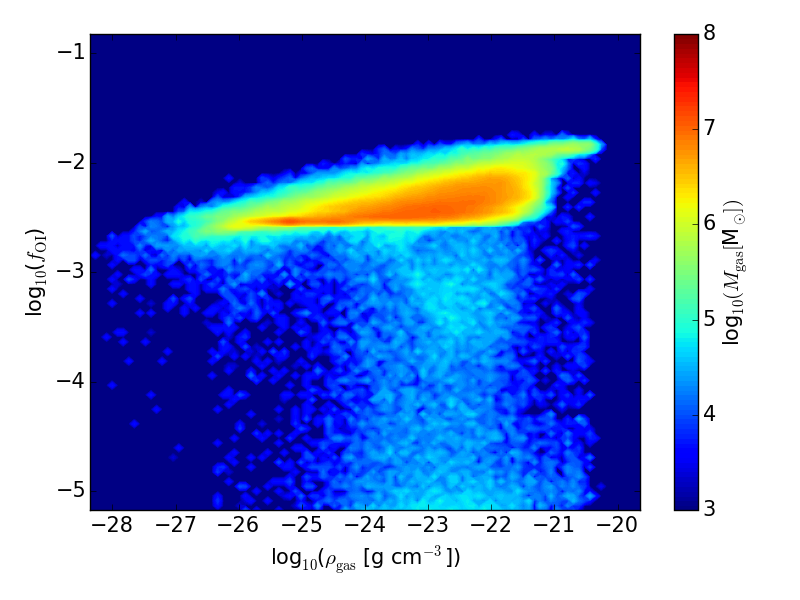}
\includegraphics[width=1.04\columnwidth,angle=0]{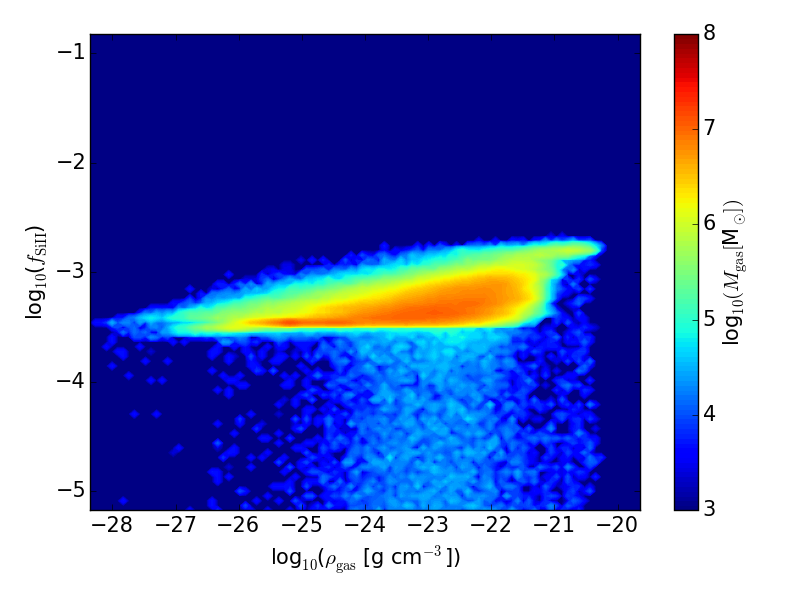}
\vspace{-0.0pt}
\caption[]{Metals as tracers of the ISM. Mass fractions at 0.4~Gyr, all as a function of gas density, for the non-eq-metals Run~01m: CI (upper-left panel), CII (upper-right), OI (lower-left), and SiII (lower-right). OI and SiII trace the entire ISM, whereas CI and CII trace different ISM regimes.
}
\label{gaso2krome_isogal:fig:metalphasediagrams_run01m}
\end{figure*}

\begin{figure*}
\centering
\vspace{-12.5pt}
\includegraphics[width=1.04\columnwidth,angle=0]{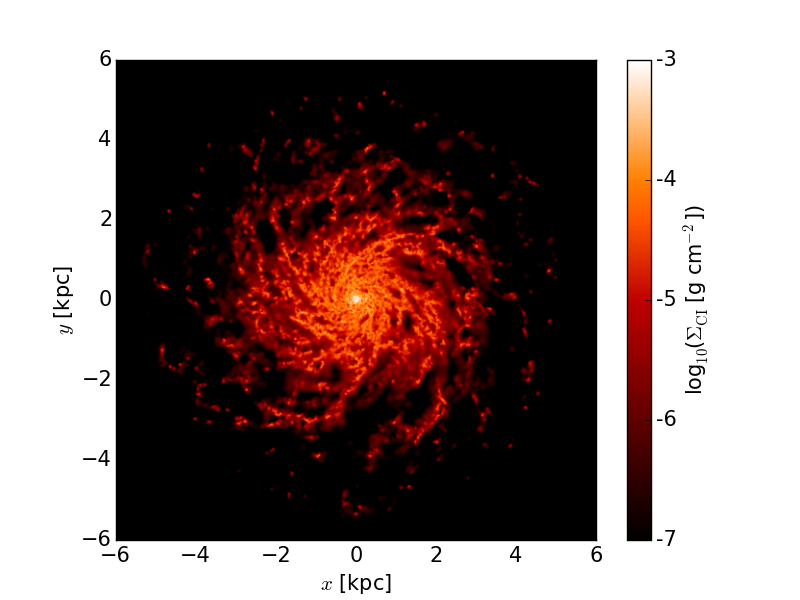}
\includegraphics[width=1.04\columnwidth,angle=0]{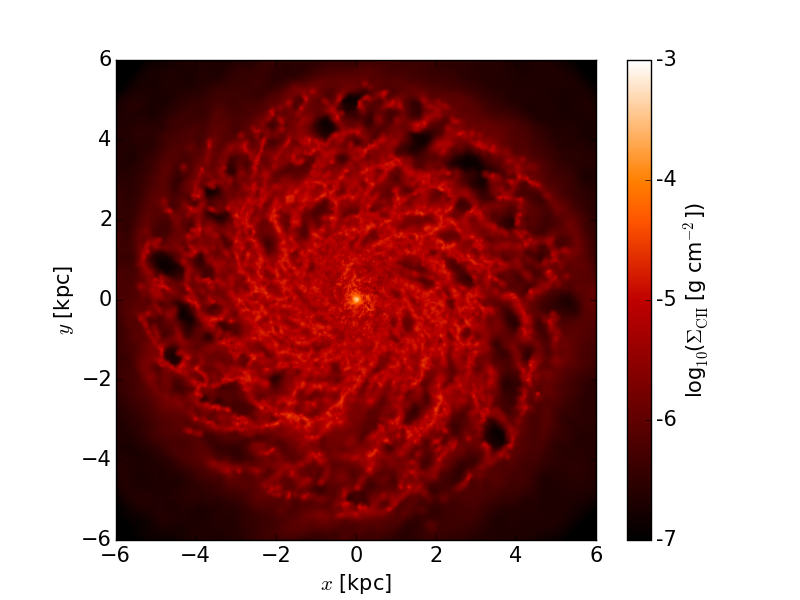}
\includegraphics[width=1.04\columnwidth,angle=0]{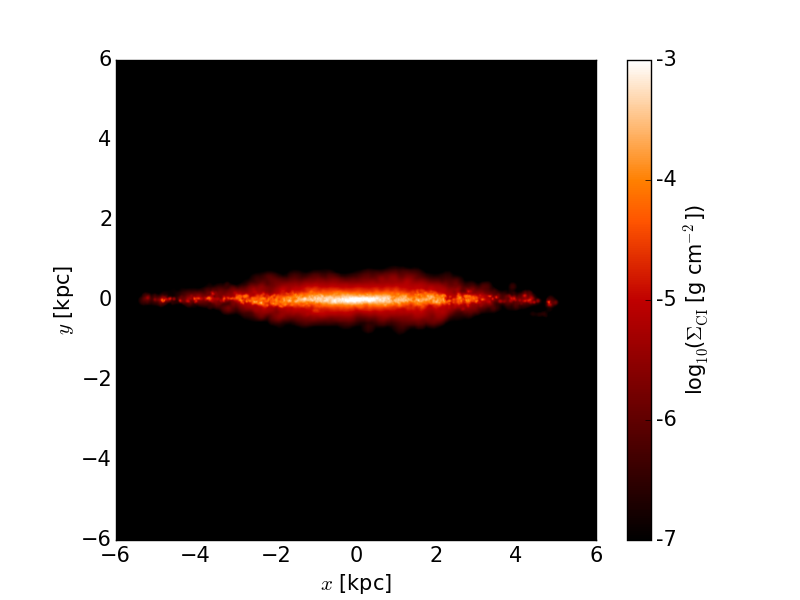}
\includegraphics[width=1.04\columnwidth,angle=0]{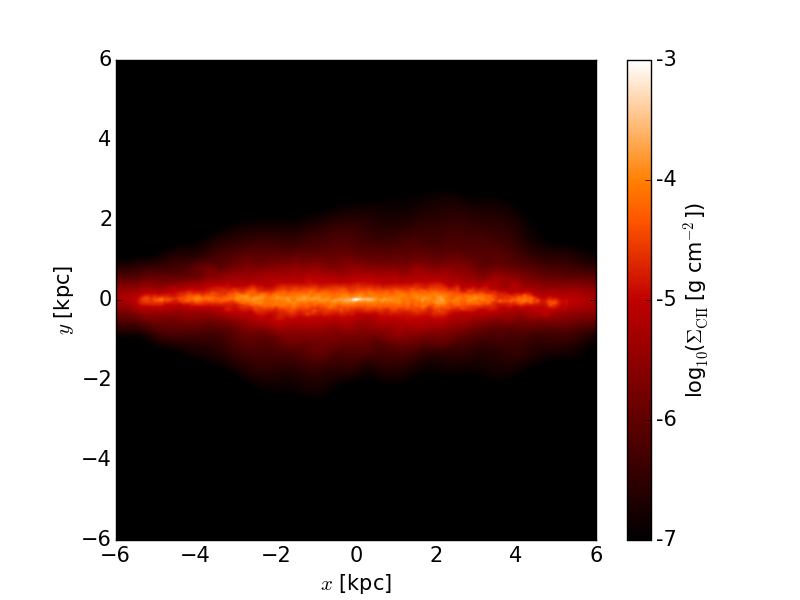}
\vspace{8.0pt}
\caption[]{Metals as tracers of the ISM. Surface density maps at 0.4~Gyr for the non-eq-metals Run~01m: face-on CI (upper-left panel), face-on CII (upper-right), edge-on CI (lower-left), and edge-on CII (lower-right).
}
\label{gaso2krome_isogal:fig:metalsurfdenmaps_run01m}
\end{figure*}

The reason why the SFR and KS diagrams are not that different despite the large differences in the gas density--temperature diagrams is because those differences mostly occur at relatively low densities. The gas structure in the SF region (defined by $T_{\rm gas} < T_{\rm SF}$ and $\rho_{\rm gas} > \rho_{\rm SF}$) is almost the same, as the gas mass fraction of potentially star-forming gas in Runs~01 and 01m is 7 and 8 per cent, respectively. Moreover, the difference in H2I abundance is not large enough to produce discernible discrepancies in the molecular KS diagrams. All of the above suggests that our model is relatively robust, when it comes to SF observables, in spite of some differences in the gas thermodynamics. Indeed, the mass of potentially star-forming gas is very similar for a large range of SF density and temperature thresholds, down to $T_{\rm SF} \sim 10^2$~K, when the gas thermodynamics differences become important and the eq-metals simulation forms more stars than the non-eq-metals run.


\subsection{Dependence on metallicity}\label{gaso2krome_isogal:sec:dependence_on_metallicity}

In this section, we study the dependence on the (initial) metallicity of the gas, by comparing Runs 01, 04, and 05. We remind the reader that, due to continuous metal injection from SNae and turbulent metal diffusion, the metallicity of the gas varies significantly with time, as it can be seen in the upper panels of Fig.~\ref{gaso2krome_isogal:fig:phasediagrams_run05vs01vs04}. We note that the magnitude of the spread in metallicity is basically the same in all runs: the amount of metals deposited into the gas from SNae is more or less the same, regardless of the initial metallicity. This is because the amount of SF (and subsequent SNae) is very similar, yielding the same increase in average $Z/Z_{\odot}$ (equal to 0.3) for all three runs. For this reason, the final metallicity simply depends on the initial metallicity, which was varied to test its effects on the gas thermodynamics and abundances and not to reproduce the observed mass--metallicity relation. Nevertheless, given the wide range of possible values of gas metallicity for a stellar mass of $2.7 \times 10^9$~M$_{\odot}$ ($Z/Z_{\odot} \sim 0.2$--1.3; see, e.g. \citealt{Sanchez_et_al_2017}), all our galaxies are within the observed range.

In our model, there are several physical phenomena linked to the amount of metals in the gas. Metal cooling increases with metallicity, because of the enhanced density of colliders (ions, electrons), and this increase is commonly modelled in PIE runs with the metal cooling scaling linearly with $Z/Z_{\odot}$. A higher metallicity also implies a higher dust-to-gas ratio, which in this work is assumed to scale as $D = D_{\odot} Z/Z_{\odot}$. Consequently, we also expect stronger gas cooling by dust, a larger abundance of molecular hydrogen, and stronger molecular-hydrogen cooling (which, however, at the densities and metallicities considered in this work, is sub-dominant with respect to metal cooling).

In agreement with the above expectations, the gas density--temperature diagrams (middle panels of Fig.~\ref{gaso2krome_isogal:fig:phasediagrams_run05vs01vs04}) show that the gas gets colder as the metallicity increases. In particular, when comparing Runs~05 (of initial metallicity $0.1 Z_{\odot}$) and 04 ($Z_{\odot}$), we observe relatively more gas in the low-metallicity run at $\sim$$10^4$~K (at low densities) and in the region of phase space that connects the $10^4$-K gas with the $10^2$-K gas (at $\sim$$10^{-25}$~g~cm$^{-3}$). Moreover, if we select gas at $10^{-26}$~g~cm$^{-3}$, the coldest gas in Runs~05 and 04 has a temperature of $\sim$$10^4$ and 300~K, respectively. The dependence on $f_{\rm metals}$, however, does not appear to be very strong: this is due to the spread in metallicity, which reduces the effective difference between runs (e.g. from the initial factor of 10 to only a few, when comparing Runs~05 and 04). This is the likely explanation why the effect is visibly stronger in \citet[][]{Richings_Schaye_2016}, who do not include any chemical enrichment from SNae (nor turbulent metal diffusion) and therefore have a constant metallicity throughout their simulations.

As expected from the direct link between metals, dust, and H2I formation, the amount of molecular gas increases with metallicity, as it can be seen in the lower panels of Fig.~\ref{gaso2krome_isogal:fig:phasediagrams_run05vs01vs04}. The gas mass fraction of gas with $f_{\rm H2I}/f_{\rm H} > 0.9$ for Runs~05, 04, and 01 is 12, 22, and 30 per cent, respectively. We note that the density at which the gas starts transitioning from atomic to molecular is more or less the same in the three runs (at $\sim$$10^{-24}$~g~cm$^{-3}$), in disagreement with previous studies (both assuming equilibrium -- \citealt{McKee_Krumholz_2010}; \citealt{Sternberg_et_al_2014} -- and not -- \citealt{Gnedin_et_al_2009}). This is caused by the fact that metal injection and diffusion smear out the differences in metallicity between the runs (for example, the ratio between the average metallicities of Runs 04 and 05 starts at 10 and ends at 3). We also note a wide spread of densities in the region of gas transitioning from atomic to molecular.

In Fig.~\ref{gaso2krome_isogal:fig:gasandH2IsurfdenmapsandKS_run05vs01vs04}, we show gas maps and KS diagrams for the same runs. We note that there is almost no difference in the total gas surface density (not shown) but there is a clear increase of H2I with metallicity (upper panels; consistent with the lower panels of Fig.~\ref{gaso2krome_isogal:fig:phasediagrams_run05vs01vs04}). This difference in H2I abundance translates into a visible change in the molecular KS law (middle panels): because the difference is larger at low densities (as at higher densities the gas saturates to fully molecular), the molecular KS law in Run~05 has a slope that matches quite well the data by \citet{Bigiel_et_al_2010}, even though the observed galaxies, being local, have a slightly higher metallicity (we remind the reader that the final average gas metallicity of Run~05 is $Z/Z_{\odot} = 0.4$). The classic KS diagrams (lower panels) do not show any difference, as expected.

When performing the same kind of comparison with the non-eq-metals runs (i.e. Runs~01m, 04m, and 05m), we obtain similar results.


\subsection{H2I formation on dust: models and clumping}\label{gaso2krome_isogal:sec:H2Iformation}

In this section, we briefly study the effects of changing the model of formation of H2I on dust (J75 versus CS09) and the value of the clumping factor ($C_{\rho} = 1$ or 10). We chose the value 10 because it is the same value used in \citet{Gnedin_et_al_2009}, \citet{Christensen_et_al_2012}, and \citet{Tomassetti_et_al_2015}. Moreover, it is also consistent with the values found in the simulations with a variable clumping factor by \citet{Lupi_et_al_2017}.

The amount of H2I varies significantly when we vary the formation model and/or the clumping factor (see Fig.~\ref{gaso2krome_isogal:fig:run01vs02vs03}). As expected, assuming a clumping factor of 10 increases the amount of H2I: the gas mass fraction of gas with $f_{\rm H2I}/f_{\rm H} > 0.9$ for Runs~01 ($C_{\rho} = 1$) and 02 ($C_{\rho} = 10$) is 22 and 50 per cent, respectively. The choice of formation model is also important, with the J75 model (Run~03) yielding less H2I than the CS09 model (15 versus 22 per cent), but not as crucial as the choice of $C_{\rho}$. This is because our simulations are all run at relatively high metallicity, where the two models of H2I formation on dust are most similar. The differences are visible also in the H2I surface density maps (not shown), even though they are quite clear only when comparing different clumping factors.

\begin{figure*}
\centering
\vspace{-6.0pt}
\includegraphics[width=1.04\columnwidth,angle=0]{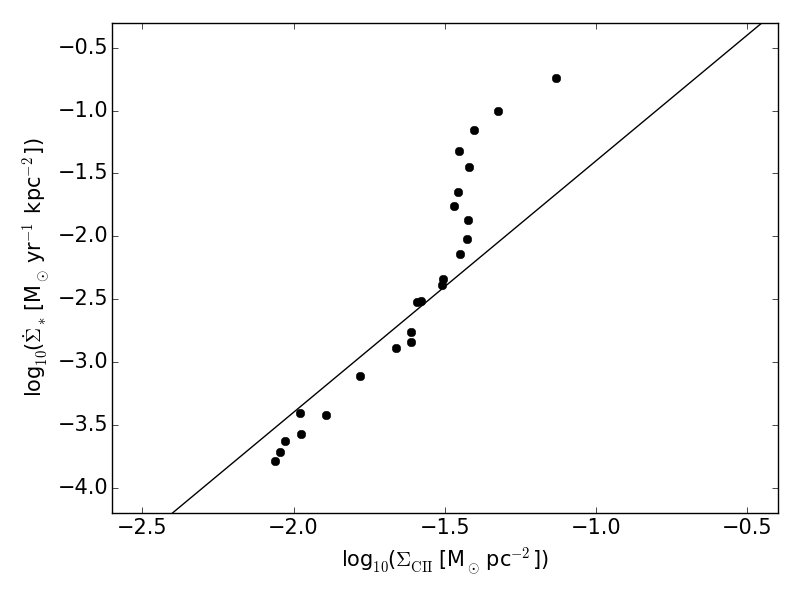}
\includegraphics[width=1.04\columnwidth,angle=0]{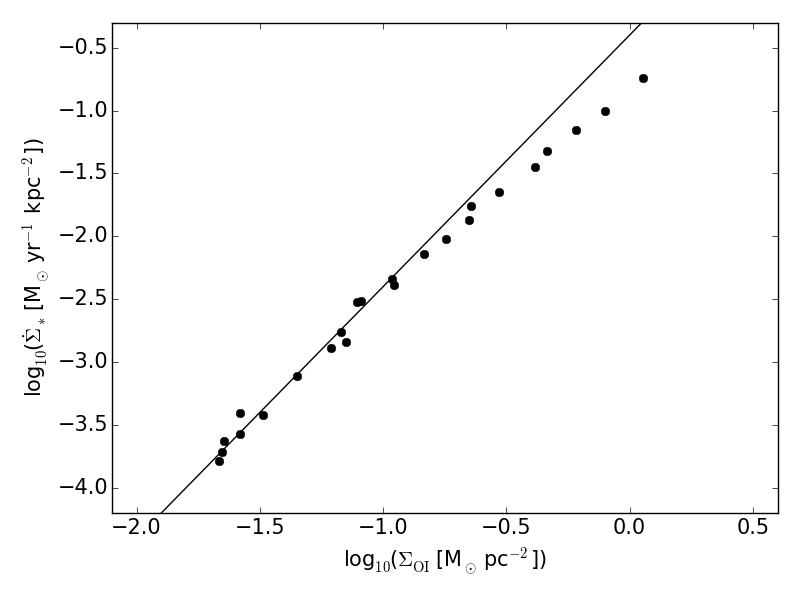}
\vspace{-0.0pt}
\caption[]{KS-like diagrams at 0.4~Gyr, for the non-eq-metals Run~01m, between CII (left-hand panel) or OI (right-hand panel) and SFR (in the last 100~Myr) surface densities, for the 50 innermost 200-pc-thick concentric cylindrical bins (black points). The solid line in the left-hand panel shows the relation from Eq.~\eqref{gaso2krome_isogal:eq:ksCII}, whereas that in the right-hand panel shows the same relation with $\Sigma_{\rm CII}$ replaced by $\Sigma_{\rm OI}$ and with the normalisation divided by 10.
}
\label{gaso2krome_isogal:fig:ks_CII_OI}
\end{figure*}

The different H2I abundance does not have a large effect on the density--temperature structure of the gas (not shown). This is due to the fact that the most important H2I-related phenomenon which could in principle affect the gas thermodynamics, molecular-hydrogen cooling, is negligible with respect to metal cooling, at the densities and metallicities considered in this work.

The lack of significant variations in the gas $\rho$--$T$ diagram explains the absence of discrepancy in the classic KS diagrams (not shown). On the other hand, because of the disparity in H2I abundances, the molecular KS diagrams are slightly different, with Runs~01 and 03 matching the observed slope marginally better than Run~02. In our simulations, we do not need a (constant) clumping factor to improve the match with the molecular KS relation. In fact, the simulations with $C_{\rho} = 1$ perform slightly better. We caution, however, that this is also the result of our choice of parameters (including the SF efficiency parameter).

We performed the same kind of comparison with the non-eq-metals runs (i.e. Runs~01m, 02m, and 03m) and obtained the same results.


\subsection{Metal species as tracers of star formation and the interstellar medium}\label{gaso2krome_isogal:sec:metals_as_tracers}

In this section, we study the role of metal species as tracers of the ISM and SFR, focussing on the non-eq-metals Run~01m.

In Fig.~\ref{gaso2krome_isogal:fig:metalphasediagrams_run01m}, we show the abundances of CI, CII, OI, and SiII as a function of gas density. CI and CII are somewhat complementary: CI is more abundant for $\rho_{\rm gas} \gtrsim 10^{-23}$~g~cm$^{-3}$, whereas most of CII is found at lower densities. OI and SiII, on the other hand, have a constant abundance at almost all densities. The same phase diagrams for OII, SiI, and SiIII (not shown) reveal instead that these species are negligible. In Fig.~\ref{gaso2krome_isogal:fig:metalsurfdenmaps_run01m}, the gas surface density maps reinforce the notion that CI traces higher-density gas (in the central regions of the galaxy). CII, on the other hand, traces the entire ISM, as do OI and SiII (not shown).

Metal lines have also been used as tracers of SFR \citep[e.g.][]{Stacey_et_al_1991}. \citet{DeLooze_et_al_2014} analysed the applicability of metal lines as tracers of the ISM in different types of galaxies, obtaining the general relation

\begin{equation}\label{gaso2krome_isogal:eq:DeLooze}
\begin{aligned}
&\log_{10} (\dot{\Sigma}_{*})= \xi + \chi \log_{10} (\Sigma_{\rm line}),
\end{aligned}
\end{equation}
\\
\noindent where $\dot{\Sigma}_{*}$ is given in M$_{\odot}$~yr$^{-1}$~kpc$^{-2}$, $\Sigma_{\rm line}$ (with line = CII or OI) is given in L$_{\odot}$~kpc$^{-2}$, and $[\chi, \xi] = [1.01 \pm 0.02, -6.99 \pm 0.14]$ and  $[1.00 \pm 0.03, -6.79 \pm 0.22]$ for CII and OI, respectively. For simplicity, hereafter we assume $\chi = 1$ and $\xi = -7$.

In order to compare our results to observations, we first translate the metal-species gas mass surface densities from our simulations into metal-line surface brightnesses. To do so, we follow \citet{Pallottini_et_al_2017b}, who find (for $n_{\rm gas} \lesssim 10^3$~cm$^{-3}$, the critical density for CII emission)

\begin{equation}\label{gaso2krome_isogal:eq:Pallottini1}
\begin{aligned}
&\Sigma_{\rm CII,lum} = \epsilon_{\rm CII} \Sigma_{\rm gas},
\end{aligned}
\end{equation}
\\
\noindent where the emissivity

\begin{equation}\label{gaso2krome_isogal:eq:Pallottini2}
\begin{aligned}
&\epsilon_{\rm CII} = 0.1 \left(\frac{n_{\rm gas}}{100 [{\rm cm}^{-3}]}\right) \frac{Z}{Z_{\odot}} [{\rm L}_{\odot}/{\rm M}_{\odot}].
\end{aligned}
\end{equation}
\\
Combining Eqs~\eqref{gaso2krome_isogal:eq:DeLooze}--\eqref{gaso2krome_isogal:eq:Pallottini2}, considering typical values from Run~01m [$\mu = 2$, $Z/Z_{\odot} = 0.5$, and $f_{\rm CII} = 10^{-3}$ (see the upper-right panel of Fig.~\ref{gaso2krome_isogal:fig:metalphasediagrams_run01m})], and assuming for simplicity that $\Sigma_{\rm CII, mass} = 2 \rho_{\rm CII} z_{\rm d}$, we obtain

\begin{equation}\label{gaso2krome_isogal:eq:ksCII}
\begin{aligned}
&\log_{10} (\dot{\Sigma}_{*})= 0.6 + 2 \log_{10} (\Sigma_{\rm CII, mass}),
\end{aligned}
\end{equation}
\\
\noindent where $\dot{\Sigma}_{*}$ is given in M$_{\odot}$~yr$^{-1}$~kpc$^{-2}$ and $\Sigma_{\rm CII}$ is given in M$_{\odot}$~pc$^{-2}$. Eq.~\eqref{gaso2krome_isogal:eq:ksCII} is plotted in the left-hand diagram of Fig.~\ref{gaso2krome_isogal:fig:ks_CII_OI}, where we show the relation between SFR and CII surface density for the non-eq-metals Run~01m.

The agreement between our data and observations is very good for $\Sigma_{\rm CII,mass} \lesssim 0.04$~M$_{\odot}$~pc$^{-2}$. The mismatch at high surface densities is due to the drop of $f_{\rm CII}$ at high gas densities (where recombination is more efficient), as shown in the upper-right panel of Fig.~\ref{gaso2krome_isogal:fig:metalphasediagrams_run01m}.

Since we do not have an equivalent of Eqs~\eqref{gaso2krome_isogal:eq:Pallottini1}--\eqref{gaso2krome_isogal:eq:Pallottini2} for OI, we made the simple assumption that a relation with the same slope of Eq.~\eqref{gaso2krome_isogal:eq:ksCII} holds. In order to match the simulation data, we then divided the normalisation factor by 10. This is equivalent to assuming an OI emissivity 10 times smaller than that of CII, which is consistent with observations \citep[e.g.][]{Malhotra_et_al_2001,Cormier_et_al_2015,Michalowski_et_al_2016}.


\section{Conclusions}\label{gaso2krome_isogal:sec:Conclusions}

We used a novel interface between the modern-SPH code \gasoline and the chemistry package \kromes, to accurately follow the hydrodynamical and chemical evolution of an isolated galaxy at $z = 3$. This is the \textit{first} work in which metal injection from SNae, turbulent metal diffusion, and a metal network with its corresponding non-equilibrium metal cooling were self-consistently modelled in a galaxy simulation.

We built a suite of ten high-resolution hydrodynamical simulations, in which we varied the chemical network, how metal cooling is modelled, the initial gas metallicity, and how H2I forms on dust, to assess the importance of each physical parameter/prescription and, in the case of the non-eq-metals runs, to study the role of metals as tracers of the ISM and SFR.

We itemize our findings below.

\begin{enumerate}

\item Our default run with half-solar metallicity gas, a primordial network (HI, HII, H$^-$, H2I, H2II, HeI, HeII, HeIII, and e$^-$) with dust, optically thin equilibrium (PIE) metal cooling, and the CS09 model of H2I formation on dust with no clumping factor is able to create a realistic galaxy which follows the classic KS relation (Fig.~\ref{gaso2krome_isogal:fig:ks_HIH2I_bigiel_run01vs01m}). Indeed, all the simulations in this work recover the classic KS relation. The simulated molecular KS relation has a slightly steeper slope than the observed one, but the simulated and observed curves intersect in the middle of the observed gas surface density range (Fig.~\ref{gaso2krome_isogal:fig:ks_H2I_bigiel_run01vs01m}). 

\item Modelling the chemical evolution of seven metal species (CI, CII, OI, OII, SiI, SiII, and SiIII) and optically thick non-equilibrium metal cooling has a substantial effect on the thermodynamics of the gas, the chemical abundances, and the appearance of the galaxy. With respect to the eq-metals gas, non-eq-metals gas is typically warmer (Fig.~\ref{gaso2krome_isogal:fig:rhoT_run01vs01m}), has a larger molecular gas mass fraction (Fig.~\ref{gaso2krome_isogal:fig:rhomu_run01vs01m}), a smoother disc (Fig.~\ref{gaso2krome_isogal:fig:gassurfdenmaps_run01vs01m}), and a slightly less extended molecular disc (Fig.~\ref{gaso2krome_isogal:fig:H2Isurfdenmaps_run01vs01m}). The differences are due to the proper account of shielding when computing metal cooling and, to a lesser extent, to different metal networks and non-equilibrium effects.

\item However, the differences in the gas thermodynamics are typically small for cold, dense gas, where stars form, implying that both the SF history (Fig.~\ref{gaso2krome_isogal:fig:sfr}) and KS diagrams (Figs~\ref{gaso2krome_isogal:fig:ks_HIH2I_bigiel_run01vs01m}--\ref{gaso2krome_isogal:fig:ks_H2I_bigiel_run01vs01m}) are similar for eq-metals and non-eq-metals simulations.

\item Varying the initial gas metallicity has direct consequences on gas cooling and H2I formation, with a higher metallicity producing a colder gas and a larger fraction of molecular gas (Fig.~\ref{gaso2krome_isogal:fig:phasediagrams_run05vs01vs04}). The differences, however, are less conspicuous than previously thought because metal enrichment from SNae and turbulent metal diffusion dilute the initial metallicity discrepancy between simulations. The low-metallicity run recovers not only the classic but also the molecular KS relation.

\item Varying the model of H2I formation on dust and the clumping factor has a direct effect on the abundance of molecular hydrogen (Fig.~\ref{gaso2krome_isogal:fig:run01vs02vs03}). Expectedly, a larger clumping factor yields more H2I. We also find that using the J75 model results in a smaller fraction of H2I than when using the CS09 model, hence producing a molecular KS diagram with a slope slightly closer to the observed one. Using a clumping factor of 10 moves the simulated molecular KS relation away from the observed one.

\item Properly following the chemical evolution of metal species is not only useful to account for the non-equilibrium metal cooling: the metal species themselves can be a valuable tracer for the ISM (Figs~\ref{gaso2krome_isogal:fig:metalphasediagrams_run01m}--\ref{gaso2krome_isogal:fig:metalsurfdenmaps_run01m}) and SFR (Fig.~\ref{gaso2krome_isogal:fig:ks_CII_OI}). We showed that our simulations agree quite well with observations which link the SFR to CII emission lines.

\end{enumerate}

In summary, an improved and more accurate prescription of optically thick non-equilibrium metal cooling, together with the possibility to model and follow individual metal species (in the context of comparisons with observations), are the main qualities of this new interface. We note here that improvements can and will be made in future work. On the hydrodynamics code side, the most important is the addition of local radiation from young stars, and we refer to \citet{Lupi_et_al_2017} for a thorough study performed on a galaxy identical to ours. On the chemistry side, we wish to add more metal and molecular species, to improve our modelling of non-equilibrium metal cooling and be able to follow more metal and molecular lines as tracers of the ISM and SFR. Additionally, we plan on constructing synthetic emission maps from selected lines (e.g. the [CII] 157.74~$\mu$m line) to provide a better comparison with current and future FIR observations.

The work presented here was the \textit{first} application of an interface between \krome and an SPH implementation of the equations of hydrodynamics, and we expect this interface to be used in several applications, e.g. from isolated simulations of galaxies, mergers of galaxies, and circumnuclear discs, to cosmological zoom-in simulations of galaxy and BH formation.

\section*{Acknowledgements}
We thank the reviewer for the useful comments that greatly improved this work. PRC thanks Robert Feldmann, Francesco Haardt, Alexander~P. Hobbs, Andrea Pallottini, Rafael Souza Lima, and James~W. Wadsley for useful exchanges, and Sijing Shen for fruitful discussions and for having provided the metal cooling table described in \citet{Shen_et_al_2010,Shen_et_al_2013}. Numerical calculations were performed on Piz Daint, at the Swiss National Supercomputing Center (CSCS). Part of the analysis was performed using the {\scshape pynbody} package \citep[][]{Pontzen_et_al_2013}. PRC acknowledges research funding by the Deutsche Forschungsgemeinschaft (DFG) Sonderforschungsbereiche (SFB) 963 and support by the Tomalla Foundation. AL acknowledges support from the European Research Council (Projects No. 267117, `DARK', and No. 614199, `BLACK'). SB and DRGS thank for funding through the DFG priority program `The Physics of the Interstellar Medium' (projects BO 4113/1-2 and SCHL 1964/1-2). DRGS acknowledges funding via Fondecyt regular (project 1161247), the `Concurso Proyectos Internacionales de Investigaci{\'o}n, Convocatoria 2015' (project PII20150171), ALMA--Conicyt (project 31160001), and the BASAL Centro de Astrof{\'i}sica y Tecnolog{\'i}as Afines (CATA) PFB-06/2007. TG acknowledges the Centre for Star and Planet Formation funded by the Danish National Research Foundation.

\scalefont{0.94}
\setlength{\bibhang}{1.6em} 
\setlength\labelwidth{0.0em}
\bibliographystyle{mnras}
\bibliography{gaso2krome_isogal}

\begin{thebibliography}{}
\makeatletter
\relax
\def\mn@urlcharsother{\let\do\@makeother \do\$\do\&\do\#\do\^\do\_\do\%\do\~}
\def\mn@doi{\begingroup\mn@urlcharsother \@ifnextchar [ {\mn@doi@}
  {\mn@doi@[]}}
\def\mn@doi@[#1]#2{\def\@tempa{#1}\ifx\@tempa\@empty \href
  {http://dx.doi.org/#2} {doi:#2}\else \href {http://dx.doi.org/#2} {#1}\fi
  \endgroup}
\def\mn@eprint#1#2{\mn@eprint@#1:#2::\@nil}
\def\mn@eprint@arXiv#1{\href {http://arxiv.org/abs/#1} {{\tt arXiv:#1}}}
\def\mn@eprint@dblp#1{\href {http://dblp.uni-trier.de/rec/bibtex/#1.xml}
  {dblp:#1}}
\def\mn@eprint@#1:#2:#3:#4\@nil{\def\@tempa {#1}\def\@tempb {#2}\def\@tempc
  {#3}\ifx \@tempc \@empty \let \@tempc \@tempb \let \@tempb \@tempa \fi \ifx
  \@tempb \@empty \def\@tempb {arXiv}\fi \@ifundefined
  {mn@eprint@\@tempb}{\@tempb:\@tempc}{\expandafter \expandafter \csname
  mn@eprint@\@tempb\endcsname \expandafter{\@tempc}}}

\bibitem[\protect\citeauthoryear{{ALMA Partnership} et~al.,}{{ALMA Partnership}
  et~al.}{2015}]{ALMA_2015}
{ALMA Partnership} et~al., 2015, \mn@doi [\apjl] {10.1088/2041-8205/808/1/L1},
  \href {http://adsabs.harvard.edu/abs/2015ApJ...808L...1A} {808, L1}

\bibitem[\protect\citeauthoryear{{Abel}, {Anninos}, {Zhang}  \&
  {Norman}}{{Abel} et~al.}{1997}]{Abel_et_al_1997}
{Abel} T.,  {Anninos} P.,  {Zhang} Y.,   {Norman} M.~L.,  1997, \mn@doi [\na]
  {10.1016/S1384-1076(97)00010-9}, \href
  {http://adsabs.harvard.edu/abs/1997NewA....2..181A} {2, 181}

\bibitem[\protect\citeauthoryear{{Adelberger}, {Steidel}, {Pettini}, {Shapley},
  {Reddy}  \& {Erb}}{{Adelberger} et~al.}{2005}]{Adelberger_et_al_2005}
{Adelberger} K.~L.,  {Steidel} C.~C.,  {Pettini} M.,  {Shapley} A.~E.,  {Reddy}
  N.~A.,   {Erb} D.~K.,  2005, \mn@doi [\apj] {10.1086/426580}, \href
  {http://adsabs.harvard.edu/abs/2005ApJ...619..697A} {619, 697}

\bibitem[\protect\citeauthoryear{{Agertz} et~al.,}{{Agertz}
  et~al.}{2007}]{Agertz_et_al_2007}
{Agertz} O.,  et~al., 2007, \mn@doi [\mnras]
  {10.1111/j.1365-2966.2007.12183.x}, \href
  {http://adsabs.harvard.edu/abs/2007MNRAS.380..963A} {380, 963}

\bibitem[\protect\citeauthoryear{{Agertz}, {Teyssier}  \& {Moore}}{{Agertz}
  et~al.}{2009}]{Agertz_et_al_2009}
{Agertz} O.,  {Teyssier} R.,   {Moore} B.,  2009, \mn@doi [\mnras]
  {10.1111/j.1745-3933.2009.00685.x}, \href
  {http://adsabs.harvard.edu/abs/2009MNRAS.397L..64A} {397, L64}

\bibitem[\protect\citeauthoryear{{Aldrovandi} \& {Pequignot}}{{Aldrovandi} \&
  {Pequignot}}{1973}]{Aldrovandi_Pequignot_1973}
{Aldrovandi} S.~M.~V.,  {Pequignot} D.,  1973, \aap, \href
  {http://adsabs.harvard.edu/abs/1973A%26A....25..137A} {25, 137}

\bibitem[\protect\citeauthoryear{{Asplund}, {Grevesse}, {Sauval}  \&
  {Scott}}{{Asplund} et~al.}{2009}]{Asplund_et_al_2009}
{Asplund} M.,  {Grevesse} N.,  {Sauval} A.~J.,   {Scott} P.,  2009, \mn@doi
  [\araa] {10.1146/annurev.astro.46.060407.145222}, \href
  {http://adsabs.harvard.edu/abs/2009ARA%26A..47..481A} {47, 481}

\bibitem[\protect\citeauthoryear{{Bakes} \& {Tielens}}{{Bakes} \&
  {Tielens}}{1994}]{Bakes_Tielens_1994}
{Bakes} E.~L.~O.,  {Tielens} A.~G.~G.~M.,  1994, \mn@doi [\apj]
  {10.1086/174188}, \href {http://adsabs.harvard.edu/abs/1994ApJ...427..822B}
  {427, 822}

\bibitem[\protect\citeauthoryear{{Barlow}}{{Barlow}}{1984}]{Barlow_1984}
{Barlow} S.~E.,  1984, PhD thesis, UNIVERSITY OF COLORADO AT BOULDER.

\bibitem[\protect\citeauthoryear{{Bellovary}, {Governato}, {Quinn}, {Wadsley},
  {Shen}  \& {Volonteri}}{{Bellovary} et~al.}{2010}]{Bellovary_et_al_2010}
{Bellovary} J.~M.,  {Governato} F.,  {Quinn} T.~R.,  {Wadsley} J.,  {Shen} S.,
   {Volonteri} M.,  2010, \mn@doi [\apjl] {10.1088/2041-8205/721/2/L148}, \href
  {http://adsabs.harvard.edu/abs/2010ApJ...721L.148B} {721, L148}

\bibitem[\protect\citeauthoryear{{Bellovary}, {Brooks}, {Volonteri},
  {Governato}, {Quinn}  \& {Wadsley}}{{Bellovary}
  et~al.}{2013}]{Bellovary_et_al_2013}
{Bellovary} J.,  {Brooks} A.,  {Volonteri} M.,  {Governato} F.,  {Quinn} T.,
  {Wadsley} J.,  2013, \mn@doi [\apj] {10.1088/0004-637X/779/2/136}, \href
  {http://adsabs.harvard.edu/abs/2013ApJ...779..136B} {779, 136}

\bibitem[\protect\citeauthoryear{{Bigiel}, {Leroy}, {Walter}, {Brinks}, {de
  Blok}, {Madore}  \& {Thornley}}{{Bigiel} et~al.}{2008}]{Bigiel_et_al_2008}
{Bigiel} F.,  {Leroy} A.,  {Walter} F.,  {Brinks} E.,  {de Blok} W.~J.~G.,
  {Madore} B.,   {Thornley} M.~D.,  2008, \mn@doi [\aj]
  {10.1088/0004-6256/136/6/2846}, \href
  {http://adsabs.harvard.edu/abs/2008AJ....136.2846B} {136, 2846}

\bibitem[\protect\citeauthoryear{{Bigiel}, {Leroy}, {Walter}, {Blitz},
  {Brinks}, {de Blok}  \& {Madore}}{{Bigiel} et~al.}{2010}]{Bigiel_et_al_2010}
{Bigiel} F.,  {Leroy} A.,  {Walter} F.,  {Blitz} L.,  {Brinks} E.,  {de Blok}
  W.~J.~G.,   {Madore} B.,  2010, \mn@doi [\aj] {10.1088/0004-6256/140/5/1194},
  \href {http://adsabs.harvard.edu/abs/2010AJ....140.1194B} {140, 1194}

\bibitem[\protect\citeauthoryear{{Bovino}, {Grassi}, {Schleicher}  \&
  {Latif}}{{Bovino} et~al.}{2014}]{Bovino_et_al_2014}
{Bovino} S.,  {Grassi} T.,  {Schleicher} D.~R.~G.,   {Latif} M.~A.,  2014,
  \mn@doi [\apjl] {10.1088/2041-8205/790/2/L35}, \href
  {http://adsabs.harvard.edu/abs/2014ApJ...790L..35B} {790, L35}

\bibitem[\protect\citeauthoryear{{Bovino}, {Grassi}, {Capelo}, {Schleicher}  \&
  {Banerjee}}{{Bovino} et~al.}{2016}]{Bovino_et_al_2016}
{Bovino} S.,  {Grassi} T.,  {Capelo} P.~R.,  {Schleicher} D.~R.~G.,
  {Banerjee} R.,  2016, \mn@doi [\aap] {10.1051/0004-6361/201628158}, \href
  {http://adsabs.harvard.edu/abs/2016A%26A...590A..15B} {590, A15}

\bibitem[\protect\citeauthoryear{{Braun} \& {Schmidt}}{{Braun} \&
  {Schmidt}}{2015}]{Braun_Schmidt_2015}
{Braun} H.,  {Schmidt} W.,  2015, \mn@doi [\mnras] {10.1093/mnras/stv1856},
  \href {http://adsabs.harvard.edu/abs/2015MNRAS.454.1545B} {454, 1545}

\bibitem[\protect\citeauthoryear{{Brook} et~al.,}{{Brook}
  et~al.}{2012}]{Brook_et_al_2012}
{Brook} C.~B.,  et~al., 2012, \mn@doi [\mnras]
  {10.1111/j.1365-2966.2012.21738.x}, \href
  {http://adsabs.harvard.edu/abs/2012MNRAS.426..690B} {426, 690}

\bibitem[\protect\citeauthoryear{{Bryan} et~al.,}{{Bryan}
  et~al.}{2014}]{Bryan_et_al_2014}
{Bryan} G.~L.,  et~al., 2014, \mn@doi [\apjs] {10.1088/0067-0049/211/2/19},
  \href {http://adsabs.harvard.edu/abs/2014ApJS..211...19B} {211, 19}

\bibitem[\protect\citeauthoryear{{Burton}, {Hollenbach}  \& {Tielens}}{{Burton}
  et~al.}{1990}]{Burton_et_al_1990}
{Burton} M.~G.,  {Hollenbach} D.~J.,   {Tielens} A.~G.~G.~M.,  1990, \mn@doi
  [\apj] {10.1086/169516}, \href
  {http://adsabs.harvard.edu/abs/1990ApJ...365..620B} {365, 620}

\bibitem[\protect\citeauthoryear{{Camm}}{{Camm}}{1950}]{Camm_1950}
{Camm} G.~L.,  1950, \mn@doi [\mnras] {10.1093/mnras/110.4.305}, \href
  {http://adsabs.harvard.edu/abs/1950MNRAS.110..305C} {110, 305}

\bibitem[\protect\citeauthoryear{{Capelo}, {Volonteri}, {Dotti}, {Bellovary},
  {Mayer}  \& {Governato}}{{Capelo} et~al.}{2015}]{Capelo_et_al_2015}
{Capelo} P.~R.,  {Volonteri} M.,  {Dotti} M.,  {Bellovary} J.~M.,  {Mayer} L.,
   {Governato} F.,  2015, \mn@doi [\mnras] {10.1093/mnras/stu2500}, \href
  {http://adsabs.harvard.edu/abs/2015MNRAS.447.2123C} {447, 2123}

\bibitem[\protect\citeauthoryear{{Capelo}, {Dotti}, {Volonteri}, {Mayer},
  {Bellovary}  \& {Shen}}{{Capelo} et~al.}{2017}]{Capelo_et_al_2017}
{Capelo} P.~R.,  {Dotti} M.,  {Volonteri} M.,  {Mayer} L.,  {Bellovary} J.~M.,
   {Shen} S.,  2017, \mn@doi [\mnras] {10.1093/mnras/stx1067}, \href
  {http://adsabs.harvard.edu/abs/2017MNRAS.469.4437C} {469, 4437}

\bibitem[\protect\citeauthoryear{{Capitelli}, {Coppola}, {Diomede}  \&
  {Longo}}{{Capitelli} et~al.}{2007}]{Capitelli_et_al_2007}
{Capitelli} M.,  {Coppola} C.~M.,  {Diomede} P.,   {Longo} S.,  2007, \mn@doi
  [\aap] {10.1051/0004-6361:20077600}, \href
  {http://adsabs.harvard.edu/abs/2007A%26A...470..811C} {470, 811}

\bibitem[\protect\citeauthoryear{{Cazaux} \& {Spaans}}{{Cazaux} \&
  {Spaans}}{2009}]{Cazaux_Spaans_2009}
{Cazaux} S.,  {Spaans} M.,  2009, \mn@doi [\aap] {10.1051/0004-6361:200811302},
  \href {http://adsabs.harvard.edu/abs/2009A%26A...496..365C} {496, 365}

\bibitem[\protect\citeauthoryear{{Cen}}{{Cen}}{1992}]{Cen_1992}
{Cen} R.,  1992, \mn@doi [\apjs] {10.1086/191630}, \href
  {http://adsabs.harvard.edu/abs/1992ApJS...78..341C} {78, 341}

\bibitem[\protect\citeauthoryear{{Christensen}, {Quinn}, {Stinson}, {Bellovary}
   \& {Wadsley}}{{Christensen} et~al.}{2010}]{Christensen_et_al_2010}
{Christensen} C.~R.,  {Quinn} T.,  {Stinson} G.,  {Bellovary} J.,   {Wadsley}
  J.,  2010, \mn@doi [\apj] {10.1088/0004-637X/717/1/121}, \href
  {http://adsabs.harvard.edu/abs/2010ApJ...717..121C} {717, 121}

\bibitem[\protect\citeauthoryear{{Christensen}, {Quinn}, {Governato}, {Stilp},
  {Shen}  \& {Wadsley}}{{Christensen} et~al.}{2012}]{Christensen_et_al_2012}
{Christensen} C.,  {Quinn} T.,  {Governato} F.,  {Stilp} A.,  {Shen} S.,
  {Wadsley} J.,  2012, \mn@doi [\mnras] {10.1111/j.1365-2966.2012.21628.x},
  \href {http://adsabs.harvard.edu/abs/2012MNRAS.425.3058C} {425, 3058}

\bibitem[\protect\citeauthoryear{{Conroy}, {Gunn}  \& {White}}{{Conroy}
  et~al.}{2009}]{Conroy_et_al_2009}
{Conroy} C.,  {Gunn} J.~E.,   {White} M.,  2009, \mn@doi [\apj]
  {10.1088/0004-637X/699/1/486}, \href
  {http://adsabs.harvard.edu/abs/2009ApJ...699..486C} {699, 486}

\bibitem[\protect\citeauthoryear{{Coppola}, {Longo}, {Capitelli}, {Palla}  \&
  {Galli}}{{Coppola} et~al.}{2011}]{Coppola_et_al_2011}
{Coppola} C.~M.,  {Longo} S.,  {Capitelli} M.,  {Palla} F.,   {Galli} D.,
  2011, \mn@doi [\apjs] {10.1088/0067-0049/193/1/7}, \href
  {http://adsabs.harvard.edu/abs/2011ApJS..193....7C} {193, 7}

\bibitem[\protect\citeauthoryear{{Cormier} et~al.,}{{Cormier}
  et~al.}{2015}]{Cormier_et_al_2015}
{Cormier} D.,  et~al., 2015, \mn@doi [\aap] {10.1051/0004-6361/201425207},
  \href {http://adsabs.harvard.edu/abs/2015A%26A...578A..53C} {578, A53}

\bibitem[\protect\citeauthoryear{{Corrigan}}{{Corrigan}}{1965}]{Corrigan_1965}
{Corrigan} S.~J.~B.,  1965, \mn@doi [\jcp] {10.1063/1.1696701}, \href
  {http://adsabs.harvard.edu/abs/1965JChPh..43.4381C} {43, 4381}

\bibitem[\protect\citeauthoryear{{Dalgarno} \& {Lepp}}{{Dalgarno} \&
  {Lepp}}{1987}]{Dalgarno_Lepp_1987}
{Dalgarno} A.,  {Lepp} S.,  1987, in {Vardya} M.~S.,  {Tarafdar} S.~P.,  eds,
  IAU Symposium Vol. 120, Astrochemistry. pp 109--118

\bibitem[\protect\citeauthoryear{{Dalla Vecchia} \& {Schaye}}{{Dalla Vecchia}
  \& {Schaye}}{2008}]{DallaVecchia_Schaye_2008}
{Dalla Vecchia} C.,  {Schaye} J.,  2008, \mn@doi [\mnras]
  {10.1111/j.1365-2966.2008.13322.x}, \href
  {http://adsabs.harvard.edu/abs/2008MNRAS.387.1431D} {387, 1431}

\bibitem[\protect\citeauthoryear{{Dalla Vecchia} \& {Schaye}}{{Dalla Vecchia}
  \& {Schaye}}{2012}]{DallaVecchia_Schaye_2012}
{Dalla Vecchia} C.,  {Schaye} J.,  2012, \mn@doi [\mnras]
  {10.1111/j.1365-2966.2012.21704.x}, \href
  {http://adsabs.harvard.edu/abs/2012MNRAS.426..140D} {426, 140}

\bibitem[\protect\citeauthoryear{{De Looze} et~al.,}{{De Looze}
  et~al.}{2014}]{DeLooze_et_al_2014}
{De Looze} I.,  et~al., 2014, \mn@doi [\aap] {10.1051/0004-6361/201322489},
  \href {http://adsabs.harvard.edu/abs/2014A%26A...568A..62D} {568, A62}

\bibitem[\protect\citeauthoryear{{Dehnen} \& {Aly}}{{Dehnen} \&
  {Aly}}{2012}]{Dehnen_Aly_2012}
{Dehnen} W.,  {Aly} H.,  2012, \mn@doi [\mnras]
  {10.1111/j.1365-2966.2012.21439.x}, \href
  {http://adsabs.harvard.edu/abs/2012MNRAS.425.1068D} {425, 1068}

\bibitem[\protect\citeauthoryear{{Dekel}, {Sari}  \& {Ceverino}}{{Dekel}
  et~al.}{2009}]{Dekel_et_al_2009}
{Dekel} A.,  {Sari} R.,   {Ceverino} D.,  2009, \mn@doi [\apj]
  {10.1088/0004-637X/703/1/785}, \href
  {http://adsabs.harvard.edu/abs/2009ApJ...703..785D} {703, 785}

\bibitem[\protect\citeauthoryear{{Diemer} \& {Kravtsov}}{{Diemer} \&
  {Kravtsov}}{2015}]{Diemer_Kravtsov_2015}
{Diemer} B.,  {Kravtsov} A.~V.,  2015, \mn@doi [\apj]
  {10.1088/0004-637X/799/1/108}, \href
  {http://adsabs.harvard.edu/abs/2015ApJ...799..108D} {799, 108}

\bibitem[\protect\citeauthoryear{{Dove}, {Rusk}, {Cribb}  \& {Martin}}{{Dove}
  et~al.}{1987}]{Dove_et_al_1987}
{Dove} J.~E.,  {Rusk} A.~C.~M.,  {Cribb} P.~H.,   {Martin} P.~G.,  1987,
  \mn@doi [\apj] {10.1086/165375}, \href
  {http://adsabs.harvard.edu/abs/1987ApJ...318..379D} {318, 379}

\bibitem[\protect\citeauthoryear{{Draine} \& {Lee}}{{Draine} \&
  {Lee}}{1984}]{Draine_Lee_1984}
{Draine} B.~T.,  {Lee} H.~M.,  1984, \mn@doi [\apj] {10.1086/162480}, \href
  {http://adsabs.harvard.edu/abs/1984ApJ...285...89D} {285, 89}

\bibitem[\protect\citeauthoryear{{Durier} \& {Dalla Vecchia}}{{Durier} \&
  {Dalla Vecchia}}{2012}]{Durier_DallaVecchia_2012}
{Durier} F.,  {Dalla Vecchia} C.,  2012, \mn@doi [\mnras]
  {10.1111/j.1365-2966.2011.19712.x}, \href
  {http://adsabs.harvard.edu/abs/2012MNRAS.419..465D} {419, 465}

\bibitem[\protect\citeauthoryear{{Dutton} \& {Macci{\`o}}}{{Dutton} \&
  {Macci{\`o}}}{2014}]{Dutton_Maccio_2014}
{Dutton} A.~A.,  {Macci{\`o}} A.~V.,  2014, \mn@doi [\mnras]
  {10.1093/mnras/stu742}, \href
  {http://adsabs.harvard.edu/abs/2014MNRAS.441.3359D} {441, 3359}

\bibitem[\protect\citeauthoryear{{Feldmann}, {Gnedin}  \&
  {Kravtsov}}{{Feldmann} et~al.}{2011}]{Feldmann_et_al_2011}
{Feldmann} R.,  {Gnedin} N.~Y.,   {Kravtsov} A.~V.,  2011, \mn@doi [\apj]
  {10.1088/0004-637X/732/2/115}, \href
  {http://adsabs.harvard.edu/abs/2011ApJ...732..115F} {732, 115}

\bibitem[\protect\citeauthoryear{{Ferland}, {Peterson}, {Horne}, {Welsh}  \&
  {Nahar}}{{Ferland} et~al.}{1992}]{Ferland_et_al_1992}
{Ferland} G.~J.,  {Peterson} B.~M.,  {Horne} K.,  {Welsh} W.~F.,   {Nahar}
  S.~N.,  1992, \mn@doi [\apj] {10.1086/171063}, \href
  {http://adsabs.harvard.edu/abs/1992ApJ...387...95F} {387, 95}

\bibitem[\protect\citeauthoryear{{Ferland}, {Korista}, {Verner}, {Ferguson},
  {Kingdon}  \& {Verner}}{{Ferland} et~al.}{1998}]{Ferland_et_al_1998}
{Ferland} G.~J.,  {Korista} K.~T.,  {Verner} D.~A.,  {Ferguson} J.~W.,
  {Kingdon} J.~B.,   {Verner} E.~M.,  1998, \mn@doi [\pasp] {10.1086/316190},
  \href {http://adsabs.harvard.edu/abs/1998PASP..110..761F} {110, 761}

\bibitem[\protect\citeauthoryear{{Ferland} et~al.,}{{Ferland}
  et~al.}{2013}]{Ferland_et_al_2013}
{Ferland} G.~J.,  et~al., 2013, \rmxaa, \href
  {http://adsabs.harvard.edu/abs/2013RMxAA..49..137F} {49, 137}

\bibitem[\protect\citeauthoryear{{Ferri{\`e}re}}{{Ferri{\`e}re}}{2001}]{Ferriere_2001}
{Ferri{\`e}re} K.~M.,  2001, \mn@doi [Reviews of Modern Physics]
  {10.1103/RevModPhys.73.1031}, \href
  {http://adsabs.harvard.edu/abs/2001RvMP...73.1031F} {73, 1031}

\bibitem[\protect\citeauthoryear{{Fryxell} et~al.,}{{Fryxell}
  et~al.}{2000}]{Fryxell_2000}
{Fryxell} B.,  et~al., 2000, \mn@doi [\apjs] {10.1086/317361}, \href
  {http://adsabs.harvard.edu/abs/2000ApJS..131..273F} {131, 273}

\bibitem[\protect\citeauthoryear{{Gabor}, {Capelo}, {Volonteri}, {Bournaud},
  {Bellovary}, {Governato}  \& {Quinn}}{{Gabor}
  et~al.}{2016}]{Gabor_et_al_2016}
{Gabor} J.~M.,  {Capelo} P.~R.,  {Volonteri} M.,  {Bournaud} F.,  {Bellovary}
  J.,  {Governato} F.,   {Quinn} T.,  2016, \mn@doi [\aap]
  {10.1051/0004-6361/201527143}, \href
  {http://adsabs.harvard.edu/abs/2016A%26A...592A..62G} {592, A62}

\bibitem[\protect\citeauthoryear{{Genel}, {Genzel}, {Bouch{\'e}}, {Naab}  \&
  {Sternberg}}{{Genel} et~al.}{2009}]{Genel_et_al_2009}
{Genel} S.,  {Genzel} R.,  {Bouch{\'e}} N.,  {Naab} T.,   {Sternberg} A.,
  2009, \mn@doi [\apj] {10.1088/0004-637X/701/2/2002}, \href
  {http://adsabs.harvard.edu/abs/2009ApJ...701.2002G} {701, 2002}

\bibitem[\protect\citeauthoryear{{Glover}}{{Glover}}{2015}]{Glover_2015}
{Glover} S.~C.~O.,  2015, \mn@doi [\mnras] {10.1093/mnras/stv1059}, \href
  {http://adsabs.harvard.edu/abs/2015MNRAS.451.2082G} {451, 2082}

\bibitem[\protect\citeauthoryear{{Glover} \& {Abel}}{{Glover} \&
  {Abel}}{2008}]{Glover_Abel_2008}
{Glover} S.~C.~O.,  {Abel} T.,  2008, \mn@doi [\mnras]
  {10.1111/j.1365-2966.2008.13224.x}, \href
  {http://adsabs.harvard.edu/abs/2008MNRAS.388.1627G} {388, 1627}

\bibitem[\protect\citeauthoryear{{Glover} \& {Jappsen}}{{Glover} \&
  {Jappsen}}{2007}]{Glover_Jappsen_2007}
{Glover} S.~C.~O.,  {Jappsen} A.-K.,  2007, \mn@doi [\apj] {10.1086/519445},
  \href {http://adsabs.harvard.edu/abs/2007ApJ...666....1G} {666, 1}

\bibitem[\protect\citeauthoryear{{Gnat} \& {Sternberg}}{{Gnat} \&
  {Sternberg}}{2007}]{Gnat_Sternberg_2007}
{Gnat} O.,  {Sternberg} A.,  2007, \mn@doi [\apjs] {10.1086/509786}, \href
  {http://adsabs.harvard.edu/abs/2007ApJS..168..213G} {168, 213}

\bibitem[\protect\citeauthoryear{{Gnedin} \& {Kravtsov}}{{Gnedin} \&
  {Kravtsov}}{2010}]{Gnedin_Kravtsov_2010}
{Gnedin} N.~Y.,  {Kravtsov} A.~V.,  2010, \mn@doi [\apj]
  {10.1088/0004-637X/714/1/287}, \href
  {http://adsabs.harvard.edu/abs/2010ApJ...714..287G} {714, 287}

\bibitem[\protect\citeauthoryear{{Gnedin} \& {Kravtsov}}{{Gnedin} \&
  {Kravtsov}}{2011}]{Gnedin_Kravtsov_2011}
{Gnedin} N.~Y.,  {Kravtsov} A.~V.,  2011, \mn@doi [\apj]
  {10.1088/0004-637X/728/2/88}, \href
  {http://adsabs.harvard.edu/abs/2011ApJ...728...88G} {728, 88}

\bibitem[\protect\citeauthoryear{{Gnedin}, {Tassis}  \& {Kravtsov}}{{Gnedin}
  et~al.}{2009}]{Gnedin_et_al_2009}
{Gnedin} N.~Y.,  {Tassis} K.,   {Kravtsov} A.~V.,  2009, \mn@doi [\apj]
  {10.1088/0004-637X/697/1/55}, \href
  {http://adsabs.harvard.edu/abs/2009ApJ...697...55G} {697, 55}

\bibitem[\protect\citeauthoryear{{Gonzalez-Perez}, {Lacey}, {Baugh}, {Lagos},
  {Helly}, {Campbell}  \& {Mitchell}}{{Gonzalez-Perez}
  et~al.}{2014}]{Gonzales-Perez_et_al_2014}
{Gonzalez-Perez} V.,  {Lacey} C.~G.,  {Baugh} C.~M.,  {Lagos} C.~D.~P.,
  {Helly} J.,  {Campbell} D.~J.~R.,   {Mitchell} P.~D.,  2014, \mn@doi [\mnras]
  {10.1093/mnras/stt2410}, \href
  {http://adsabs.harvard.edu/abs/2014MNRAS.439..264G} {439, 264}

\bibitem[\protect\citeauthoryear{{Grassi}, {Krstic}, {Merlin}, {Buonomo},
  {Piovan}  \& {Chiosi}}{{Grassi} et~al.}{2011}]{Grassi_et_al_2011}
{Grassi} T.,  {Krstic} P.,  {Merlin} E.,  {Buonomo} U.,  {Piovan} L.,
  {Chiosi} C.,  2011, \mn@doi [\aap] {10.1051/0004-6361/200913779}, \href
  {http://adsabs.harvard.edu/abs/2011A%26A...533A.123G} {533, A123}

\bibitem[\protect\citeauthoryear{{Grassi}, {Bovino}, {Schleicher}, {Prieto},
  {Seifried}, {Simoncini}  \& {Gianturco}}{{Grassi}
  et~al.}{2014}]{Grassi_et_al_2014}
{Grassi} T.,  {Bovino} S.,  {Schleicher} D.~R.~G.,  {Prieto} J.,  {Seifried}
  D.,  {Simoncini} E.,   {Gianturco} F.~A.,  2014, \mn@doi [\mnras]
  {10.1093/mnras/stu114}, \href
  {http://adsabs.harvard.edu/abs/2014MNRAS.439.2386G} {439, 2386}

\bibitem[\protect\citeauthoryear{{Grassi}, {Bovino}, {Haugb{\o}lle}  \&
  {Schleicher}}{{Grassi} et~al.}{2017}]{Grassi_et_al_2017}
{Grassi} T.,  {Bovino} S.,  {Haugb{\o}lle} T.,   {Schleicher} D.~R.~G.,  2017,
  \mn@doi [\mnras] {10.1093/mnras/stw2871}, \href
  {http://adsabs.harvard.edu/abs/2017MNRAS.466.1259G} {466, 1259}

\bibitem[\protect\citeauthoryear{{Haardt} \& {Madau}}{{Haardt} \&
  {Madau}}{2012}]{Haardt_Madau_2012}
{Haardt} F.,  {Madau} P.,  2012, \mn@doi [\apj] {10.1088/0004-637X/746/2/125},
  \href {http://adsabs.harvard.edu/abs/2012ApJ...746..125H} {746, 125}

\bibitem[\protect\citeauthoryear{{Hernquist}}{{Hernquist}}{1990}]{Hernquist_1990}
{Hernquist} L.,  1990, \mn@doi [\apj] {10.1086/168845}, \href
  {http://adsabs.harvard.edu/abs/1990ApJ...356..359H} {356, 359}

\bibitem[\protect\citeauthoryear{{Hindmarsh}}{{Hindmarsh}}{1983}]{Hindmarsh_1983}
{Hindmarsh} A.~C.,  1983, IMACS Transactions on Scientific Computation, 1, 55

\bibitem[\protect\citeauthoryear{{Hollenbach} \& {McKee}}{{Hollenbach} \&
  {McKee}}{1979}]{Hollenbach_McKee_1979}
{Hollenbach} D.,  {McKee} C.~F.,  1979, \mn@doi [\apjs] {10.1086/190631}, \href
  {http://adsabs.harvard.edu/abs/1979ApJS...41..555H} {41, 555}

\bibitem[\protect\citeauthoryear{{Hopkins}}{{Hopkins}}{2015}]{Hopkins_2015}
{Hopkins} P.~F.,  2015, \mn@doi [\mnras] {10.1093/mnras/stv195}, \href
  {http://adsabs.harvard.edu/abs/2015MNRAS.450...53H} {450, 53}

\bibitem[\protect\citeauthoryear{{Hopkins}, {Narayanan}  \& {Murray}}{{Hopkins}
  et~al.}{2013}]{Hopkins_et_al_2013}
{Hopkins} P.~F.,  {Narayanan} D.,   {Murray} N.,  2013, \mn@doi [\mnras]
  {10.1093/mnras/stt723}, \href
  {http://adsabs.harvard.edu/abs/2013MNRAS.432.2647H} {432, 2647}

\bibitem[\protect\citeauthoryear{{Hopkins} et~al.,}{{Hopkins}
  et~al.}{2017}]{Hopkins_et_al_2017}
{Hopkins} P.~F.,  et~al., 2017, preprint, \href
  {http://adsabs.harvard.edu/abs/2017arXiv170707010H} {} (\mn@eprint {arXiv}
  {1707.07010})

\bibitem[\protect\citeauthoryear{{Hu}, {Naab}, {Walch}, {Glover}  \&
  {Clark}}{{Hu} et~al.}{2016}]{Hu_et_al_2016}
{Hu} C.-Y.,  {Naab} T.,  {Walch} S.,  {Glover} S.~C.~O.,   {Clark} P.~C.,
  2016, \mn@doi [\mnras] {10.1093/mnras/stw544}, \href
  {http://adsabs.harvard.edu/abs/2016MNRAS.458.3528H} {458, 3528}

\bibitem[\protect\citeauthoryear{{Hu}, {Naab}, {Glover}, {Walch}  \&
  {Clark}}{{Hu} et~al.}{2017}]{Hu_et_al_2017}
{Hu} C.-Y.,  {Naab} T.,  {Glover} S.~C.~O.,  {Walch} S.,   {Clark} P.~C.,
  2017, \mn@doi [\mnras] {10.1093/mnras/stx1773}, \href
  {http://adsabs.harvard.edu/abs/2017MNRAS.471.2151H} {471, 2151}

\bibitem[\protect\citeauthoryear{{Janev}, {Langer}  \& {Evans}}{{Janev}
  et~al.}{1987}]{Janev_et_al_1987}
{Janev} R.~K.,  {Langer} W.~D.,   {Evans} K.,  1987, {Elementary processes in
  Hydrogen-Helium plasmas - Cross sections and reaction rate coefficients}.
Springer

\bibitem[\protect\citeauthoryear{{Jeans}}{{Jeans}}{1902}]{Jeans_1902}
{Jeans} J.~H.,  1902, \mn@doi [Phil. Trans. Roy. Soc.]
  {10.1098/rsta.1902.0012}, \href
  {http://adsabs.harvard.edu/abs/1902RSPTA.199....1J} {199, 1}

\bibitem[\protect\citeauthoryear{{Jenkins}}{{Jenkins}}{2009}]{Jenkins_2009}
{Jenkins} E.~B.,  2009, \mn@doi [\apj] {10.1088/0004-637X/700/2/1299}, \href
  {http://adsabs.harvard.edu/abs/2009ApJ...700.1299J} {700, 1299}

\bibitem[\protect\citeauthoryear{{Jimenez}, {Flynn}, {MacDonald}  \&
  {Gibson}}{{Jimenez} et~al.}{2003}]{Jimenez_et_al_2003}
{Jimenez} R.,  {Flynn} C.,  {MacDonald} J.,   {Gibson} B.~K.,  2003, \mn@doi
  [Science] {10.1126/science.1080866}, \href
  {http://adsabs.harvard.edu/abs/2003Sci...299.1552J} {299, 1552}

\bibitem[\protect\citeauthoryear{{Jura}}{{Jura}}{1975}]{Jura_1975}
{Jura} M.,  1975, \mn@doi [\apj] {10.1086/153545}, \href
  {http://adsabs.harvard.edu/abs/1975ApJ...197..575J} {197, 575}

\bibitem[\protect\citeauthoryear{{Karpas}, {Anicich}  \& {Huntress}}{{Karpas}
  et~al.}{1979}]{Karpas_et_al_1979}
{Karpas} Z.,  {Anicich} V.,   {Huntress} W.~T.,  1979, \mn@doi [\jcp]
  {10.1063/1.437823}, \href {http://adsabs.harvard.edu/abs/1979JChPh..70.2877K}
  {70, 2877}

\bibitem[\protect\citeauthoryear{{Katz}}{{Katz}}{1992}]{Katz_1992}
{Katz} N.,  1992, \mn@doi [\apj] {10.1086/171366}, \href
  {http://adsabs.harvard.edu/abs/1992ApJ...391..502K} {391, 502}

\bibitem[\protect\citeauthoryear{{Katz}, {Sijacki}  \& {Haehnelt}}{{Katz}
  et~al.}{2015}]{Katz_et_al_2015}
{Katz} H.,  {Sijacki} D.,   {Haehnelt} M.~G.,  2015, \mn@doi [\mnras]
  {10.1093/mnras/stv1048}, \href
  {http://adsabs.harvard.edu/abs/2015MNRAS.451.2352K} {451, 2352}

\bibitem[\protect\citeauthoryear{{Keller}, {Wadsley}, {Benincasa}  \&
  {Couchman}}{{Keller} et~al.}{2014}]{Keller_et_al_2014}
{Keller} B.~W.,  {Wadsley} J.,  {Benincasa} S.~M.,   {Couchman} H.~M.~P.,
  2014, \mn@doi [\mnras] {10.1093/mnras/stu1058}, \href
  {http://adsabs.harvard.edu/abs/2014MNRAS.442.3013K} {442, 3013}

\bibitem[\protect\citeauthoryear{{Kennicutt}}{{Kennicutt}}{1989}]{Kennicutt_1989}
{Kennicutt} Jr. R.~C.,  1989, \mn@doi [\apj] {10.1086/167834}, \href
  {http://adsabs.harvard.edu/abs/1989ApJ...344..685K} {344, 685}

\bibitem[\protect\citeauthoryear{{Kennicutt}}{{Kennicutt}}{1998}]{Kennicutt_1998}
{Kennicutt} Jr. R.~C.,  1998, \mn@doi [\apj] {10.1086/305588}, \href
  {http://adsabs.harvard.edu/abs/1998ApJ...498..541K} {498, 541}

\bibitem[\protect\citeauthoryear{{Kennicutt}, {Tamblyn}  \&
  {Congdon}}{{Kennicutt} et~al.}{1994}]{Kennicutt_et_al_1994}
{Kennicutt} Jr. R.~C.,  {Tamblyn} P.,   {Congdon} C.~E.,  1994, \mn@doi [\apj]
  {10.1086/174790}, \href {http://adsabs.harvard.edu/abs/1994ApJ...435...22K}
  {435, 22}

\bibitem[\protect\citeauthoryear{{Kimm}, {Cen}, {Devriendt}, {Dubois}  \&
  {Slyz}}{{Kimm} et~al.}{2015}]{Kimm_et_al_2015}
{Kimm} T.,  {Cen} R.,  {Devriendt} J.,  {Dubois} Y.,   {Slyz} A.,  2015,
  \mn@doi [\mnras] {10.1093/mnras/stv1211}, \href
  {http://adsabs.harvard.edu/abs/2015MNRAS.451.2900K} {451, 2900}

\bibitem[\protect\citeauthoryear{{Kimura}, {Lane}, {Dalgarno}  \&
  {Dixson}}{{Kimura} et~al.}{1993}]{Kimura_et_al_1993}
{Kimura} M.,  {Lane} N.~F.,  {Dalgarno} A.,   {Dixson} R.~G.,  1993, \mn@doi
  [\apj] {10.1086/172410}, \href
  {http://adsabs.harvard.edu/abs/1993ApJ...405..801K} {405, 801}

\bibitem[\protect\citeauthoryear{{Kingdon} \& {Ferland}}{{Kingdon} \&
  {Ferland}}{1996}]{Kingdon_Ferland_1996}
{Kingdon} J.~B.,  {Ferland} G.~J.,  1996, \mn@doi [\apjs] {10.1086/192335},
  \href {http://adsabs.harvard.edu/abs/1996ApJS..106..205K} {106, 205}

\bibitem[\protect\citeauthoryear{{K{\"o}rtgen}, {Bovino}, {Schleicher},
  {Giannetti}  \& {Banerjee}}{{K{\"o}rtgen} et~al.}{2017}]{Kortgen_et_al_2017}
{K{\"o}rtgen} B.,  {Bovino} S.,  {Schleicher} D.~R.~G.,  {Giannetti} A.,
  {Banerjee} R.,  2017, \mn@doi [\mnras] {10.1093/mnras/stx1005}, \href
  {http://adsabs.harvard.edu/abs/2017MNRAS.469.2602K} {469, 2602}

\bibitem[\protect\citeauthoryear{{Kreckel}, {Bruhns}, {{\v C}{\'{\i}}{\v z}ek},
  {Glover}, {Miller}, {Urbain}  \& {Savin}}{{Kreckel}
  et~al.}{2010}]{Kreckel_et_al_2010}
{Kreckel} H.,  {Bruhns} H.,  {{\v C}{\'{\i}}{\v z}ek} M.,  {Glover} S.~C.~O.,
  {Miller} K.~A.,  {Urbain} X.,   {Savin} D.~W.,  2010, \mn@doi [Science]
  {10.1126/science.1187191}, \href
  {http://adsabs.harvard.edu/abs/2010Sci...329...69K} {329, 69}

\bibitem[\protect\citeauthoryear{{Kroupa}}{{Kroupa}}{2001}]{Kroupa_2001}
{Kroupa} P.,  2001, \mn@doi [\mnras] {10.1046/j.1365-8711.2001.04022.x}, \href
  {http://adsabs.harvard.edu/abs/2001MNRAS.322..231K} {322, 231}

\bibitem[\protect\citeauthoryear{{Krumholz}, {Dekel}  \& {McKee}}{{Krumholz}
  et~al.}{2012}]{Krumholz_et_al_2012}
{Krumholz} M.~R.,  {Dekel} A.,   {McKee} C.~F.,  2012, \mn@doi [\apj]
  {10.1088/0004-637X/745/1/69}, \href
  {http://adsabs.harvard.edu/abs/2012ApJ...745...69K} {745, 69}

\bibitem[\protect\citeauthoryear{{Le Teuff}, {Millar}  \& {Markwick}}{{Le
  Teuff} et~al.}{2000}]{LeTeuff_et_al_2000}
{Le Teuff} Y.~H.,  {Millar} T.~J.,   {Markwick} A.~J.,  2000, \mn@doi [\aaps]
  {10.1051/aas:2000265}, \href
  {http://adsabs.harvard.edu/abs/2000A%26AS..146..157L} {146, 157}

\bibitem[\protect\citeauthoryear{{Lenzuni}, {Chernoff}  \&
  {Salpeter}}{{Lenzuni} et~al.}{1991}]{Lenzuni_et_al_1991}
{Lenzuni} P.,  {Chernoff} D.~F.,   {Salpeter} E.~E.,  1991, \mn@doi [\apjs]
  {10.1086/191580}, \href {http://adsabs.harvard.edu/abs/1991ApJS...76..759L}
  {76, 759}

\bibitem[\protect\citeauthoryear{{Lupi}, {Bovino}, {Capelo}, {Volonteri}  \&
  {Silk}}{{Lupi} et~al.}{2017}]{Lupi_et_al_2017}
{Lupi} A.,  {Bovino} S.,  {Capelo} P.~R.,  {Volonteri} M.,   {Silk} J.,  2017,
  preprint, \href {http://adsabs.harvard.edu/abs/2017arXiv171001315L} {}
  (\mn@eprint {arXiv} {1710.01315})

\bibitem[\protect\citeauthoryear{{Maio}, {Dolag}, {Ciardi}  \&
  {Tornatore}}{{Maio} et~al.}{2007}]{Maio_et_al_2007}
{Maio} U.,  {Dolag} K.,  {Ciardi} B.,   {Tornatore} L.,  2007, \mn@doi [\mnras]
  {10.1111/j.1365-2966.2007.12016.x}, \href
  {http://adsabs.harvard.edu/abs/2007MNRAS.379..963M} {379, 963}

\bibitem[\protect\citeauthoryear{{Malhotra} et~al.,}{{Malhotra}
  et~al.}{2001}]{Malhotra_et_al_2001}
{Malhotra} S.,  et~al., 2001, \mn@doi [\apj] {10.1086/323046}, \href
  {http://adsabs.harvard.edu/abs/2001ApJ...561..766M} {561, 766}

\bibitem[\protect\citeauthoryear{{Mathis}, {Rumpl}  \& {Nordsieck}}{{Mathis}
  et~al.}{1977}]{Mathis_et_al_1977}
{Mathis} J.~S.,  {Rumpl} W.,   {Nordsieck} K.~H.,  1977, \mn@doi [\apj]
  {10.1086/155591}, \href {http://adsabs.harvard.edu/abs/1977ApJ...217..425M}
  {217, 425}

\bibitem[\protect\citeauthoryear{{McKee} \& {Krumholz}}{{McKee} \&
  {Krumholz}}{2010}]{McKee_Krumholz_2010}
{McKee} C.~F.,  {Krumholz} M.~R.,  2010, \mn@doi [\apj]
  {10.1088/0004-637X/709/1/308}, \href
  {http://adsabs.harvard.edu/abs/2010ApJ...709..308M} {709, 308}

\bibitem[\protect\citeauthoryear{{McKee} \& {Ostriker}}{{McKee} \&
  {Ostriker}}{1977}]{McKee_Ostriker_1977}
{McKee} C.~F.,  {Ostriker} J.~P.,  1977, \mn@doi [\apj] {10.1086/155667}, \href
  {http://adsabs.harvard.edu/abs/1977ApJ...218..148M} {218, 148}

\bibitem[\protect\citeauthoryear{{Micha{\l}owski} et~al.,}{{Micha{\l}owski}
  et~al.}{2016}]{Michalowski_et_al_2016}
{Micha{\l}owski} M.~J.,  et~al., 2016, \mn@doi [\aap]
  {10.1051/0004-6361/201629441}, \href
  {http://adsabs.harvard.edu/abs/2016A%26A...595A..72M} {595, A72}

\bibitem[\protect\citeauthoryear{{Mitchell} \& {Deveau}}{{Mitchell} \&
  {Deveau}}{1983}]{Mitchell_Deveau_1983}
{Mitchell} G.~F.,  {Deveau} T.~J.,  1983, \mn@doi [\apj] {10.1086/160812},
  \href {http://adsabs.harvard.edu/abs/1983ApJ...266..646M} {266, 646}

\bibitem[\protect\citeauthoryear{{Mo}, {Mao}  \& {White}}{{Mo}
  et~al.}{1998}]{Mo_et_al_1998}
{Mo} H.~J.,  {Mao} S.,   {White} S.~D.~M.,  1998, \mn@doi [\mnras]
  {10.1046/j.1365-8711.1998.01227.x}, \href
  {http://adsabs.harvard.edu/abs/1998MNRAS.295..319M} {295, 319}

\bibitem[\protect\citeauthoryear{{Monaghan}}{{Monaghan}}{1992}]{Monaghan_1992}
{Monaghan} J.~J.,  1992, \mn@doi [\araa] {10.1146/annurev.astro.30.1.543},
  \href {http://adsabs.harvard.edu/abs/1992ARA%26A..30..543M} {30, 543}

\bibitem[\protect\citeauthoryear{{Nahar}}{{Nahar}}{1995}]{Nahar_1995}
{Nahar} S.~N.,  1995, \mn@doi [\apjs] {10.1086/192248}, \href
  {http://adsabs.harvard.edu/abs/1995ApJS..101..423N} {101, 423}

\bibitem[\protect\citeauthoryear{{Nahar}}{{Nahar}}{1996}]{Nahar_1996}
{Nahar} S.~N.,  1996, \mn@doi [\apjs] {10.1086/192336}, \href
  {http://adsabs.harvard.edu/abs/1996ApJS..106..213N} {106, 213}

\bibitem[\protect\citeauthoryear{{Nahar}}{{Nahar}}{1999}]{Nahar_1999}
{Nahar} S.~N.,  1999, \mn@doi [\apjs] {10.1086/313173}, \href
  {http://adsabs.harvard.edu/abs/1999ApJS..120..131N} {120, 131}

\bibitem[\protect\citeauthoryear{{Nahar}}{{Nahar}}{2000}]{Nahar_2000}
{Nahar} S.~N.,  2000, \mn@doi [\apjs] {10.1086/313307}, \href
  {http://adsabs.harvard.edu/abs/2000ApJS..126..537N} {126, 537}

\bibitem[\protect\citeauthoryear{{Nahar} \& {Pradhan}}{{Nahar} \&
  {Pradhan}}{1997}]{Nahar_Pradhan_1997}
{Nahar} S.~N.,  {Pradhan} A.~K.,  1997, \mn@doi [\apjs] {10.1086/313013}, \href
  {http://adsabs.harvard.edu/abs/1997ApJS..111..339N} {111, 339}

\bibitem[\protect\citeauthoryear{{Navarro}, {Frenk}  \& {White}}{{Navarro}
  et~al.}{1996}]{Navarro_et_al_1996}
{Navarro} J.~F.,  {Frenk} C.~S.,   {White} S.~D.~M.,  1996, \mn@doi [\apj]
  {10.1086/177173}, \href {http://adsabs.harvard.edu/abs/1996ApJ...462..563N}
  {462, 563}

\bibitem[\protect\citeauthoryear{{Nickerson}, {Teyssier}  \&
  {Rosdahl}}{{Nickerson} et~al.}{2017}]{Nickerson_et_al_2017}
{Nickerson} S.,  {Teyssier} R.,   {Rosdahl} J.,  2017, in prep.

\bibitem[\protect\citeauthoryear{{Omukai}}{{Omukai}}{2000}]{Omukai_2000}
{Omukai} K.,  2000, \mn@doi [\apj] {10.1086/308776}, \href
  {http://adsabs.harvard.edu/abs/2000ApJ...534..809O} {534, 809}

\bibitem[\protect\citeauthoryear{{Omukai}, {Tsuribe}, {Schneider}  \&
  {Ferrara}}{{Omukai} et~al.}{2005}]{Omukai_et_al_2005}
{Omukai} K.,  {Tsuribe} T.,  {Schneider} R.,   {Ferrara} A.,  2005, \mn@doi
  [\apj] {10.1086/429955}, \href
  {http://adsabs.harvard.edu/abs/2005ApJ...626..627O} {626, 627}

\bibitem[\protect\citeauthoryear{{Oppenheimer} \& {Schaye}}{{Oppenheimer} \&
  {Schaye}}{2013a}]{Oppenheimer_Schaye_2013a}
{Oppenheimer} B.~D.,  {Schaye} J.,  2013a, \mn@doi [\mnras]
  {10.1093/mnras/stt1043}, \href
  {http://adsabs.harvard.edu/abs/2013MNRAS.434.1043O} {434, 1043}

\bibitem[\protect\citeauthoryear{{Oppenheimer} \& {Schaye}}{{Oppenheimer} \&
  {Schaye}}{2013b}]{Oppenheimer_Schaye_2013b}
{Oppenheimer} B.~D.,  {Schaye} J.,  2013b, \mn@doi [\mnras]
  {10.1093/mnras/stt1150}, \href
  {http://adsabs.harvard.edu/abs/2013MNRAS.434.1063O} {434, 1063}

\bibitem[\protect\citeauthoryear{{Pakmor} \& {Springel}}{{Pakmor} \&
  {Springel}}{2013}]{Pakmor_Springel_2013}
{Pakmor} R.,  {Springel} V.,  2013, \mn@doi [\mnras] {10.1093/mnras/stt428},
  \href {http://adsabs.harvard.edu/abs/2013MNRAS.432..176P} {432, 176}

\bibitem[\protect\citeauthoryear{{Pallottini}, {Ferrara}, {Gallerani},
  {Vallini}, {Maiolino}  \& {Salvadori}}{{Pallottini}
  et~al.}{2017a}]{Pallottini_et_al_2017a}
{Pallottini} A.,  {Ferrara} A.,  {Gallerani} S.,  {Vallini} L.,  {Maiolino} R.,
    {Salvadori} S.,  2017a, \mn@doi [\mnras] {10.1093/mnras/stw2847}, \href
  {http://adsabs.harvard.edu/abs/2017MNRAS.465.2540P} {465, 2540}

\bibitem[\protect\citeauthoryear{{Pallottini}, {Ferrara}, {Bovino}, {Vallini},
  {Gallerani}, {Maiolino}  \& {Salvadori}}{{Pallottini}
  et~al.}{2017b}]{Pallottini_et_al_2017b}
{Pallottini} A.,  {Ferrara} A.,  {Bovino} S.,  {Vallini} L.,  {Gallerani} S.,
  {Maiolino} R.,   {Salvadori} S.,  2017b, \mn@doi [\mnras]
  {10.1093/mnras/stx1792}, \href
  {http://adsabs.harvard.edu/abs/2017MNRAS.471.4128P} {471, 4128}

\bibitem[\protect\citeauthoryear{{Papadopoulos} \& {Thi}}{{Papadopoulos} \&
  {Thi}}{2013}]{Papadopoulos_Thi_2013}
{Papadopoulos} P.~P.,  {Thi} W.-F.,  2013, in {Torres} D.~F.,  {Reimer} O.,
  eds,  Astrophysics and Space Science Proceedings Vol. 34, Cosmic Rays in
  Star-Forming Environments. p.~41 (\mn@eprint {arXiv} {1207.2048}),
  \mn@doi{10.1007/978-3-642-35410-6_5}

\bibitem[\protect\citeauthoryear{{Pilbratt} et~al.,}{{Pilbratt}
  et~al.}{2010}]{Pilbratt_et_al_2010}
{Pilbratt} G.~L.,  et~al., 2010, \mn@doi [\aap] {10.1051/0004-6361/201014759},
  \href {http://adsabs.harvard.edu/abs/2010A%26A...518L...1P} {518, L1}

\bibitem[\protect\citeauthoryear{{Pontzen}, {Ro{\v s}kar}, {Stinson}  \&
  {Woods}}{{Pontzen} et~al.}{2013}]{Pontzen_et_al_2013}
{Pontzen} A.,  {Ro{\v s}kar} R.,  {Stinson} G.,   {Woods} R.,  2013, {pynbody:
  N-Body/SPH analysis for python}, Astrophysics Source Code Library (\mn@eprint
  {ascl} {1305.002})

\bibitem[\protect\citeauthoryear{{Poulaert}, {Brouillard}, {Claeys}, {McGowan}
  \& {Van Wassenhove}}{{Poulaert} et~al.}{1978}]{Poulaert_et_al_1978}
{Poulaert} G.,  {Brouillard} F.,  {Claeys} W.,  {McGowan} J.~W.,   {Van
  Wassenhove} G.,  1978, \mn@doi [Journal of Physics B Atomic Molecular
  Physics] {10.1088/0022-3700/11/21/006}, \href
  {http://adsabs.harvard.edu/abs/1978JPhB...11L.671P} {11, L671}

\bibitem[\protect\citeauthoryear{{Raiteri}, {Villata}  \& {Navarro}}{{Raiteri}
  et~al.}{1996}]{Raiteri_et_al_1996}
{Raiteri} C.~M.,  {Villata} M.,   {Navarro} J.~F.,  1996, \aap, \href
  {http://adsabs.harvard.edu/abs/1996A%26A...315..105R} {315, 105}

\bibitem[\protect\citeauthoryear{{Richings} \& {Schaye}}{{Richings} \&
  {Schaye}}{2016}]{Richings_Schaye_2016}
{Richings} A.~J.,  {Schaye} J.,  2016, \mn@doi [\mnras] {10.1093/mnras/stw327},
  \href {http://adsabs.harvard.edu/abs/2016MNRAS.458..270R} {458, 270}

\bibitem[\protect\citeauthoryear{{Richings}, {Schaye}  \&
  {Oppenheimer}}{{Richings} et~al.}{2014a}]{Richings_et_al_2014a}
{Richings} A.~J.,  {Schaye} J.,   {Oppenheimer} B.~D.,  2014a, \mn@doi [\mnras]
  {10.1093/mnras/stu525}, \href
  {http://adsabs.harvard.edu/abs/2014MNRAS.440.3349R} {440, 3349}

\bibitem[\protect\citeauthoryear{{Richings}, {Schaye}  \&
  {Oppenheimer}}{{Richings} et~al.}{2014b}]{Richings_et_al_2014b}
{Richings} A.~J.,  {Schaye} J.,   {Oppenheimer} B.~D.,  2014b, \mn@doi [\mnras]
  {10.1093/mnras/stu1046}, \href
  {http://adsabs.harvard.edu/abs/2014MNRAS.442.2780R} {442, 2780}

\bibitem[\protect\citeauthoryear{{Ritchie} \& {Thomas}}{{Ritchie} \&
  {Thomas}}{2001}]{Ritchie_Thomas_2001}
{Ritchie} B.~W.,  {Thomas} P.~A.,  2001, \mn@doi [\mnras]
  {10.1046/j.1365-8711.2001.04268.x}, \href
  {http://adsabs.harvard.edu/abs/2001MNRAS.323..743R} {323, 743}

\bibitem[\protect\citeauthoryear{{Robertson} \& {Kravtsov}}{{Robertson} \&
  {Kravtsov}}{2008}]{Robertson_Kravtsov_2008}
{Robertson} B.~E.,  {Kravtsov} A.~V.,  2008, \mn@doi [\apj] {10.1086/587796},
  \href {http://adsabs.harvard.edu/abs/2008ApJ...680.1083R} {680, 1083}

\bibitem[\protect\citeauthoryear{{Rodenbeck} \& {Schleicher}}{{Rodenbeck} \&
  {Schleicher}}{2016}]{Rodenbeck_Schleicher_2016}
{Rodenbeck} K.,  {Schleicher} D.~R.~G.,  2016, \mn@doi [\aap]
  {10.1051/0004-6361/201527393}, \href
  {http://adsabs.harvard.edu/abs/2016A%26A...593A..89R} {593, A89}

\bibitem[\protect\citeauthoryear{{Ro{\v s}kar}, {Fiacconi}, {Mayer},
  {Kazantzidis}, {Quinn}  \& {Wadsley}}{{Ro{\v s}kar}
  et~al.}{2015}]{Roskar_et_al_2015}
{Ro{\v s}kar} R.,  {Fiacconi} D.,  {Mayer} L.,  {Kazantzidis} S.,  {Quinn}
  T.~R.,   {Wadsley} J.,  2015, \mn@doi [\mnras] {10.1093/mnras/stv312}, \href
  {http://adsabs.harvard.edu/abs/2015MNRAS.449..494R} {449, 494}

\bibitem[\protect\citeauthoryear{{Safranek-Shrader}, {Krumholz}, {Kim},
  {Ostriker}, {Klein}, {Li}, {McKee}  \& {Stone}}{{Safranek-Shrader}
  et~al.}{2017}]{Safranek-Shrader_et_al_2017}
{Safranek-Shrader} C.,  {Krumholz} M.~R.,  {Kim} C.-G.,  {Ostriker} E.~C.,
  {Klein} R.~I.,  {Li} S.,  {McKee} C.~F.,   {Stone} J.~M.,  2017, \mn@doi
  [\mnras] {10.1093/mnras/stw2647}, \href
  {http://adsabs.harvard.edu/abs/2017MNRAS.465..885S} {465, 885}

\bibitem[\protect\citeauthoryear{{S{\'a}nchez} et~al.,}{{S{\'a}nchez}
  et~al.}{2017}]{Sanchez_et_al_2017}
{S{\'a}nchez} S.~F.,  et~al., 2017, \mn@doi [\mnras] {10.1093/mnras/stx808},
  \href {http://adsabs.harvard.edu/abs/2017MNRAS.469.2121S} {469, 2121}

\bibitem[\protect\citeauthoryear{{Savin}, {Krsti{\'c}}, {Haiman}  \&
  {Stancil}}{{Savin} et~al.}{2004}]{Savin_et_al_2004}
{Savin} D.~W.,  {Krsti{\'c}} P.~S.,  {Haiman} Z.,   {Stancil} P.~C.,  2004,
  \mn@doi [\apjl] {10.1086/421108}, \href
  {http://adsabs.harvard.edu/abs/2004ApJ...606L.167S} {606, L167}

\bibitem[\protect\citeauthoryear{{Scannapieco} et~al.,}{{Scannapieco}
  et~al.}{2012}]{Scannapieco_et_al_2012}
{Scannapieco} C.,  et~al., 2012, \mn@doi [\mnras]
  {10.1111/j.1365-2966.2012.20993.x}, \href
  {http://adsabs.harvard.edu/abs/2012MNRAS.423.1726S} {423, 1726}

\bibitem[\protect\citeauthoryear{{Schaye} et~al.,}{{Schaye}
  et~al.}{2010}]{Schaye_et_al_2010}
{Schaye} J.,  et~al., 2010, \mn@doi [\mnras]
  {10.1111/j.1365-2966.2009.16029.x}, \href
  {http://adsabs.harvard.edu/abs/2010MNRAS.402.1536S} {402, 1536}

\bibitem[\protect\citeauthoryear{{Schaye} et~al.,}{{Schaye}
  et~al.}{2015}]{Schaye_et_al_2015}
{Schaye} J.,  et~al., 2015, \mn@doi [\mnras] {10.1093/mnras/stu2058}, \href
  {http://adsabs.harvard.edu/abs/2015MNRAS.446..521S} {446, 521}

\bibitem[\protect\citeauthoryear{{Schmidt}}{{Schmidt}}{1959}]{Schmidt_1959}
{Schmidt} M.,  1959, \mn@doi [\apj] {10.1086/146614}, \href
  {http://adsabs.harvard.edu/abs/1959ApJ...129..243S} {129, 243}

\bibitem[\protect\citeauthoryear{{Schmidt}}{{Schmidt}}{1963}]{Schmidt_1963}
{Schmidt} M.,  1963, \mn@doi [\apj] {10.1086/147553}, \href
  {http://adsabs.harvard.edu/abs/1963ApJ...137..758S} {137, 758}

\bibitem[\protect\citeauthoryear{{Schneider}, {Dulieu}, {Giusti-Suzor}  \&
  {Roueff}}{{Schneider} et~al.}{1994}]{Schneider_et_al_1994}
{Schneider} I.~F.,  {Dulieu} O.,  {Giusti-Suzor} A.,   {Roueff} E.,  1994,
  \mn@doi [\apj] {10.1086/173948}, \href
  {http://adsabs.harvard.edu/abs/1994ApJ...424..983S} {424, 983}

\bibitem[\protect\citeauthoryear{{Schuessler} \& {Schmitt}}{{Schuessler} \&
  {Schmitt}}{1981}]{Schuessler_Schmitt_1981}
{Schuessler} I.,  {Schmitt} D.,  1981, \aap, \href
  {http://adsabs.harvard.edu/abs/1981A%26A....97..373S} {97, 373}

\bibitem[\protect\citeauthoryear{{Segers}, {Oppenheimer}, {Schaye}  \&
  {Richings}}{{Segers} et~al.}{2017}]{Segers_et_al_2017}
{Segers} M.~C.,  {Oppenheimer} B.~D.,  {Schaye} J.,   {Richings} A.~J.,  2017,
  \mn@doi [\mnras] {10.1093/mnras/stx1633}, \href
  {http://adsabs.harvard.edu/abs/2017MNRAS.471.1026S} {471, 1026}

\bibitem[\protect\citeauthoryear{{Seifried} \& {Walch}}{{Seifried} \&
  {Walch}}{2016}]{Seifried_Walch_2016}
{Seifried} D.,  {Walch} S.,  2016, \mn@doi [\mnras] {10.1093/mnrasl/slw035},
  \href {http://adsabs.harvard.edu/abs/2016MNRAS.459L..11S} {459, L11}

\bibitem[\protect\citeauthoryear{{Shapiro} \& {Kang}}{{Shapiro} \&
  {Kang}}{1987}]{Shapiro_Kang_1987}
{Shapiro} P.~R.,  {Kang} H.,  1987, \mn@doi [\apj] {10.1086/165350}, \href
  {http://adsabs.harvard.edu/abs/1987ApJ...318...32S} {318, 32}

\bibitem[\protect\citeauthoryear{{Shen}, {Wadsley}  \& {Stinson}}{{Shen}
  et~al.}{2010}]{Shen_et_al_2010}
{Shen} S.,  {Wadsley} J.,   {Stinson} G.,  2010, \mn@doi [\mnras]
  {10.1111/j.1365-2966.2010.17047.x}, \href
  {http://adsabs.harvard.edu/abs/2010MNRAS.407.1581S} {407, 1581}

\bibitem[\protect\citeauthoryear{{Shen}, {Madau}, {Guedes}, {Mayer},
  {Prochaska}  \& {Wadsley}}{{Shen} et~al.}{2013}]{Shen_et_al_2013}
{Shen} S.,  {Madau} P.,  {Guedes} J.,  {Mayer} L.,  {Prochaska} J.~X.,
  {Wadsley} J.,  2013, \mn@doi [\apj] {10.1088/0004-637X/765/2/89}, \href
  {http://adsabs.harvard.edu/abs/2013ApJ...765...89S} {765, 89}

\bibitem[\protect\citeauthoryear{{Spitzer}}{{Spitzer}}{1942}]{Spitzer_1942}
{Spitzer} Jr. L.,  1942, \mn@doi [\apj] {10.1086/144407}, \href
  {http://adsabs.harvard.edu/abs/1942ApJ....95..329S} {95, 329}

\bibitem[\protect\citeauthoryear{{Springel} \& {White}}{{Springel} \&
  {White}}{1999}]{Springel_White_1999}
{Springel} V.,  {White} S.~D.~M.,  1999, \mn@doi [\mnras]
  {10.1046/j.1365-8711.1999.02613.x}, \href
  {http://adsabs.harvard.edu/abs/1999MNRAS.307..162S} {307, 162}

\bibitem[\protect\citeauthoryear{{Springel}, {Di Matteo}  \&
  {Hernquist}}{{Springel} et~al.}{2005a}]{Springel_et_al_2005b}
{Springel} V.,  {Di Matteo} T.,   {Hernquist} L.,  2005a, \mn@doi [\mnras]
  {10.1111/j.1365-2966.2005.09238.x}, \href
  {http://adsabs.harvard.edu/abs/2005MNRAS.361..776S} {361, 776}

\bibitem[\protect\citeauthoryear{{Springel} et~al.,}{{Springel}
  et~al.}{2005b}]{Springel_et_al_2005a}
{Springel} V.,  et~al., 2005b, \mn@doi [\nat] {10.1038/nature03597}, \href
  {http://adsabs.harvard.edu/abs/2005Natur.435..629S} {435, 629}

\bibitem[\protect\citeauthoryear{{Stacey}, {Geis}, {Genzel}, {Lugten},
  {Poglitsch}, {Sternberg}  \& {Townes}}{{Stacey}
  et~al.}{1991}]{Stacey_et_al_1991}
{Stacey} G.~J.,  {Geis} N.,  {Genzel} R.,  {Lugten} J.~B.,  {Poglitsch} A.,
  {Sternberg} A.,   {Townes} C.~H.,  1991, \mn@doi [\apj] {10.1086/170062},
  \href {http://adsabs.harvard.edu/abs/1991ApJ...373..423S} {373, 423}

\bibitem[\protect\citeauthoryear{{Stadel}}{{Stadel}}{2001}]{Stadel_2001}
{Stadel} J.~G.,  2001, PhD thesis, UNIVERSITY OF WASHINGTON

\bibitem[\protect\citeauthoryear{{Stancil} et~al.,}{{Stancil}
  et~al.}{1998}]{Stancil_et_al_1998}
{Stancil} P.~C.,  et~al., 1998, \mn@doi [\apj] {10.1086/305937}, \href
  {http://adsabs.harvard.edu/abs/1998ApJ...502.1006S} {502, 1006}

\bibitem[\protect\citeauthoryear{{Stancil}, {Schultz}, {Kimura}, {Gu}, {Hirsch}
   \& {Buenker}}{{Stancil} et~al.}{1999}]{Stancil_et_al_1999}
{Stancil} P.~C.,  {Schultz} D.~R.,  {Kimura} M.,  {Gu} J.-P.,  {Hirsch} G.,
  {Buenker} R.~J.,  1999, \mn@doi [\aaps] {10.1051/aas:1999419}, \href
  {http://adsabs.harvard.edu/abs/1999A%26AS..140..225S} {140, 225}

\bibitem[\protect\citeauthoryear{{Stenrup}, {Larson}  \& {Elander}}{{Stenrup}
  et~al.}{2009}]{Stenrup_et_al_2009}
{Stenrup} M.,  {Larson} {\AA}.,   {Elander} N.,  2009, preprint, \href
  {http://adsabs.harvard.edu/abs/2009arXiv0902.1900S} {} (\mn@eprint {arXiv}
  {0902.1900})

\bibitem[\protect\citeauthoryear{{Sternberg}, {Le Petit}, {Roueff}  \& {Le
  Bourlot}}{{Sternberg} et~al.}{2014}]{Sternberg_et_al_2014}
{Sternberg} A.,  {Le Petit} F.,  {Roueff} E.,   {Le Bourlot} J.,  2014, \mn@doi
  [\apj] {10.1088/0004-637X/790/1/10}, \href
  {http://adsabs.harvard.edu/abs/2014ApJ...790...10S} {790, 10}

\bibitem[\protect\citeauthoryear{{Stinson}, {Seth}, {Katz}, {Wadsley},
  {Governato}  \& {Quinn}}{{Stinson} et~al.}{2006}]{Stinson_et_al_2006}
{Stinson} G.,  {Seth} A.,  {Katz} N.,  {Wadsley} J.,  {Governato} F.,   {Quinn}
  T.,  2006, \mn@doi [\mnras] {10.1111/j.1365-2966.2006.11097.x}, \href
  {http://adsabs.harvard.edu/abs/2006MNRAS.373.1074S} {373, 1074}

\bibitem[\protect\citeauthoryear{{Sutherland} \& {Dopita}}{{Sutherland} \&
  {Dopita}}{1993}]{Sutherland_Dopita_1993}
{Sutherland} R.~S.,  {Dopita} M.~A.,  1993, \mn@doi [\apjs] {10.1086/191823},
  \href {http://adsabs.harvard.edu/abs/1993ApJS...88..253S} {88, 253}

\bibitem[\protect\citeauthoryear{{Tacconi} et~al.,}{{Tacconi}
  et~al.}{2010}]{Tacconi_et_al_2010}
{Tacconi} L.~J.,  et~al., 2010, \mn@doi [\nat] {10.1038/nature08773}, \href
  {http://adsabs.harvard.edu/abs/2010Natur.463..781T} {463, 781}

\bibitem[\protect\citeauthoryear{{Teyssier}}{{Teyssier}}{2002}]{Teyssier_2002}
{Teyssier} R.,  2002, \mn@doi [\aap] {10.1051/0004-6361:20011817}, \href
  {http://adsabs.harvard.edu/abs/2002A%26A...385..337T} {385, 337}

\bibitem[\protect\citeauthoryear{{Thielemann}, {Nomoto}  \&
  {Yokoi}}{{Thielemann} et~al.}{1986}]{Thielemann_et_al_1986}
{Thielemann} F.-K.,  {Nomoto} K.,   {Yokoi} K.,  1986, \aap, \href
  {http://adsabs.harvard.edu/abs/1986A%26A...158...17T} {158, 17}

\bibitem[\protect\citeauthoryear{{Tielens} \& {Hollenbach}}{{Tielens} \&
  {Hollenbach}}{1985}]{Tielens_Hollenbach_1985}
{Tielens} A.~G.~G.~M.,  {Hollenbach} D.,  1985, \mn@doi [\apj]
  {10.1086/163112}, \href {http://adsabs.harvard.edu/abs/1985ApJ...291..747T}
  {291, 747}

\bibitem[\protect\citeauthoryear{{Tomassetti}, {Porciani}, {Romano-D{\'{\i}}az}
   \& {Ludlow}}{{Tomassetti} et~al.}{2015}]{Tomassetti_et_al_2015}
{Tomassetti} M.,  {Porciani} C.,  {Romano-D{\'{\i}}az} E.,   {Ludlow} A.~D.,
  2015, \mn@doi [\mnras] {10.1093/mnras/stu2273}, \href
  {http://adsabs.harvard.edu/abs/2015MNRAS.446.3330T} {446, 3330}

\bibitem[\protect\citeauthoryear{{Toomre}}{{Toomre}}{1964}]{Toomre_1964}
{Toomre} A.,  1964, \mn@doi [\apj] {10.1086/147861}, \href
  {http://adsabs.harvard.edu/abs/1964ApJ...139.1217T} {139, 1217}

\bibitem[\protect\citeauthoryear{{Truelove}, {Klein}, {McKee}, {Holliman},
  {Howell}  \& {Greenough}}{{Truelove} et~al.}{1997}]{Truelove_et_al_1997}
{Truelove} J.~K.,  {Klein} R.~I.,  {McKee} C.~F.,  {Holliman} II J.~H.,
  {Howell} L.~H.,   {Greenough} J.~A.,  1997, \mn@doi [\apjl] {10.1086/310975},
  \href {http://adsabs.harvard.edu/abs/1997ApJ...489L.179T} {489, L179}

\bibitem[\protect\citeauthoryear{{Verner} \& {Ferland}}{{Verner} \&
  {Ferland}}{1996}]{Verner_Ferland_1996}
{Verner} D.~A.,  {Ferland} G.~J.,  1996, \mn@doi [\apjs] {10.1086/192284},
  \href {http://adsabs.harvard.edu/abs/1996ApJS..103..467V} {103, 467}

\bibitem[\protect\citeauthoryear{{Verner}, {Ferland}, {Korista}  \&
  {Yakovlev}}{{Verner} et~al.}{1996}]{Verner_et_al_1996}
{Verner} D.~A.,  {Ferland} G.~J.,  {Korista} K.~T.,   {Yakovlev} D.~G.,  1996,
  \mn@doi [\apj] {10.1086/177435}, \href
  {http://adsabs.harvard.edu/abs/1996ApJ...465..487V} {465, 487}

\bibitem[\protect\citeauthoryear{{Vitvitska}, {Klypin}, {Kravtsov}, {Wechsler},
  {Primack}  \& {Bullock}}{{Vitvitska} et~al.}{2002}]{Vitvitska_et_al_2002}
{Vitvitska} M.,  {Klypin} A.~A.,  {Kravtsov} A.~V.,  {Wechsler} R.~H.,
  {Primack} J.~R.,   {Bullock} J.~S.,  2002, \mn@doi [\apj] {10.1086/344361},
  \href {http://adsabs.harvard.edu/abs/2002ApJ...581..799V} {581, 799}

\bibitem[\protect\citeauthoryear{{Voronov}}{{Voronov}}{1997}]{Voronov_1997}
{Voronov} G.~S.,  1997, \mn@doi [Atomic Data and Nuclear Data Tables]
  {10.1006/adnd.1997.0732}, \href
  {http://adsabs.harvard.edu/abs/1997ADNDT..65....1V} {65, 1}

\bibitem[\protect\citeauthoryear{{Wadsley}, {Stadel}  \& {Quinn}}{{Wadsley}
  et~al.}{2004}]{Wadsley_et_al_2004}
{Wadsley} J.~W.,  {Stadel} J.,   {Quinn} T.,  2004, \mn@doi [\na]
  {10.1016/j.newast.2003.08.004}, \href
  {http://adsabs.harvard.edu/abs/2004NewA....9..137W} {9, 137}

\bibitem[\protect\citeauthoryear{{Wadsley}, {Veeravalli}  \&
  {Couchman}}{{Wadsley} et~al.}{2008}]{Wadsley_et_al_2008}
{Wadsley} J.~W.,  {Veeravalli} G.,   {Couchman} H.~M.~P.,  2008, \mn@doi
  [\mnras] {10.1111/j.1365-2966.2008.13260.x}, \href
  {http://adsabs.harvard.edu/abs/2008MNRAS.387..427W} {387, 427}

\bibitem[\protect\citeauthoryear{{Wadsley}, {Keller}  \& {Quinn}}{{Wadsley}
  et~al.}{2017}]{Wadsley_et_al_2017}
{Wadsley} J.~W.,  {Keller} B.~W.,   {Quinn} T.~R.,  2017, \mn@doi [\mnras]
  {10.1093/mnras/stx1643}, \href
  {http://adsabs.harvard.edu/abs/2017MNRAS.471.2357W} {471, 2357}

\bibitem[\protect\citeauthoryear{{Weidemann}}{{Weidemann}}{1987}]{Weidemann_1987}
{Weidemann} V.,  1987, \aap, \href
  {http://adsabs.harvard.edu/abs/1987A%26A...188...74W} {188, 74}

\bibitem[\protect\citeauthoryear{{Weingartner} \& {Draine}}{{Weingartner} \&
  {Draine}}{2001}]{Weingartner_Draine_2001b}
{Weingartner} J.~C.,  {Draine} B.~T.,  2001, \mn@doi [\apj] {10.1086/324035},
  \href {http://adsabs.harvard.edu/abs/2001ApJ...563..842W} {563, 842}

\bibitem[\protect\citeauthoryear{{Wendland}}{{Wendland}}{1995}]{Wendland_1995}
{Wendland} H.,  1995, \mn@doi [Adv. in Comput. Math.] {10.1007/BF02123482}, 4,
  389

\bibitem[\protect\citeauthoryear{{Wolcott-Green}, {Haiman}  \&
  {Bryan}}{{Wolcott-Green} et~al.}{2011}]{Wolcott-Green_et_al_2011}
{Wolcott-Green} J.,  {Haiman} Z.,   {Bryan} G.~L.,  2011, \mn@doi [\mnras]
  {10.1111/j.1365-2966.2011.19538.x}, \href
  {http://adsabs.harvard.edu/abs/2011MNRAS.418..838W} {418, 838}

\bibitem[\protect\citeauthoryear{{Wolfire}, {Hollenbach}, {McKee}, {Tielens}
  \& {Bakes}}{{Wolfire} et~al.}{1995}]{Wolfire_et_al_1995}
{Wolfire} M.~G.,  {Hollenbach} D.,  {McKee} C.~F.,  {Tielens} A.~G.~G.~M.,
  {Bakes} E.~L.~O.,  1995, \mn@doi [\apj] {10.1086/175510}, \href
  {http://adsabs.harvard.edu/abs/1995ApJ...443..152W} {443, 152}

\bibitem[\protect\citeauthoryear{{Wolfire}, {McKee}, {Hollenbach}  \&
  {Tielens}}{{Wolfire} et~al.}{2003}]{Wolfire_et_al_2003}
{Wolfire} M.~G.,  {McKee} C.~F.,  {Hollenbach} D.,   {Tielens} A.~G.~G.~M.,
  2003, \mn@doi [\apj] {10.1086/368016}, \href
  {http://adsabs.harvard.edu/abs/2003ApJ...587..278W} {587, 278}

\bibitem[\protect\citeauthoryear{{Woosley} \& {Weaver}}{{Woosley} \&
  {Weaver}}{1995}]{Woosley_Weaver_1995}
{Woosley} S.~E.,  {Weaver} T.~A.,  1995, \mn@doi [\apjs] {10.1086/192237},
  \href {http://adsabs.harvard.edu/abs/1995ApJS..101..181W} {101, 181}

\bibitem[\protect\citeauthoryear{{Yamasawa}, {Habe}, {Kozasa}, {Nozawa},
  {Hirashita}, {Umeda}  \& {Nomoto}}{{Yamasawa}
  et~al.}{2011}]{Yamasawa_et_al_2011}
{Yamasawa} D.,  {Habe} A.,  {Kozasa} T.,  {Nozawa} T.,  {Hirashita} H.,
  {Umeda} H.,   {Nomoto} K.,  2011, \mn@doi [\apj]
  {10.1088/0004-637X/735/1/44}, \href
  {http://adsabs.harvard.edu/abs/2011ApJ...735...44Y} {735, 44}

\bibitem[\protect\citeauthoryear{{Yoshida}, {Omukai}, {Hernquist}  \&
  {Abel}}{{Yoshida} et~al.}{2006}]{Yoshida_et_al_2006}
{Yoshida} N.,  {Omukai} K.,  {Hernquist} L.,   {Abel} T.,  2006, \mn@doi [\apj]
  {10.1086/507978}, \href {http://adsabs.harvard.edu/abs/2006ApJ...652....6Y}
  {652, 6}

\bibitem[\protect\citeauthoryear{{Zhao} et~al.,}{{Zhao}
  et~al.}{2004}]{Zhao_et_al_2004}
{Zhao} L.~B.,  et~al., 2004, \mn@doi [\apj] {10.1086/424729}, \href
  {http://adsabs.harvard.edu/abs/2004ApJ...615.1063Z} {615, 1063}

\bibitem[\protect\citeauthoryear{{Zhukovska}, {Gail}  \&
  {Trieloff}}{{Zhukovska} et~al.}{2008}]{Zhukovska_et_al_2008}
{Zhukovska} S.,  {Gail} H.-P.,   {Trieloff} M.,  2008, \mn@doi [\aap]
  {10.1051/0004-6361:20077789}, \href
  {http://adsabs.harvard.edu/abs/2008A%26A...479..453Z} {479, 453}

\bibitem[\protect\citeauthoryear{{de Jong}}{{de Jong}}{1972}]{deJong_1972}
{de Jong} T.,  1972, \aap, \href
  {http://adsabs.harvard.edu/abs/1972A%26A....20..263D} {20, 263}

\makeatother
\end{thebibliography}
\normalsize

\clearpage

\appendix

\section{The importance of shielding}\label{gaso2krome_isogal:sec:The_importance_of_shielding}

\begin{figure}
\centering
\vspace{-6.0pt}
\includegraphics[width=1.12\columnwidth,angle=0]{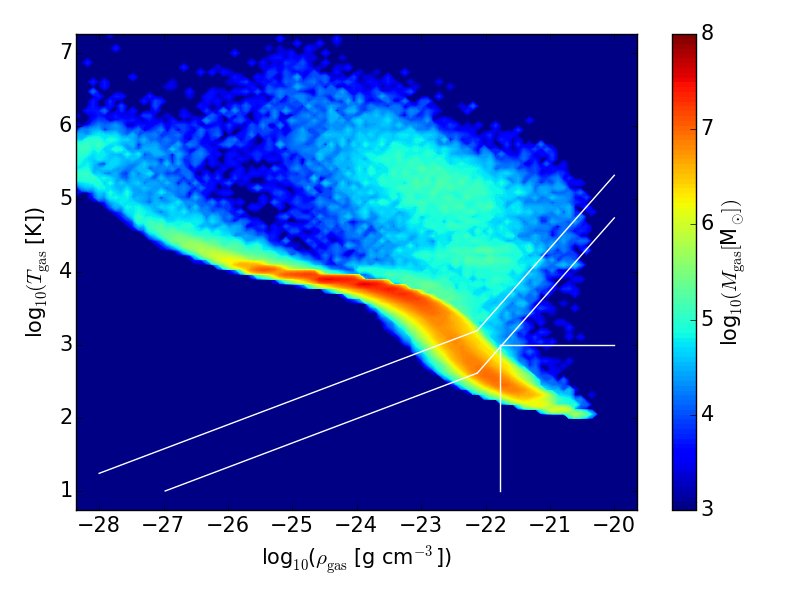}
\includegraphics[width=1.12\columnwidth,angle=0]{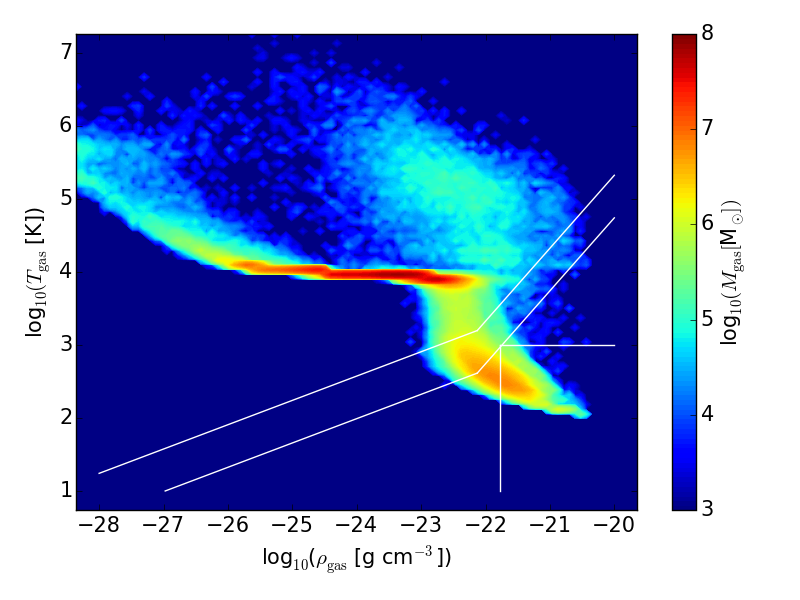}
\vspace{-6.0pt}
\caption[]{Dependence on the modelling of metals, when assuming no shielding. Gas $\rho$--$T$ diagrams at 0.4~Gyr, for the eq-metals Run~01 (upper panel) and the non-eq-metals Run~01m (lower panel) with no shielding. The lines are the same described in Fig.~\ref{gaso2krome_isogal:fig:rhoT_run01vs01m}. The differences between the upper and lower diagrams are negligible, when compared to the discrepancies between the panels of Fig.~\ref{gaso2krome_isogal:fig:rhoT_run01vs01m}.
}
\label{gaso2krome_isogal:fig:rhoT_run01vs01m_noshield}
\end{figure}

In this Section, we show the effect of not including any shielding from the photoionizing UV background in Runs~01 and 01m. In Fig.~\ref{gaso2krome_isogal:fig:rhoT_run01vs01m_noshield}, we show the gas $\rho$--$T$ diagrams for the eq-metals Run~01 (upper panel) and the non-eq-metals Run~01m (lower panel) with no shielding. The two diagrams are very similar to each other, except mostly for the density range $10^{-23} \lesssim \rho_{\rm gas} \lesssim 10^{-22}$~g~cm$^{-3}$, in which the non-eq-metals is slightly warmer than the eq-metals gas. This implies that the major differences observed between the two panels of Fig.~\ref{gaso2krome_isogal:fig:rhoT_run01vs01m} are likely due to the implementation of shielding, with the minor remaining differences due to a combination of different metal networks and non-equilibrium effects (see the discussion in Section~\ref{gaso2krome_isogal:sec:eqmetvsnoneqmet}).

\section{Reaction rates employed in the present work}\label{gaso2krome_isogal:sec:Reactions}

The chemical network used in this work is very similar to that described in \citeauthor{Bovino_et_al_2016} (\citeyear{Bovino_et_al_2016}; a slightly updated version of that network, named \textit{react\_galaxy\_ism}, is publicly available with the \krome package under the folder \textsc{krome/networks}). However, since the publication of \citet{Bovino_et_al_2016}, a few typos were found and there have been a number of changes, mostly involving the temperature limits of validity of the reactions (we note that, in our simulations, the gas mass fraction of particles with $T_{\rm gas} > 10^7$~K is $\lesssim$ 0.01 per cent and there are zero particles with $T_{\rm gas} > 10^8$~K). Therefore, for definiteness, we list the reaction rates in Tables~\ref{tab:rates}--\ref{tab:photo} and describe the changes from \citet{Bovino_et_al_2016} in the notes. In this section only, we use the usual chemical notation: H = HI; H$^+$ = HII, H$_2$ = H2I, H$_2^+$ = H2II, He = HeI, He$^+$ = HeII, He$^{++}$ = HeIII, C = CI, C$^+$ = CII, O = OI, O$^+$ = OII, Si = SiI, Si$^+$ = SiII, and Si$^{++}$ = SiIII.

	\begin{table*}
	\caption{List of reactions and rates included in our chemical network.}\label{tab:rates}
	\def\arraystretch{1.15}
        \begin{tabular}{@{}lllc}
        	\hline\hline
	Reaction & Rate coefficient (cm$^3$ s$^{-1}$) & temp. range & Ref. \\
	\hline
	
	1\dots\dots ~~H + e$^-$ $\rightarrow$ H$^+$ + 2e$^-$  & $k_1$ = exp[-32.713 967 86+13.536 556 0 ln $T_{\rm e}$ & & 1 \\
	& $\phantom{k_{1}=}$ - 5.739 328 75 (ln $T_{\rm e}$)$^2$+1.563 154 98 (ln $T_{\rm e}$)$^3$ & & \\
	& $\phantom{k_{1}=}$ - 0.287 705 60 (ln $T_{\rm e}$)$^4$+3.482 559 77 $\times$ 10$^{-2}$(ln $T_{\rm e}$)$^5$ & & \\
	& $\phantom{k_{1}=}$ - 2.631 976 17 $\times$ 10$^{-3}$(ln $T_{\rm e}$)$^6$+1.119 543 95 $\times$ 10$^{-4}$(ln $T_{\rm e}$)$^7$ & \\
	& $\phantom{k_{1}=}$ - 2.039 149 85 $\times$ 10$^{-6}$(ln $T_{\rm e}$)$^8$] & \\
        
	2\dots\dots ~~H$^+$ + e$^-$ $\rightarrow$ H  + $\gamma$ & $k_2$ = 3.92 $\times$ 10$^{-13}$ $T_{\rm e}$ $^{-0.6353}$ & $T \le 5500$ K & 2 \\
	& $\phantom{k_{2}}$ = $\exp$[-28.613 033 806 892 32 & $T > 5500$ K & \\
	& $\phantom{k_{2}=}$ - 7.241 125 657 826 851 $\times$ 10$^{-1}$ ln $T_{\rm e}$ \\
	& $\phantom{k_{2}=}$ - 2.026 044 731 984 691 $\times$ 10$^{-2}$ (ln $T_{\rm e}$)$^2$ \\
	& $\phantom{k_{2}=}$ - 2.380 861 877 349 834 $\times$ 10$^{-3}$ (ln $T_{\rm e}$)$^3$ \\
	& $\phantom{k_{2}=}$ - 3.212 605 213 188 796 $\times$ 10$^{-4}$ (ln $T_{\rm e}$)$^4$ \\
	& $\phantom{k_{2}=}$ - 1.421 502 914 054 107 $\times$ 10$^{-5}$ (ln $T_{\rm e}$)$^5$ \\
	& $\phantom{k_{2}=}$ + 4.989 108 920 299 513  $\times$ 10$^{-6}$ (ln $T_{\rm e}$)$^6$ \\
	& $\phantom{k_{2}=}$ + 5.755 614 137 575 758  $\times$ 10$^{-7}$ (ln $T_{\rm e}$)$^7$ \\
	& $\phantom{k_{2}=}$ - 1.856 767 039 775 261  $\times$ 10$^{-8}$  (ln $T_{\rm e}$)$^8$ \\
	& $\phantom{k_{2}=}$ - 3.071 135 243 196 595  $\times$ 10$^{-9}$  (ln $T_{\rm e}$)$^9$] \\

	3\dots\dots ~~He + e$^-$ $\rightarrow$ He$^+$ + 2e$^-$  & $k_3$ = $\exp$[-44.098 648 86 + 23.915 965 63 ln $T_{\rm e}$ & & 1 \\
	& $\phantom{k_{3}=}$ - 10.753 230 2 (ln $T_{\rm e}$)$^2$ \\
	& $\phantom{k_{3}=}$ + 3.058 038 75 (ln $T_{\rm e}$)$^3$ \\
	& $\phantom{k_{3}=}$ - 5.685 118 9 $\times$ 10$^{-1}$ (ln $T_{\rm e}$)$^4$ \\
	& $\phantom{k_{3}=}$ + 6.795 391 23 $\times$ 10$^{-2}$ (ln $T_{\rm e}$)$^5$ \\
	& $\phantom{k_{3}=}$ - 5.009 056 10 $\times$ 10$^{-3}$ (ln $T_{\rm e}$)$^6$ \\
	& $\phantom{k_{3}=}$ + 2.067 236 16 $\times$ 10$^{-4}$ (ln $T_{\rm e}$)$^7$ \\
	& $\phantom{k_{3}=}$ - 3.649 161 41 $\times$ 10$^{-6}$ (ln $T_{\rm e}$)$^8$]  & \\
	
	4\dots\dots ~~He$^+$ + e$^-$ $\rightarrow$ He + $\gamma$ & $k_{4}$ =  3.92 $\times$ 10$^{-13}$ $T_{\rm e}^{-0.6353}$ &  $T_{\rm e} \le 0.8$ eV & 3 \\
	& $\phantom{k_{4}}$ =  3.92 $\times$ 10$^{-13}$ $T_{\rm e}^{-0.6353}$ & $T_{\rm e} > 0.8$ eV \\
	& $\phantom{k_{4}=}$ + 1.54 $\times$ 10$^{-9}$ $T_{\rm e}^{-1.5}$ [1.0 + 0.3 / $\exp$(8.099 328 789 667/$T_{\rm e}$)] \\
	& $\phantom{k_{4}=}$ /[$\exp$(40.496 643 948 336 62/$T_{\rm e}$)] \\
	
	5\dots\dots ~~He$^+$ + e$^-$ $\rightarrow$ He$^{++}$ + 2e$^-$ & $k_5$ = $\exp$[-68.710 409 902 120 01 + 43.933 476 326 35 ln $T_{\rm e}$ &  & 4 \\
	& $\phantom{k_{5}=}$ - 18.480 669 935 68 (ln $T_{\rm e}$)$^2$ \\
	& $\phantom{k_{5}=}$ + 4.701 626 486 759 002 (ln $T_{\rm e}$)$^3$ \\
	& $\phantom{k_{5}=}$ - 7.692 466 334 492 $\times$ 10$^{-1}$ (ln $T_{\rm e}$)$^4$ \\
	& $\phantom{k_{5}=}$ + 8.113 042 097 303 $\times$ 10$^{-2}$ (ln $T_{\rm e}$)$^5$ \\
	& $\phantom{k_{5}=}$ - 5.324 020 628 287 001 $\times$ 10$^{-3}$ (ln $T_{\rm e}$)$^6$ \\
	& $\phantom{k_{5}=}$ + 1.975 705 312 221 $\times$ 10$^{-4}$ (ln $T_{\rm e}$)$^7$ \\
	& $\phantom{k_{5}=}$ - 3.165 581 065 665 $\times$ 10$^{-6}$ (ln $T_{\rm e}$)$^8$] & \\
	
	6\dots\dots ~~He$^{++}$ + e$^-$ $\rightarrow$ He$^+$ + $\gamma$ & $k_6$ = 1.891 $\times$ 10$^{-10}$ [(1.0+$\sqrt{T/9.37})^{0.2476}$ & & 5 \\
	& $\phantom{k_{6}=}$ $\times$ (1.0+$\sqrt{T/(2.774\times 10^6)})^{1.7524}$ ($\sqrt{T/9.37})]^{-1}$ & \\
	
	7\dots\dots ~~H + e$^-$ $\rightarrow$ H$^-$ + $\gamma$ & $k_7$ = $1.4 \times 10^{-18} T^{0.928}\exp(-T/16200)$ & & 6 \\
	
	8\dots\dots ~~H$^-$ + H $\rightarrow$ H$_2$ + e$^-$ & $k_8$ = $a_1(T^{a_2}+a_3 T^{a_4}+a_5T^{a_6})/(1.0+a_7T^{a_8}+a_9T^{a_{10}}+a_{11}T^{a_{12}})$ & & 7 \\
	& $a_1 = 1.35 \times 10^{-9}$ \\
	& $a_2 = 9.8493 \times 10^{-2}$ \\                                                                       
	& $a_3 = 3.2852 \times 10^{-1}$ \\                                                                      
	& $a_4 = 5.5610 \times 10^{-1}$ \\                                                                       
	& $a_5 = 2.7710 \times 10^{-7}$ \\                                                                       
	& $a_6 = 2.1826$ \\                                            
	& $a_7 = 6.1910 \times 10^{-3}$ \\                                                                       
	& $a_8 = 1.0461$ \\                                                                       
	& $a_9 = 8.9712 \times 10^{-11}$ \\                                                                       
	& $a_{10} = 3.0424$ \\                                                                      
	& $a_{11} = 3.2576 \times 10^{-14}$ \\                                                                      
	& $a_{12} = 3.7741$ \\
	
	\hline
	\multicolumn{4}{p{\linewidth}}{
        References -- 1: \citet{Janev_et_al_1987}; 2:  \citet{Abel_et_al_1997} fit by data from \citet{Ferland_et_al_1992}; 3: \citet{Cen_1992,Aldrovandi_Pequignot_1973}; 4: Aladdin database \url{https://www-amdis.iaea.org/ALADDIN/}, see \citet{Abel_et_al_1997}; 5: \citet{Verner_Ferland_1996}; 6: \citet{deJong_1972}; 7: \citet{Kreckel_et_al_2010}. Notes -- In this and the next tables, $T = T_{\rm gas}$; $T_{\rm e} = k_{\rm B}T$, where $k_{\rm B}$ is the Boltzmann constant, is the gas temperature in eV. Reaction~6 and one coefficient in Reaction~8 were correct in the code used by \citet{Bovino_et_al_2016} but incorrect in the published paper.
      	}
	\end{tabular}
	\end{table*}
         
	\begin{table*}
	\contcaption{List of reactions and rates included in our chemical network.}
	\def\arraystretch{1.15}
	\begin{tabular}{@{}lllc}
	\hline\hline
	Reaction & Rate coefficient (cm$^3$ s$^{-1}$) & temp. range & Ref. \\
	\hline
	
	9\dots\dots ~~ H + H$^+$ $\rightarrow$ H$_2^+$ + $\gamma$ & $k_9$ =  $2.10\times 10^{-20}(T/30.)^{-0.15}$ & $T < 30 $ K &  8 \\
	& $\phantom{k_{9}}$ = dex$[-18.20-3.194\log_{10}T+1.786(\log_{10} T)^{2}-0.2072(\log_{10}T)^3]$ & $T \geq 30$ K \\
        
	10\dots\dots ~~H$_2^+$ + H $\rightarrow$ H$_2$ + H$^+$ & $k_{10}$ = 6.0 $\times$ 10$^{-10}$ & & 9 \\
        
	11\dots\dots ~~H$_2$ + H$^+$ $\rightarrow$ H$_2^+$ + H & $k_{11}$ = dex$[\sum_{i=0}^{7} a_i [\log_{10}(T)]^i] $ \;\;\;\;\;\;\;\;\;\;\;\;\;\;(see Table~\ref{fitkrstic}) & $T \leq 10^8$ K & 10 \\
        
	12\dots\dots ~~H$^-$ + e$^-$ $\rightarrow$ H + 2e$^-$  & $k_{12}$ = $\exp$[-18.018 493 342 73 & & 1 \\
	& $\phantom{k_{12}=}$ + 2.360 852 208 681 ln $T_{\rm e}$ \\
	& $\phantom{k_{12}=}$ - 2.827 443 061 704 $\times$ 10$^{-1}$ (ln $T_{\rm e}$)$^2$ \\
	& $\phantom{k_{12}=}$ + 1.623 316 639 567 $\times$ 10$^{-2}$ (ln $T_{\rm e}$)$^3$ \\
	& $\phantom{k_{12}=}$ - 3.365 012 031 362 999 $\times$ 10$^{-2}$ (ln $T_{\rm e}$)$^4$ \\
	& $\phantom{k_{12}=}$ + 1.178 329 782 711  $\times$ 10$^{-2}$ (ln $T_{\rm e}$)$^5$ \\
	& $\phantom{k_{12}=}$ - 1.656 194 699 504  $\times$ 10$^{-3}$ (ln $T_{\rm e}$)$^6$ \\
	& $\phantom{k_{12}=}$ + 1.068 275 202 678  $\times$ 10$^{-4}$ (ln $T_{\rm e}$)$^7$ \\
	& $\phantom{k_{12}=}$ - 2.631 285 809 207  $\times$ 10$^{-6}$ (ln $T_{\rm e}$)$^8$]& \\
	
	13\dots\dots ~~H$^-$ + H $\rightarrow$ 2H +  e$^-$ & $k_{13}$ = 2.56 $\times$ 10$^{-9}$ $T_{\rm e}^{1.781 86}$ & $T_{\rm e} \le 0.1$ eV & 11 \\
	& $\phantom{k_{13}}$ =  $\exp$[-20.372 608 965 333 24 & $T_{\rm e} > 0.1$ eV \\
	& $\phantom{k_{13}=}$ + 1.139 449 335 841 631 ln $T_{\rm e}$ \\
	& $\phantom{k_{13}=}$ - 1.421 013 521 554 148 $\times$ 10$^{-1}$ (ln $T_{\rm e}$)$^2$ \\
	& $\phantom{k_{13}=}$ + 8.464 455 386 63 $\times$ 10$^{-3}$ (ln $T_{\rm e}$)$^3$ \\
	& $\phantom{k_{13}=}$ - 1.432 764 121 299 2 $\times$ 10$^{-3}$ (ln $T_{\rm e}$)$^4$ \\
	& $\phantom{k_{13}=}$ +2.012 250 284 791 $\times$ 10$^{-4}$ (ln $T_{\rm e}$)$^5$ \\
	& $\phantom{k_{13}=}$ + 8.663 963 243 09 $\times$ 10$^{-5}$ (ln $T_{\rm e}$)$^6$ \\
	& $\phantom{k_{13}=}$ - 2.585 009 680 264 $\times$ 10$^{-5}$ (ln $T_{\rm e}$)$^7$ \\
	& $\phantom{k_{13}=}$ + 2.455 501 197 039 2 $\times$ 10$^{-6}$ (ln $T_{\rm e}$)$^8$ \\
	& $\phantom{k_{13}=}$ - 8.068 382 461 18 $\times$ 10$^{-8}$ (ln $T_{\rm e}$)$^9$] \\
        
	14\dots\dots ~~H$^-$ + H$^+$ $\rightarrow$ 2H + $\gamma$ & $k_{14}$ = $2.96 \times 10^{-6}/\sqrt{T}-1.73\times10^{-9}+2.50\times 10^{-10}\sqrt{T}$ & $T > 10$~K & 12 \\
	&$\phantom{k_{14}=}$ - $7.77\times10^{-13}$  \\
        
	15\dots\dots ~~H$^-$ + H$^+$ $\rightarrow$ H$_2^+$ + e$^-$ & $k_{15}$ = 10$^{-8} \times T^{-0.4}$& & 13 \\
        
	16\dots\dots ~~H$_2^+$ + e$^-$ $\rightarrow$ 2H + $\gamma$ & $ k_{16}$ = $1.0\times 10^{-8}$ & $T \le 617$ K & 14 \\
 	&$\phantom{k_{16}}$ = $1.32\times 10^{-6}T^{-0.76}$ & $T > 617$ K \\
 
	17\dots\dots ~~H$_2^+$ + H$^-$ $\rightarrow$ H + H$_2$ & $k_{17}$ = 5.0 $\times$ 10$^{-7}$ (10$^2 / T$)$^{0.5}$ & & 15 \\
        
	18\dots\dots ~~H$_2$ + H $\rightarrow$ 3H & $k_{18} = 6.67\times 10^{-12}\sqrt{T}\exp[-(1+63 593/T)]$  & & 16 \\
        
	19\dots\dots ~~H$_2$ + H$_2$ $\rightarrow$ H$_2$ + H + H & $k_{19} = 5.996\times 10^{-30}T^{4.1881}(1+6.761\times 10^{-6}T)^{-5.6881}$ & & 16 \\
	&$\phantom{k_{19}=}$ $\times \exp(-54 657.4/T)$ \\
  
	20\dots\dots ~~He$^+$ + H $\rightarrow$ He + H$^+$ & $k_{20} = 1.20\times10^{-15}(T/300)^{0.25}$ & & 17 \\

	21\dots\dots ~~He + H$^+$ $\rightarrow$ He$^+$ + H & $k_{21} = 1.26\times10^{-9}T^{-0.75}\exp(-1.275\times10^5/T)$ & $T \leq 10^4$ K & 18 \\
	& $\phantom{k_{21}}$ = $4\times10^{-37}T^{4.74}$ & $T > 10^4$ K \\

	22\dots\dots ~~H$_2$ + e$^-$ $\rightarrow$ H + H + e$^-$ & $k_{22} = 4.38 \times10^{-10}T^{0.35}\exp(-102 000/T)$ & & 19 \\

	23\dots\dots ~~H$_2$ + e$^-$ $\rightarrow$ H + H$^-$ & $k_{23} = 35.5\times T^{-2.28} \exp(-46 707/T)$ & & 20 \\

	24\dots\dots ~~H$_2$ + He $\rightarrow$ H + H + He & $k_{24} =$ dex$[-27.029 + 3.801\log_{10}(T)-29 487/T]$ & & 21 \\

	25\dots\dots ~~H$_2$ + He$^+$ $\rightarrow$ He + H + H$^+$ & $k_{25} = 3.7\times 10^{-14}\,\exp(-35.0/T)$ & & 22 \\

	26\dots\dots ~~H$_2$ + He$^+$ $\rightarrow$ H$_2^+$ + He & $k_{26} = 7.2\times 10^{-15}$ & & 22 \\

	27\dots\dots ~~H + H $\rightarrow$ H + H$^+$ + e$^-$ & $k_{27} = 1.2\times 10^{-17}T^{1.2}\exp(-157 800/T)$ & & 24 \\

	28\dots\dots ~~H + He $\rightarrow$ He + H$^+$ + e$^-$ & $k_{28} = 1.75\times 10^{-17}T^{1.3}\exp(-157 800/T)$ & & 24 \\

	29\dots\dots ~~H$^+$ + e$^-$ $\xrightarrow{\rm dust}$ H & $k_{29} = 1.225\times 10^{-13}(Z/Z_\odot) [1.0+8.074\times 10^{-6}\psi^{1.378}$ & & 25 \\
	& $\phantom{k_{29}=} \times (1+5.087\times 10^2 T^{0.015 86}\psi^{-0.4723-1.102\times 10^{-5}\ln(T)}]^{-1}$ & & \\
 
	30\dots\dots ~~He$^+$ + e$^-$ $\xrightarrow{\rm dust}$ He & $k_{30} = 5.572\times 10^{-14}(Z/Z_\odot) [1.0+6.089\times 10^{-3}\psi^{1.1728}$ & & 25 \\
	& $\phantom{k_{30}=} \times (1+4.331\times 10^2 T^{0.04 845}\psi^{-0.8120-1.333\times 10^{-4}\ln(T)}]^{-1}$ & & \\

	31\dots\dots ~~H + H $\xrightarrow{\rm dust}$ H$_2$ & see text & & 26 \\
        
	\hline
	\multicolumn{4}{p{\linewidth}}{
        References -- 8: \citet{Coppola_et_al_2011}; 9 \citet{Karpas_et_al_1979}; 10: \citet{Savin_et_al_2004} and \citet{Grassi_et_al_2011}, see table \ref{fitkrstic} for the coefficients; 11: \citet{Abel_et_al_1997}, based on \citet{Janev_et_al_1987}; 12: \citet{Stenrup_et_al_2009}; 13: \citet{Poulaert_et_al_1978}; 14: \citet{Abel_et_al_1997},  fit from data by \citet{Schneider_et_al_1994}; 15: \citet{Dalgarno_Lepp_1987}; 16: \citet{Glover_Abel_2008}; 17: \citet{Yoshida_et_al_2006}; 18: \citet{Kimura_et_al_1993}; 19: \citet{Mitchell_Deveau_1983}, fit of data by \citet{Corrigan_1965}; 20: \citet{Capitelli_et_al_2007}; 21: \citet{Dove_et_al_1987}; 22: \citet{Barlow_1984}; 24: \citet{Lenzuni_et_al_1991}; 25: \citet{Weingartner_Draine_2001b}; 26: \citet{Cazaux_Spaans_2009}, see also \citet{Grassi_et_al_2014}. Notes -- The parameter $\psi = G_0 T^{1/2}/n_{\rm e}$, where $G_0$ is the Habing flux and $n_{\rm e}$ is the electron number density, is the charging parameter. Reaction~17 and the temperature ranges of Reaction~13 were correct in the code used by \citet{Bovino_et_al_2016} but incorrect in the published paper. Reactions~14 and 25 and the temperature ranges of Reactions~11 (see also Table~\ref{fitkrstic}) and 14 were changed from the code used by \citet{Bovino_et_al_2016}.
      	}
	\end{tabular}
	
	\end{table*}

\clearpage

	\begin{table*}
        \contcaption{List of reactions and rates included in our chemical network.}
        \def\arraystretch{1.15}
        \begin{tabular}{@{}lllc}
	\hline\hline
	Reaction & Rate coefficient (cm$^3$ s$^{-1}$) & temp. range & Ref. \\
	\hline

	32\dots\dots ~~$\cp + \me^- \rightarrow \mC  + \gamma$ & $k_{32} = 4.67 \times 10^{-12}  \left(\frac{T}{300}\right)^{-0.6}$ & $T \le 7950 \: {\rm K}$ & 27 \\
	& $\phantom{k_{32} } =1.23 \times 10^{-17}  \left(\frac{T}{300}\right)^{2.49}  \exp \left(\frac{21 845.6}{T} \right)$ & $ 7950 < T \le 21 140 \: {\rm K}$ & \\
	& $\phantom{k_{32}} = 9.62 \times 10^{-8} \left(\frac{T}{300}\right)^{-1.37} \exp \left(\frac{-115 786.2}{T} \right)$ & $T > 21 140 \: {\rm K}$ & \\
	
	33\dots\dots ~~$\sip + \me^- \rightarrow \mSi + \gamma$ & $k_{33} =  7.5 \times 10^{-12} \left(\frac{T}{300}\right)^{-0.55}$  & $T \le 2000 \: {\rm K}$ & 28 \\
	& $\phantom{k_{33}}= 4.86 \times 10^{-12} \left(\frac{T}{300}\right)^{-0.32}$ & $2000 < T \le 10^{4} \: {\rm K}$ & \\
	& $\phantom{k_{33}}= 9.08 \times 10^{-14} \left(\frac{T}{300}\right)^{0.818}$ & $T > 10^{4} \: {\rm K}$ & \\

	34\dots\dots ~~$\op + \me^-  \rightarrow \mO + \gamma$ & $k_{34} = 1.30 \times 10^{-10} T^{-0.64}$ &  $T \le 400 \: {\rm K}$ & 29 \\
	& $\phantom{k_{34}} = 1.41 \times 10^{-10} T^{-0.66} + 7.4 \times 10^{-4}  T^{-1.5}$ & \\
	& $\phantom{k_{34}=} \mbox{} \times \exp \left(-\frac{175 000}{T}\right) [1.0 + 0.062 \times \exp \left(-\frac{145 000}{T}\right) ]$ & $T > 400 \: {\rm K}$ & \\

	35\dots\dots ~~$\mC  + \me^-  \rightarrow \cp  + 2\me^- $ & $k_{35} = 6.85 \times 10^{-8} (0.193 + u)^{-1} u^{0.25} e^{-u}$ & $u = 11.26 / T_{\rm e}$ & 30 \\

	36\dots\dots ~~$\mathrm{Si} + \me^-  \rightarrow \sip + 2\me^-$ & $k_{36} = 1.88 \times 10^{-7} (1.0 + u^{0.5}) (0.376 + u)^{-1} u^{0.25} e^{-u}$ & $ u = 8.2 / T_{\rm e}$ & 30 \\

	37\dots\dots ~~$\mO  + \me^-  \rightarrow \op  + 2\me^-$ & $k_{37} = 3.59 \times 10^{-8} (0.073 + u)^{-1} u^{0.34} e^{-u}$ & $u = 13.6 / T_{\rm e}$ & 30 \\

	38\dots\dots ~~$\op  + \mH  \rightarrow \mO  + \Hp$ & $ k_{38} = 4.99 \times 10^{-11} T^{0.405} + 7.54 \times 10^{-10} T^{-0.458} $ & & 31 \\

	39\dots\dots ~~$\mO  + \Hp  \rightarrow \op  + \mH$ & $k_{39} = [1.08 \times 10^{-11} T^{0.517}+ 4.00 \times 10^{-10} T^{0.006 69}] \exp  \left(-\frac{227}{T}\right) $ & & 32 \\

	40\dots\dots ~~$\mO + \Hep \rightarrow \op + \He$ & $k_{40} = 4.991 \times 10^{-15} \left(\frac{T}{10000}\right)^{0.3794} \expf{-}{T}{112 100 0}$ & $T > 10$~K & 33 \\
	& $\phantom{k_{40} = } \mbox{} + 2.780 \times 10^{-15} \left(\frac{T}{10000}\right)^{-0.2163} \expf{}{T}{815 800}$ & \\

	41\dots\dots ~~$\mC  + \Hp  \rightarrow \cp  + \mH$ & $k_{41} = 3.9 \times 10^{-16} T^{0.213}$ & & 32 \\

	42\dots\dots ~~$\cp  + \mH  \rightarrow \mC  + \Hp$ & $k_{42} = 6.08 \times 10^{-14}  \left(\frac{T}{10000}\right)^{1.96} \expf{-}{170 000}{T}$ & & 32 \\

	43\dots\dots ~~$\mC + \Hep \rightarrow \cp + \He$ & $k_{43} = 8.58 \times 10^{-17}  T^{0.757}$ & $T \leq 200 \: {\rm K}$ & 34 \\
	& $\phantom{k_{43}} = 3.25 \times 10^{-17} T^{0.968}$ & $200 < T \leq 2000 \: {\rm K}$ & \\
	& $\phantom{k_{43}} = 2.77 \times 10^{-19} T^{1.597}$ & $T > 2000 \: {\rm K}$ & \\

	44\dots\dots ~~$\mathrm{Si} + \Hp  \rightarrow \sip + \mH$ & $k_{44} = 5.88 \times 10^{-13} T^{0.848}$ & $T \le 10^{4} \: {\rm K}$ & 35 \\
	& $\phantom{k_{44}} = 1.45 \times 10^{-13} T$ & $T > 10^{4} \: {\rm K}$ & \\

	45\dots\dots ~~$\mathrm{Si} + \Hep \rightarrow \sip + \He$ & $k_{45} = 3.3 \times 10^{-9}$ & & 36 \\

	46\dots\dots ~~$\cp  + \mSi \rightarrow \mC + \sip$ & $k_{46} = 2.1 \times 10^{-9}$ & & 36 \\

	47\dots\dots ~~$\sip + \Hp \rightarrow \sipp + \mH$ & $k_{47} = 4.10 \times 10^{-10} \left( \frac{T}{10000} \right)^{0.24}$ & & 35 \\
	& $\phantom{k_{47} = } \mbox{} \times \left[1.0 + 3.17 \,\expf{}{-T}{2.39 \times 10^{6}} \right] \expf{-}{3.178}{T_{\rm e}}$ & & \\

	48\dots\dots ~~$\sipp + \mH \rightarrow \sip + \Hp$ & $k_{48} = 1.23 \times 10^{-9} \left(\frac{T}{10000}\right)^{0.24}$ & & 35 \\ 
	& $\phantom{k_{48} =} \mbox{} \times \left[1.0 + 3.17 \,\expf{}{-T}{2.39 \times 10^{6}} \right] $ & & \\ 

	49\dots\dots ~~$\sipp + \me^- \rightarrow \sip + \gamma$ & $k_{49} = 1.75 \times 10^{-12} \left( \frac{T}{10000} \right)^{-0.6346}$ & & 37 \\ 

	50\dots\dots ~~$\cp + \me^- \xrightarrow{dust} \mC$ & $k_{50} = 4.558 \times 10^{-13} (Z/Z_{\odot}) [1.0 + 6.089 \times 10^{-3} \psi^{1.128} $ & & 25 \\
	& $\phantom{k_{50} =} \times (1.0 + 4.331 \times 10^{2} T^{0.048 45} \psi^{-0.8120 - 1.333 \times 10^{-4}\ln T})]^{-1}$ & \\

	51\dots\dots ~~$\op + \me^- \xrightarrow{dust} \mO$ & $k_{51} = \frac{1}{4} k_{29}$ & & 25,38 \\

	52\dots\dots ~~$\sip + \me^- \xrightarrow{dust} \mSi$ & $k_{52} = 2.166 \times 10^{-14} (Z/Z_{\odot}) [1.0 + 5.678 \times 10^{-8} \psi^{1.874} $ & & 25 \\
	& $\phantom{k_{51}=} \times (1.0 + 4.375 \times 10^{4} T^{1.635\times10^{-6}} \psi^{-0.8964 - 7.538 \times 10^{-5}\ln T})]^{-1}$ & \\
	
	\hline
	\multicolumn{4}{p{\linewidth}}{
	References -- 27: \citet{Nahar_Pradhan_1997}; 28: \citet{Nahar_2000}; 29: \citet{Nahar_1999}; 30: \citet{Voronov_1997}; 31: \citet{Stancil_et_al_1999}; 32: \citet{Stancil_et_al_1998}; 33: \citet{Zhao_et_al_2004}; 34: \citet{Kimura_et_al_1993}; 35: \citet{Kingdon_Ferland_1996}; 36: \citet{LeTeuff_et_al_2000}; 37: \citet{Nahar_1995,Nahar_1996}; 38: \citet{Glover_Jappsen_2007}. Notes -- Reactions~47 and 48 and the temperature ranges of Reactions ~40, 47, and 48 were changed from the code used by \citet{Bovino_et_al_2016}. The functions and temperature ranges given for Reactions~40, 47, and 48 are slightly different from what given in the references.
      	} 
	\end{tabular}

	\end{table*}

\clearpage

\begin{table*}
\begin{center}
\caption{Fitting coefficients for Reaction~11 (see Table~\ref{tab:rates}). Coefficients are in the form $a(b)=a\times 10^b$.} \label{fitkrstic}
\vspace{1mm}
\begin{tabular*}{.35\textwidth}{c |r@{.}l r@{.}l}
\hline
$a_i$ &\multicolumn{2}{c}{$T<10^5\ \mathrm K$} &\multicolumn{2}{c}{$10^5\le T \le 10^8\ \mathrm K$} \T \B \\
\hline
$a_0$ & $-1$&$9153214(+2)$	& $-8$&$8755774(+3)$	\T \B \\
$a_1$ & $ 4$&$0129114(+2)$	& $ 1$&$0081246(+4)$	\T \B \\
$a_2$ & $-3$&$7446991(+2)$	& $-4$&$8606622(+3)$	\T \B \\
$a_3$ & $ 1$&$9078410(+2)$	& $ 1$&$2889659(+3)$	\T \B \\
$a_4$ & $-5$&$7263467(+1)$	& $-2$&$0319575(+2)$	\T \B \\
$a_5$ & $ 1$&$0133210(+1)$	& $ 1$&$9057493(+1)$	\T \B \\
$a_6$ & $-9$&$8012853(-1)$	& $-9$&$8530668(-1)$	\T \B \\
$a_7$ & $ 4$&$0023414(-2)$	& $ 2$&$1675387(-2)$	\T \B \\
\hline
\end{tabular*}
\end{center}
\end{table*}

	\begin{table*}
        \caption{List of photo-processes included in our chemical network.}\label{tab:photo}
        \begin{tabular}{@{}lllc}
	\hline\hline
	Reaction & Cross-section (cm$^2$) & energy range (eV) & Ref.\T \B \\
	\hline
	P1\dots\dots ~~H + $\gamma$ $\rightarrow$ H$^+$ + e$^-$  & \krome database   & & 1 \T \B \\
	P2\dots\dots ~~He + $\gamma$ $\rightarrow$ He$^+$ + e$^-$  & \krome database   & & 1 \T \B \\
	P3\dots\dots ~~He$^+$ + $\gamma$ $\rightarrow$ He$^{++}$ + e$^-$  & \krome database   & & 1 \T \B \\
	P4\dots\dots ~~H$^-$ + $\gamma$ $\rightarrow$ H + e$^-$  & SWRI database   & $E > 0.755$ & 2 \T \B \\
	P5\dots\dots ~~H$_2$ + $\gamma$ $\rightarrow$ H$_2^+$ + e$^-$  & SWRI database   & $E > 15.4$ & 2 \T \B \\
	P6\dots\dots ~~H$_2^+$ + $\gamma$ $\rightarrow$ H$^+$ + H  & LEIDEN database   & $E > 2.65$ & 3 \T \B \\
	P7\dots\dots ~~H$_2^+$ + $\gamma$ $\rightarrow$ H$^+$ + H$^+$ + e$^-$  &   dex$[-16.926-4.528\times 10^{-2} E+2.238\times 10^{-4} E^2+4.245\times10^{-7} E^3]$  & $30 < E < 90$ & 4 \T \B \\
	P8\dots\dots ~~H$_2$ + $\gamma$ $\rightarrow$ H + H  & Direct & $14.159 < E < 17.7$ & 5 \T \B \\
	P9\dots\dots ~~H$_2$ + $\gamma$ $\rightarrow$ H$_2^*$ $\rightarrow$ H + H  & Solomon &  & 6 \T \B \\
	P10\dots\dots ~~C + $\gamma$ $\rightarrow$ C$^+$ + e$^-$  & \krome database  &  & 1 \T \B \\
	P11\dots\dots ~~O + $\gamma$ $\rightarrow$ O$^+$ + e$^-$  & \krome database  &  & 1 \T \B \\
	P12\dots\dots ~~Si + $\gamma$ $\rightarrow$ Si$^+$ + e$^-$  & \krome database  &  & 1 \T \B \\
	P13\dots\dots ~~Si$^+$ + $\gamma$ $\rightarrow$ Si$^{++}$ + e$^-$  & \krome database  &  & 1 \T \B \\     
	\hline
		\multicolumn{4}{p{\linewidth}}{
	References -- 1: \citet{Verner_Ferland_1996,Verner_et_al_1996}; 2: \url{http://phidrates.space.swri.edu}; 3: \url{http://home.strw.leidenuniv.nl/~ewine/photo/}; 4: \citet{Shapiro_Kang_1987}; 5: \citet{Abel_et_al_1997}; 6: this work, based on the formula reported by \citet{Glover_Jappsen_2007}.
      	} 
	\end{tabular}

	\end{table*}

\label{lastpage}
\end{document}